\documentclass[11pt]{amsart}

\usepackage{amsmath}
\usepackage{amsfonts}
\usepackage{amssymb}
\usepackage{amscd}
\usepackage{amsthm}
\usepackage{framed}
\usepackage{fullpage}
\usepackage{graphicx}
\usepackage{latexsym}
\usepackage[numbers]{natbib}  
\usepackage{multirow}
\usepackage{tikz}
\usetikzlibrary{positioning}
\usetikzlibrary{arrows}

\allowdisplaybreaks[4]

\usepackage{hyperref}
\hypersetup{colorlinks,linkcolor={red},citecolor={blue},urlcolor={blue}}

\newtheorem{theorem}{Theorem}[section]

\newtheorem{conj}[theorem]{Conjecture}

\theoremstyle{definition}

\numberwithin{equation}{section}

\newcommand{\CC}{\mathbb C}
\newcommand{\AAA}{\mathbb A}
\newcommand{\HH}{\mathbb H}
\newcommand{\OO}{\mathbb O}

\newcommand{\NN}{\mathbb N}

\newcommand{\SSS}{\mathbb S}

\newcommand{\QQ}{\mathbb Q}
\newcommand{\RR}{\mathbb R}
\newcommand{\ZZ}{\mathbb Z}

\newcommand{\rank}{\mathop{\mathrm {rk}}\nolimits}
\newcommand{\latt}[1]{{\langle{#1}\rangle}}

\def\dim{\operatorname{dim}}

\newenvironment{psmallmatrix}
  {\left(\begin{smallmatrix}}
{\end{smallmatrix}\right)}

\newcommand{\Eh}{E_{7+1/2}}
\newcommand{\Dh}{D_{6+1/2}}
\newcommand{\Dhh}{D_{6+1/2+1/2}}
\newcommand{\Ah}{A_{5+1/2}}
\newcommand{\Ch}{C_{3+1/2}}
\newcommand{\AD}{AD_{3+1/2}}
\newcommand{\AG}{AG_{1+1/2}}
\newcommand{\UA}{U\!A_{1+1/2}}

\newcommand{\mg}{\mathfrak{g}}

\def\hv{{h^\vee}}
\newcommand{\T}{\mathsf{T}}

\usepackage{multirow}

\newcommand{\cF}{\mathcal F}

\newcommand{\TF}{\mathsf{T}^{\rm F}}

\begin{document}

 \hfill{KIAS-P24006, UUITP-06/24\\[+10mm]}

\title[On Intermediate Exceptional Series]{On Intermediate Exceptional Series}

\author{Kimyeong Lee}

\address{Korea Institute for Advanced Study, 85 Hoegiro, Dongdaemun-gu, Seoul 02455, Korea}

\email{klee@kias.re.kr}

\author{Kaiwen Sun}

\address{Department of Physics and Astronomy, Uppsala University, 75120, Uppsala, Sweden\newline
\indent Department of Mathematics, Uppsala University, 75106, Uppsala, Sweden}

\email{kaiwen.sun@physics.uu.se}

\author{Haowu Wang}

\address{School of Mathematics and Statistics, Wuhan University, Wuhan 430072, Hubei, China}

\email{haowu.wangmath@whu.edu.cn}

\subjclass[2020]{}

\date{\today}

\keywords{vertex operator (super)algebra, intermediate Lie algebra, modular form, modular linear differential equation, Hecke operator, Jacobi form, theta block}

\begin{abstract} 
The Freudenthal--Tits magic square $\mathfrak{m}(\mathbb{A}_1,\mathbb{A}_2)$ for $\mathbb{A}=\mathbb{R},\mathbb{C},\mathbb{H},\mathbb{O}$ of semi-simple Lie algebras can be extended by including the sextonions $\mathbb{S}$. A series of non-reductive Lie algebras naturally appear in the new row associated with the sextonions, which we will call the \textit{intermediate exceptional series}, with the largest one as the intermediate Lie algebra $\Eh$ constructed by Landsberg--Manivel. We study various aspects of the intermediate vertex operator (super)algebras associated with the intermediate exceptional series, including rationality, coset constructions, irreducible modules, (super)characters and modular linear differential equations. For all $\mathfrak{g}_I$ belonging to the intermediate exceptional series, the intermediate VOA $L_1(\mathfrak{g}_I)$  has characters of irreducible modules coinciding with those of the simple rational $C_2$-cofinite $W$-algebra $W_{-\hv/6}(\mathfrak{g},f_\theta)$ studied by Kawasetsu, with $\mathfrak{g} $ belonging to the Cvitanovi\'c--Deligne exceptional series. We propose some new intermediate VOA $L_k(\mathfrak{g}_I)$ with integer level $k$ and investigate their properties. For example, for the intermediate Lie algebra $D_{6+1/2}$ between $D_6$ and $E_7$ in the subexceptional series and also in Vogel's projective plane, we find that the intermediate VOA $L_2(D_{6+1/2})$  has a simple current extension to a SVOA with four irreducible Neveu--Schwarz modules, and the supercharacters can be realized by a fermionic Hecke operator on the $N=1$ Virasoro minimal model $L(c_{16,2},0)$. We also provide some (super) coset constructions such as $L_2(E_7)/L_2(D_{6+1/2})$ and $L_1(D_{6+1/2})^{\otimes2}\!/L_2(D_{6+1/2})$. In the end, we find that the theta blocks associated with the intermediate exceptional series produce some new holomorphic Jacobi forms of critical weight and lattice index.
\end{abstract}

\maketitle
\begin{small}
\tableofcontents
\end{small}

\section{Introduction}
We study the \textit{intermediate exceptional series} of non-reductive Lie algebras
\begin{align}\label{eq:ies}
\UA  \subset  AG_{1+1/2} \subset  AD_{3+1/2} \subset  C_{3+ {1}/{2}} \subset  A_{5+{1}/{2}} \subset  D_{6+{1}/{2}} \subset  E_{7+{1}/{2}},
\end{align}
with the corresponding Landsberg--Manivel's parameter $a  =-1,-2/3,0,1,2,4,8.$ This novel series is intermediate between the 
renowned \textit{Cvitanovi\'c--Deligne exceptional series} of simple Lie algebras \cite{Cvitanovic:2008zz,Deligne}
\begin{align} \label{DC}
A_1\subset A_2 \subset G_2 \subset D_4 \subset F_4 \subset E_6 \subset E_7 \subset E_8  
\end{align}
and the \textit{subexceptional series} \cite{Cvitanovic:2008zz,LMseries,DG}
\begin{align} \label{u1DCsub}
U_1 \subset A_1 \subset  A_1^3 \subset  C_3 \subset  A_5 \subset D_6 \subset E_7.   
\end{align}
This indicates that there exist subalgebra relations $E_7\subset \Eh\subset E_8$, and $D_6\subset \Dh\subset E_7$ and so on. 
The dual Coxeter numbers of the intermediate Lie algebras in  \eqref{eq:ies} are given by 
\begin{equation}\label{iesh}
\hv={5a}/{2}+4,    
\end{equation} 
while the dimensions are given by
\begin{align}\label{eq:Idim}
    \dim(\mathfrak{g})= \frac{12(2a+3)(a+2)}{(a+4)}  =\frac{12(4\hv-1)(\hv+1)}{5(\hv+6)}   .
\end{align}
The subscript $+1/2$ in \eqref{eq:ies} means that the dual Coxeter number is exactly the average of those of the corresponding algebras in the Cvitanovi\'c--Deligne exceptional \eqref{DC} and the subexceptional series \eqref{u1DCsub}. Remark that the Lie algebra $A_1$ in \eqref{u1DCsub} has the dual Coxeter number $2/3$, which is normalized by the universal formula $\hv=2a+2$ for the subexceptional series.  The last four algebras of \eqref{eq:ies} can be understood from an extension of the \textit{Freudenthal--Tits magic square} $\mathfrak{m}(\mathbb{A}_1,\mathbb{A}_2)$ for $\mathbb{A}=\mathbb{R},\mathbb{C},\mathbb{H},\mathbb{O}$ by including the sextonions $\mathbb{S}$, while the first three rely on a further extension to the \textit{magic triangle} studied by Cvitanovi\'c \cite{Cvitanovic:2008zz}, Deligne--Gross \cite{DG} and Landsberg--Manivel \cite{LMseries}. Some preliminary aspects such as some distinguished modules and dimension formulas for the intermediate Lie algebras in \eqref{eq:ies} have been studied by Barton--Sudbery \cite{BS}, Westbury \cite{Westbury} and Landsberg--Manivel \cite{LM}.

The aim of the present paper is twofold. Firstly, we want to systematically study the irreducible modules for the intermediate Lie algebras in \eqref{eq:ies} including their quadratic Casimir invariants, dimensions, decompositions in subalgebras and tensor products and so on. Secondly, we want to study the vertex algebras associated with the intermediate Lie algebras in \eqref{eq:ies}, as an analogy of the affine VOA $L_k(\mg)$ associated with simple Lie algebra $\mg$ at positive integer level $k$.  These are sometimes called intermediate VOAs in recent CFT literature. The precise nature of these vertex algebras or putative 2d CFTs is still unclear, but there are still many aspects one can explore such as the central charge, conformal weights and characters. To shorten the name, we will just call them as affine VOA $L_k(\mg_I)$ and omit the term intermediate. Most techniques we will use in this paper follow from our previous work on $\Eh$ \cite{Lee:2023owa}. The intermediate Lie algebra $\Eh$ was independently found in physics \cite{Mathur:1988na} and in mathematics \cite{Cohen} and it has the best properties among the intermediate exceptional series.

The intermediate Lie algebras are closely related to $W$-algebras. In fact, the affine VOA associated with the intermediate exceptional series at level $1$ can be understood by the affine $W$-algebra 
$W_{-\hv/6}(\mathfrak{g}_{\rm CD},f_\theta)$ studied in 
 \cite{Kawasetsu:2018irs,Kawasetsu:2018tzs}. Let $\mg_I$ be the intermediate Lie algebra in $\mg_{\rm sub}\subset \mg_I\subset \mg_{\rm CD}$. Then $L_1(\mg_I)$ has the same characters as those of $W_{-\hv/6}(\mathfrak{g}_{\rm CD},f_\theta)$, though the central charges are different. The characters of these affine $W$-algebras satisfy a one-parameter $s$ family of 4th order holormophic modular linear differential equations (MLDE), where $s$ is just the central charge of the $W$-algebra \cite{Kawasetsu:2018irs,Kawasetsu:2018tzs}.   
The central charge $c$ of $L_1(\mathfrak{g}_I)$ is related to the parameter $s$ in \cite{Kawasetsu:2018irs,Kawasetsu:2018tzs} by 
$     c=s+{6}/{5}.$ 
As we will see this $6/5$ comes from the difference between the central charge $-3/5$ of minimal model $M(5,3)$ and the central charge $3/5$ of effective minimal model $M_{\rm eff}(5,3)$.

It is well-known that all affine VOAs associated with simple Lie algebras at positive integer levels are rational and $C_2$-cofinite.  A natural question is whether this also holds for the affine VOA $L_k(\mg_I)$. Interestingly, it was noticed in \cite{Lee:2023owa} from the viewpoint of MLDE indices that $L_k(\Eh)$ is rational only for $k=1,2,3,4,5$ and ceases to be rational for higher levels. In this paper, we find a similar phenomenon for the intermediate exceptional series. For $\Dh$ which is intermediate between $D_6$ and $E_7$,  computational evidence leads us to the following conjecture:
\begin{conj}\label{conj2}
There exists rational affine VOA $L_k(\Dh)$ if and only if $k=1,2,3$. For these levels, $L_k(\Dh)$ has central charge $99k/(14+k)$, and the number $r(k)$ of its irreducible modules and the MLDE index $l(k)$ for its characters are as follows
\begin{table}[h]
  \centering
  \begin{tabular}{c|c|c|c}\hline
    $k$ & $1$ & $2$ & $3$    \\ \hline
$r(k)$  & $4$ & $13$ & $32 $     \\ \hline
$l(k)$ & $0$ & $ 36$ & $340 $     \\ \hline
     \end{tabular}
\end{table}

\noindent
The conformal weights of $L_k(\Dh)$ for $k=1,2,3$ are formulated in \eqref{eq:Dhlevel1}, \eqref{eq:Dhlevel2} and \eqref{eq:Dhlevel3} respectively. Moreover, $L_2(\Dh)$ has a simple current extension to a $N=1$ SVOA with four irreducible Neveu--Schwarz modules, and admits two super coset constructions
\begin{align}
    L_2(\Dh)=\frac{F\otimes L_2(E_7)}{L_{\rm sub}(c_{80,2},0)}\quad \textrm{and}\quad {L_2(\Dh)} = \frac{L_1(\Dh)\otimes L_1(\Dh)}{L_{\rm sub}(c_{16,10},0) }.
\end{align}
Here $F$ is the free fermion SVOA of central charge $1/2$, $L_2(E_7)$ is the affine VOA associated with $E_7$ at level $2$, $L_{\rm sub}(c_{80,2})$ is a non-diagonal modular invariant of the $N=1$ supersymmetric minimal model $L(c_{80,2},0)$ constructed in \eqref{SM802sub}, and $L_{\rm sub}(c_{16,10})$ is a non-diagonal modular invariant of the $N=1$ supersymmetric minimal model $L(c_{16,10},0)$ constructed in \eqref{SM1610sub}.
\end{conj}
The MLDE index $l$ is an important quantity measuring how meromorphic a MLDE is. For the characters of a rational VOA which form a vector-valued modular form of weight $0$ on $SL(2,\ZZ)$, the MLDE index must be a non-negative integer. We use this property to show $ L_k(\Dh)$ cannot be rational for $k>3$. More details will be discussed in Section \ref{sec:Dhhighlevel}. The super structure in VOA is often called the \textit{fermionization} $\cF$ in physics literature by the \textit{boson-fermion correspondence}. Following the notation of \cite{Lee:2022yic}, the super cosets for $L_2(\Dh)$ can be written as 
\begin{align}
\frac{ \cF(E_7)_2}{\cF(\Dh)_2} =F^{-1}\otimes SM_{\rm sub}(80,2) \quad \textrm{and }\quad  \frac{\cF(\Dh)_1^{\otimes 2}}{\cF(\Dh)_2} = SM_{\rm sub}(16,10)  .
\end{align}
The $L_2(\Dh)$ also allows a fermionic Hecke operator interpretation from the effective $N=1$ supersymmetric minimal model $L(c_{16,2},0)$. Following the notation of \cite{Lee:2022yic}, it can be written as
\begin{align}
    \cF(\Dh)_2=\TF_{11} SM_{\rm eff}(16,2)  .
\end{align}
Here the $\TF_{p}$ is the fermionic Hecke operator defined in \cite{Lee:2022yic} to map the Neveu--Schwarz characters of a SVOA of central $c$ to the  Neveu--Schwarz characters of another SVOA of central $pc$. It is the super version of the Hecke operator $\T_p$ for VOA characters recently defined by Harvey--Wu \cite{Harvey:2018rdc}.

We also study the affine VOA $L_k(\mathfrak{g})$ for other intermediate Lie algebras. For example, for $\Ah$ which is intermediate between $A_5$ and $E_6$ we have the following conjecture again from the analysis on the MLDE indices:
\begin{conj}\label{conjA}
There exists rational affine VOA $L_k(\Ah)$ if and only if $k=1,2$. For these levels, $L_k(\Ah)$ has central charge $56k/(9+k)$, and the number $r(k)$ of distinct characters and the MLDE index $l(k)$ are as follows
\begin{table}[h]
  \centering
  \begin{tabular}{c|c|c}\hline
    $k$ & $1$ & $2$      \\ \hline
$r(k)$  & $4$ & $13$       \\ \hline
$l(k)$ & $0$ & $ 36$      \\ \hline
     \end{tabular}
\end{table}

\noindent
The conformal weights of $L_k(\Ah)$ for $k=1,2$ are formulated in \eqref{eq:Ahlevel1} and \eqref{eq:Ahlevel2} respectively.
\end{conj}
For the remaining intermediate Lie algebras, we only study their affine VOAs at level $1$. The $L_k(\mathfrak{g}_I)$ are not traditional VOAs. The naive fusion rules computed from Verlinde formula have negative coefficients. They seem to give an different interpretation for the affine $W$-algebras $W_{-h^\vee/6-1+k}(\mathfrak{g}_{\rm CD},f_\theta)$, though the precise relation for $k\ge 2$ still needs to be studied. A rigorous treatment of intermediate exceptional series may follow the finite $W$-algebra approach in \cite{Desole}.

The series \eqref{eq:ies} can often be added in the beginning by a formal intermediate minimal model $IM$ with $a=-4/3$. We define the associated  VOA of $IM$ at level $k$ as the effective Virasoro minimal model $M_{\rm eff}(3k+2,3)$ with central charge $k/(k+2/3)$. The well-definedness will be discussed in Section \ref{sec:IM}. This is very similar with the Lee-Yang model often added in the beginning of Cvitanovi\'c--Deligne exceptional series, whose level $k$ VOA is defined as $M_{\rm eff}(2k+3,2)$.

It is known that the denominator identities of affine Lie algebras induce particular holomorphic Jacobi forms of singular weight, which are called theta blocks by Gritsenko--Skoruppa--Zagier \cite{GSZ19}. Moreover, there is a one-to-one correspondence between finite-dimensional semi-simple Lie algebras and holomorphic theta blocks of singular weight \cite{Wan23}. Following the philosophy of denominator identities, in this paper we introduce the theta blocks associated with intermediate Lie algebras and propose the following conjecture (see Section \ref{sec:theta-blocks} for a precise version). 
\begin{conj}\label{conjthetablock}
The theta blocks associated with the intermediate Lie algebras in \eqref{eq:ies} are holomorphic Jacobi forms of critical weight.
\end{conj}
This conjecture has been verified for the first five cases in \eqref{eq:ies}, and indicates that there is a deep connection between intermediate Lie algebras and holomorphic theta blocks of critical weight. 

The intermediate Lie algebras of classical types inspired by the early works \cite{GZ,GZ2,proctor} have been studied by Shtepin  \cite{ShtepinAC,ShtepinB,ShtepinD0,ShtepinD} in the last three decades including character formulas and dimension formulas of quasi-highest weight modules.  It seems that the intermediate Lie algebras we study here have exceptional features and are quite different from Shtepin's. For example, the intermediate Lie algebra we denote as $\Dh$ is intermediate between $D_6$ and $E_7$, while Shtepin's $\mathfrak{d}_{6+1/2}$ is intermediate between $\mathfrak{d}_{6}$ and $\mathfrak{d}_{7}$ \cite{ShtepinD0,ShtepinD}. It seems more challenging to find proper generalizations of Weyl and Kac--Weyl character formulas for the algebras in the intermediate exceptional series. Besides, the key feature of Shtepin's intermediate Lie algebras, i.e. the multiplicity-free filtration is not strictly satisfied by the algebras in  the intermediate exceptional series.

Some algebras in  the intermediate exceptional series were suggested to have real forms related to the non-compact global symmetries of the circle dimensional reduction of some supergravity theories \cite{Borsten:2017uoi}.  We hope our new results on the irreducible modules of the intermediate exceptional series can have some applications on this subject. It was also suggested in \cite{Borsten:2017uoi} that there exist some even exotic algebras associated with the trinions. It is intriguing to consider whether there exist some vertex algebras associated with these exotic algebras. 

Notation: our Lie algebra convention, especially the name of representations, follows the Lieart package of Mathematica. 

\section{Magic square, exceptional series  and Vogel's projective plane}
The Freudenthal--Tits magic square associates a Lie algebra to any pair of division algebras, denoted as $\mathfrak{m}(,)$. It realizes the philosophy that all exceptional Lie algebras exist owing to the octonions $\OO$. Apart from $G_2$ which is the automorphism group of $\OO$, all remaining four exceptional Lie algebras $F_4,E_6,E_7,E_8$ can be constructed uniformly as $\mathfrak{m}(\mathbb{A},\OO)$ for $\mathbb{A}=\mathbb{R},\mathbb{C},\mathbb{H},\mathbb{O}$. There exist several ways to construct the operation $\mathfrak{m}$, including Tits' original one, Vinberg's symmetric one and by triality. It can be proved that $\mathfrak{m}(\mathbb{A}_1,\mathbb{A}_2)$ is isomorphic to  $\mathfrak{m}(\mathbb{A}_2,\mathbb{A}_1)$ for $\mathbb{A}_{1,2}=\mathbb{R},\mathbb{C},\mathbb{H},\mathbb{O}$, i.e., the matrix is nontrivially symmetric, thus magic. We refer to the good review on the magic square in \cite{BS}. The magic square can be extended by including the sextonion $\SSS$. Then each row is added by an intermediate Lie algebra. The  algebras of extended magic square are collected in Table \ref{tab:EFT}. We denote the one from $\mathfrak{m}(\SSS,\SSS)$ as $\Dhh$, which will be discussed in Appendix \ref{app:4}. As many good properties of the algebras in \eqref{eq:ies} do not hold for $\Dhh$, we do not include it in the intermediate exceptional series.

\begin{table}[ht]
 \begin{center}
\caption{The extended Freudenthal--Tits magic square $\mathfrak{m}(\mathbb{A}_1,\mathbb{A}_2)$ over $\mathbb{A}=\RR, \CC,  \HH, \SSS, \OO$. }\label{tab:EFT}
\begin{tabular}{c|ccccccc|c}
\hline
  && $\RR$ &$\CC$   & $\HH$ & $\SSS$ & $\OO$ && exceptional series\\
 \hline
   $\RR$ && $A_1$ & $A_2$    &$C_3$ & $C_{3+{1}/{2}}$ & $F_4$ && Severi-section \\ 
  $\CC$ && $A_2$ & $A_2^2$    & $A_5$ & $A_{5+{1}/{2}}$ & $E_6$  &&  {Severi} \\ 
  $\HH$ && $C_3$ & $A_5$  &   $D_{6}$ &  $D_{6+{1}/{2}}$ & $E_7$ &&  {sub}  \\ 
  $\SSS$ && $C_{3+ {1}/{2}}$ & $A_{5+{1}/{2}}$   & $D_{6+{1}/{2}}$ & ${D}_{6+{1}/{2}+{1}/{2}}$ & $E_{7+{1}/{2}}$  && intermediate \\ 
   $\OO$ && $F_4$ & $E_6$  & $E_7$ & $E_{7+{1}/{2}}$ & $E_8$ && {Cvitanovi\'c--Deligne} \\
   \hline
\end{tabular}
 \end{center}
\end{table}

The magic square can be further extended to the \textit{magic triangle}, in which each row is extended to an \textit{exceptional series}. From bottom to top in  Table \ref{tab:EFT}, the rows are extended to the Cvitanovi\'c--Deligne, intermediate, sub, Severi, Severi-section exceptional series respectively.  For each exceptional series, there are 1) a linear relation between the dual Coxeter numbers and Landsberg--Manivel's parameter $a$, 2) the dimension as a simple rational function of  $a$, 3) some distinguished modules whose dimensions are rational functions of  $a$. In fact, an universal formula was found for the magic square in \cite{LM} for the dual Coxeter number and the dimension of $\mathfrak{m}(\AAA_1,\AAA_2)$  as
\begin{align}
    \hv=\frac{ab}{4}+a+b-2,
\end{align}
and 
\begin{align}
    \dim \mathfrak{g}(a,b)= \frac{3 (a b+2 a+2 b) (a b+4 a+4 b-4)}{(a+4) (b+4)}.
\end{align}
Here $a$ and $b$ are the Landsberg--Manivel's parameters for $\AAA_1$ and $\AAA_2$. As we will see, this formula can be extended to include $\AAA_2=\SSS$ with $b=6$, which gives $h^\vee=5a/2+4$ as in \eqref{iesh} for the intermediate exceptional series.

In the following we briefly explain each exceptional series. It is well-known that for any simple Lie algebra, there exists a universal decomposition for the tensor product 
\begin{align}   \label{g2decomp}
\mathrm{Sym}^2\mathfrak{g}&=Y_2+Y_2'+Y_2''+\mathbf{1},\\
    \mathrm{Alt}^2\mathfrak{g}&=X_2+\mathfrak{g}.
\end{align}
The famous Cvitanovi\'c--Deligne exceptional series \eqref{DC} has the special property that $Y_2''=0$ in   \eqref{g2decomp} and a uniform dimension formula
$$
\dim(\mathfrak{g}_{\rm CD})=  \frac{2 (3 a+7) (5 a+8)}{a+4} .$$
Here the Landsberg--Manivel's parameter $a  =-4/3,-1,-2/3,0,1,2,4,8.$
This exceptional series has made many appearances in number theory, representation theory and physics. In some sense, they have the maximal symmetry among simple Lie algebras, that is they have fewer irreducible modules in tensor product decompositions.  The Cvitanovi\'c--Deligne exceptional series has many distinguished modules whose dimensions can be expressed as rational functions of $a$ \cite{Deligne,Cohen}. 

The subexceptional series \eqref{u1DCsub} can be defined by the condition that each $\mathfrak{g}_{\rm sub}\times A_1$ is a maximal subalgebra of $\mg_{\rm CD}$ in \eqref{DC} with the same parameter $a$. It has dimension formula
$$
\dim(\mathfrak{g}_{\rm sub})= \frac{3(2a+3)(3a+4)}{
a+4} .$$
It has distinguished modules $V$ and $V_2$ with dimensions $6a + 8$ and  $9(a+ 1)(2a + 3)$ respectively \cite{LM}. Each of $V$, $\mathfrak{g}_{\rm sub}$ and $V_2$ corresponds to one fundamental weight. As we will see, this is also the reason why $\Dh$ has exactly three bosonic fundamental weights.

The Severi series of reductive Lie algebras
$$
U_1^2\subset A_2 \subset A_2^2\subset A_5\subset E_6
$$
is related to the Severi varieties \cite{LM}. The Lie algebras have dimension formula 
$$
\dim(\mathfrak{g}_{\rm Severi})= \frac{4 (a+1) (3 a+2)}{a+4}.
$$
The Severi series has two distinguished modules of dimension $3a + 3$. They two are isomorphic, thus are denoted as $V$ and $V^*$. Each of them corresponds to one fundamental weight. 
As we will see, this is also the reason why $\Ah$ has exactly two bosonic fundamental weights.

The Severi-section series \cite{LM} of simple Lie algebras
$$
A_1\subset A_2\subset C_3\subset F_4
$$
does not have extra algebras beyond those in the magic square \eqref{tab:EFT}. It has dimension formula 
$$
\dim(\mathfrak{g}_{\rm Severi-sec})= \frac{3 a (3 a+2)}{a+4}  .
$$
It has a  distinguished module $V$ of dimension $3a + 2$.  This module also exists for the intermediate Lie algebra $\Ch$. The Severi-section series was also notably studied in \cite{westbury2006universal}.

The exceptional series reviewed above are one-parameter families of Lie algebras. It has close connection with \textit{Vogel's universal Lie algebra} \cite{vogel1999universal}. Vogel found that any simple Lie algebra can be parameterized by a point in $\mathbb{P}^2$, denoted as $\mathfrak{g}[\alpha,\beta,\gamma]$, where $\alpha,\beta,\gamma$ are the homogeneous coordinates. The $\mathbb{P}^2$ is often called \textit{Vogel's projective plane}. Vogel's universal formulas for the dimension and the dual Coxeter number of simple Lie algebras are
\begin{align}\label{vogeldim}
    \dim(\mathfrak{g})=-\frac{(2t-\alpha)(2t-\beta)(2t-\gamma) }{\alpha\beta\gamma},\qquad t=\alpha+\beta+\gamma=\hv.
\end{align}
As the plane is projective, the $\alpha$ is often normalized as $-2$ such that $\beta$ and $\gamma$ are mostly integers. Vogel found that the irreducible modules in the tensor product decomposition \eqref{g2decomp} have the following universal dimension formulas
\begin{align}
  \dim(  Y_2) =-\frac{t(2t - \beta)(2t - \gamma)(t + \beta)(t + \gamma)(3\alpha - 2t)}{
\alpha^2\beta \gamma (\alpha -\beta)(\alpha -\gamma)},
\end{align}
\begin{align}
    \dim(Y_2')=\dim(Y_2)|_{\alpha\leftrightarrow\beta},\qquad \dim(Y_2'')=\dim(Y_2)|_{\alpha\leftrightarrow\gamma},
\end{align}
and 
\begin{align}
\dim(X_2) = \frac{(2t -\alpha)(2t -\beta )(2t -\gamma)(t +\alpha)(t +\beta)(t +\gamma)}{\alpha^2\beta^2\gamma^2}.
\end{align}
The quadratic Casimir invariants of $Y_2,Y_2',Y_2'',X_2$ are $2t-\alpha,2t-\beta,2t-\gamma, 2t$ respectively. Vogel also found the dimensional formulas for the modules appearing in the decomposition of $\mathfrak{g}^{\otimes 3}$.  Besides, the decomposition of $\mathfrak{g}^{\otimes 4}$ for Vogel's universal Lie algebra has been recently studied in \cite{Avetisyan:2023xdx}. The arbitrary $k$-th symmetric power of $\mathfrak{g}$ denoted as $Y_k$ has dimension formula given in \cite{LMuni}. Vogel's universal Lie algebra can also be extended to some Lie superalgebras and intermediate Lie algebras. 

Remarkably, the Cvitanovi\'c--Deligne exceptional and subexceptional series have distinguished lines in the Vogel's plane. The Severi and Severi-section series  do not form lines in Vogel's plane. For the intermediate exceptional series, it is easy to check that in \eqref{eq:ies}, only $\Dh$ and $\Eh$ satisfy the dimension formula \eqref{vogeldim}. The $\Dh$ has $(\beta,\gamma)=(6,10)$ and $\Eh$ has $(\beta,\gamma)=(10,16)$. 
We summarize Vogel's parameters for the exceptional series in the following Table \ref{tab:vogel}. We can see that for $\Dh$ and $\Eh$, the Vogel's parameters are exactly in the middle of those of the algebras in the CD and subexceptional series. Vogel's dimension formulas and quadratic Casimir invariant formulas are very useful to cross-check with our predictions for $\Dh$.

\begin{table}[ht]
\def\arraystretch{1.1}
 \begin{center}
\caption{Vogel's parameters $(\beta,\gamma)$ for sub, intermediate and Cvitanovi\'c--Deligne exceptional series with $\alpha=-2$.}\label{tab:vogel}
\begin{tabular}{c|cccccccccc}
\hline
$a$    &  $ -1$ & $-2/3$ & $0$ &  $1$ &  $2$ & $4$  &   $8$ \\ \hline

$\mathfrak{g}$ &   $U_{1} $  & $A_{1} $  &$A_{1}^3$  &  $C_{3 }$ & $A_{5 }$   & $D_{6 }$   & $E_{7 }$   \\  
   
$ (\beta,\gamma) $   & $  (-1,3)$ &  $ (-\frac{2}{3},\frac{10}{3}) $ & $(0,4)  $  &  $ (1,5) $ &  $  (2,6)  $  &  $   ( 4,8) $   & $ (8,12)  $       \\ \hline

  $\mathfrak{g}$   & $U\!A_{1+1/2} $  & $AG_{1+1/2} $  &$AD_{3+1/2}$  &  $C_{3+ {1}/{2}}$ & $A_{5+{1}/{2}}$   & $D_{6+{1}/{2}}$   & $E_{7+{1}/{2}}$   \\

$ (\beta,\gamma) $     & $ - $ &  $  -$ & $ - $  &  $ - $ &  $ -   $  &  $  (6,10)  $   & $  (10,16) $       \\ \hline

$\mathfrak{g}$    & $A_{2} $  & $G_{2} $  &$D_{4}$  &  $F_{4}$ & $E_{6}$   & $E_7$   & $E_{8}$   \\

$ (\beta,\gamma) $   & $(3,2)  $ &  $(\frac{10}{3},\frac{8}{3})  $ & $(4,4)  $  &  $(5,6)  $ &  $  (6,8)  $  &  $ (8,12)   $   & $(12,20)   $       \\ \hline

\end{tabular}

 \end{center}
\end{table}

\section{Universal properties of intermediate exceptional series}
\subsection{Algebraic properties}
Consider the minimal gradation for $\mathfrak{g}$ belonging to the Cvitanovi\'c--Deligne exceptional series
\begin{equation}
    \mg=\mg_{-1}\oplus \mg_{-1/2}\oplus \mg_{0}\oplus \mg_{1/2}\oplus \mg_{1}.
\end{equation}
This is a five step grading defined by the highest root $\theta$. It is equivalent to the decomposition $\mathfrak{g}_{\rm sub}\times SU(2)\subset \mathfrak{g}$ where one has the universal form
\begin{align}\label{eq:unidecomp}
    \mathfrak{g}=(\mathfrak{g}_{\rm sub},\mathbf{1})\oplus (R_{\rm sub},\mathbf{2})\oplus (\mathbf{1},\mathbf{3}).
\end{align}
Here $R_{\rm sub}$ is a module of $\mathfrak{g}_{\rm sub}$, and is irreducible in most cases. The gradation and decomposition is connected by  $\mg_{0}=\mathfrak{g}_{\rm sub}\oplus \mathbb{C}$, $\mg_{1/2}=\mg_{-1/2}=R_{\rm sub}$ and $\mg_1=\mg_{-1}=\mathbb{C}$. The intermediate Lie algebra $\mg_I$ is defined as a subalgebra of $\mg$ by the decomposition
\begin{align}\label{intdef}
 \mg_I= \mathfrak{g}_{\rm sub}\oplus R_{\rm sub}\oplus \mathbb{C}  .
\end{align}
Although this construction exists for any simple Lie algebra $\mathfrak{g}$, the intermediate Lie algebra $\mg_I$ is most interesting when $\mg$ belongs to the Cvitanovi\'c--Deligne exceptional series. 

The decomposition \eqref{intdef} in fact gives all positive roots of intermediate Lie algebra $\mg_I$, which are the union of all positive roots of $\mathfrak{g}_{\rm sub}$ and half of the weights of $R_{\rm sub}$. The half weights can be picked out by requiring they have positive intersection with the highest root $\theta$ of $\mathfrak{g}_{\rm sub}$, i.e.
\begin{align}\label{eq:positive-roots}
    \Delta_+( \mg_I)=\Delta_+(\mathfrak{g}_{\rm sub} )+\{w\in  R_{\rm sub} |w\cdot \theta >0\}
\end{align}
Once all positive roots of  $\mg_I$ are known, one can define the Weyl vector of $\mg_I$ as
\begin{align}
    \rho= \sum_{\alpha\in \Delta_+( \mg_I)} \alpha.
\end{align}
For simple Lie algebras, the Weyl vector in the fundamental weight basis is always $(1,1,\dots,1)$. For intermediate Lie algebras, the Weyl vector in the fundamental weight basis of $\mg_{\rm sub}$ has some components bigger than $1$. It was noticed in \cite{Lee:2023owa} that the Freudenthal--de Vries strange formula
\begin{align}
    |\rho|^2=\frac{h^\vee \dim(\mathfrak{g})}{12}
\end{align}
for simple Lie algebras still holds for $\Eh$. We find that the strange formula also holds for $\Dh$ and $\Ah$. Here the norm of $\rho$ is computed by the bilinear form of $\mg_{\rm sub}$. We also checked that for smaller intermediate Lie algebras in \eqref{eq:ies}, the strange formula no longer holds. This is not so surprising knowing that the properties of  intermediate Lie algebras in \eqref{eq:ies} become worse and worse from $\Eh$ to $\UA$.

The rank $r$ of $\mg_I$ is the same as that of $\mathfrak{g}_{\rm sub}$. Then the dual Coxeter number of $\mg_I$ can be defined similar with the simple Lie algebra as
\begin{align}\label{intdualCox}
    h^\vee =\frac{1}{2r}\sum_{\alpha\in \Delta( \mg_I) }\langle \alpha,\alpha \rangle_{\mg_{\rm sub}}.
\end{align}
One can check that the dual Coxeter number defined in this way is the same with $5a/2+4$ with $a$ being Landsberg--Manivel's parameter. One can also check that this dual Coxeter number is always the average of those of $\mathfrak{g}_{\rm sub} $ and $\mg_{\rm CD}$.

The analogy of the Dynkin diagram seems to exist at least for  $\Ch$, $\Ah$, $\Dh$ and $\Eh$. In these cases, the Dynkin diagram for intermediate Lie algebra $\mg_I$ is the same with that of $\mg_{\rm sub}$, though the comarks of each node may be different.  For some nodes, the comark multiplies   such that the dual Coxeter number 
\begin{align}
    h^\vee =\sum_{i=0}^{r}a^\vee_i=1+ \sum_{i=1}^{r}\rho_i a^\vee_{i,\mathrm{sub}}
\end{align}
increase from that of $\mg_{\rm sub}$ to the expected value of $\mg_I$. The multiplicity here is just the component of the Weyl vector $\rho$ of $\mg_I$. One can check that this dual Coxeter number is again consistent with \eqref{intdualCox}. Though lacking rigorous treatment, the concept of highest weight modules seems to still  work for intermediate Lie algebras and these modules are associated with the Dynkin labels just like $\mg_{\rm sub}$. Therefore, we use the Dynkin labels and levels of $\mg_{\rm sub}$ to describe the highest weight modules of $\mg_I$. It is natural to consider whether these modules enjoy Weyl-type dimension formula and character formula. It turns out for some special types of highest weight modules, we do find correct generalizations of Weyl dimension formulas. To discuss these special types of modules, we need to introduce the \textit{fermionic fundamental weights} for intermediate Lie algebras.

The concept of {fermionic fundamental weight} was introduced for $\Eh$ in our previous work \cite{Lee:2023owa}  to distinguish the bosonic and fermionic integrable highest-weight modules of $\Eh$. Originally, they are the fundamental weights with multiplicity $2$ in the Weyl vector of the intermediate Lie algebra.  
The comarks of the fermionic fundamental weights are doubled in the intermediate Lie algebras such that the dual Coxeter number increases to the required value. When the highest weight contains an even/odd number of fermionic fundamental weights, the module is called bosonic/fermionic respectively. When the highest weight contains no fermionic fundamental weight, the module is called purely bosonic. For purely bosonic modules, the direct generalization of the Weyl dimension formula holds by just enlarging the set of positive roots. Interestingly, in \cite{Lee:2023owa} a modified Weyl dimension formula was found for the irreducible modules with highest weight containing one single fermionic fundamental weight. 

The concept of the fermionic fundamental weight and many features described above turn out to persist for many big algebras in the intermediate exceptional series \eqref{eq:ies}. Most importantly, we observe that there exists \textit{three} fermionic fundamental weights for each of them. For $\Ah$ and $\Dh$, the concept of fermionic fundamental weights is most clear and useful. Compared to $A_5$ and $D_6$, the Weyl-vector components and the comarks of the fermionic fundamental weights of $\Ah$ and $\Dh$ get doubled, such that the dual Coxeter numbers increase to the required values. The fermionic fundamental weights also play important roles in generalizing the Weyl dimension formula for these two intermediate Lie algebras, which will be discussed in detail later. 
For $\AD$ and $\Ch$, as the rank of the algebra is $3$, all three fundamental weights should be regarded as fermionic. 
For $\UA$ and $\AG$, even though the rank is $1$, the level $1$ affine VOA still has four characters, such that we can think there still exist three fundamental weights, two of which are pseudo, and all of which are fermionic. 

An important quantity of the highest weight module $R_\lambda$ is the quadratic Casimir invariant. For simple Lie algebras, it is well-known   $C_2(R_\lambda)=\langle\lambda+2\rho,\lambda\rangle$. This formula fails for intermediate Lie algebras because of the absence of suitable bilinear form. Instead, some corrections are needed to match the conformal weights of intermediate VOAs. A conjectural simple formula was found for these  corrections  for $\Eh$ in \cite{Lee:2023owa} and called the \textit{odd corrections}. The name comes from the observation that the correction only involves fermionic fundamental weights, but not the bosonic ones. Inspired by this, we conjecture that at least for $\Eh$, $\Dh$ and $\Ah$, the following type of quadratic Casimir invariant formula exist
\begin{align}\label{eq:C2general}
C_2(R_\lambda)=\langle\lambda+2\rho,\lambda\rangle_{\mg_{\rm sub}}+\langle\lambda,\lambda\rangle_{\rm odd}.
\end{align}
The precise form of the correction term $\langle\lambda,\lambda\rangle_{\rm odd}$ for $\Dh$ and $\Ah$ will be given for each case, which only involves fermionic fundamental weights like the $\Eh$ case. 

The index $I$ of a module $R$ for simple Lie algebra $\mathfrak{g}$ is related to the quadratic Casimir invariant and dimension by
\begin{align}\label{eq:defindex}
    I(R)=\frac{ |R|\times C_2(R)}{2\dim(\mathfrak{g})}.
\end{align}
We regard this as a definition of index for intermediate Lie algebras. Then we can compute the index of modules of intermediate Lie algebra $\mg_I$ once the dimension and quadratic Casimir invariant are known.  The indices are very useful in  the study on the module decompositions under $\mg_{\rm sub}\subset \mg_I\subset \mg_{\rm CD}$, as both sides of a module decomposition should have the same total indices, besides the total dimensions. We will use this property to determine many module decompositions for each intermediate Lie algebra in \eqref{eq:ies}.

Each algebra in the intermediate exceptional series except for $\Eh$ has three level-one modules which are denoted as $R_1,R_2$ and $R_3$.  These modules appear as the leading terms of the characters of affine VOA $L_1(\mathfrak{g}_I)$, which will be discussed in the next subsection.
The $R_3$ is related to the $R_{\rm sub}$ in \eqref{eq:unidecomp} by  $R_3=R_{\rm sub}\oplus \CC$ and has dimension $3(2a+3)$. We did not find dimension formulas for $R_1$ and $R_2$. But we notice that they are in pairs and related to the fundamental module $f$ in the Cvitanovi\'c--Deligne exceptional series by
\begin{align}
    f_{\rm CD}=R_1+R_2.
\end{align}
This is one of the consequences of the coset \eqref{cosetLY} discussed in the next subsection. 
For $\Eh$, $R_3$ is just the fundamental module $\bf 57$ and there are no $R_1$ or $R_2$ modules.

Intermediate Lie algebra $\mg_I$ has an automorphism group inherited from both $\mg_{\rm sub}$ and $\mg_{\rm CD}$. It can be trivial, $\ZZ_2$ or $\ZZ_3$ for the intermediate exceptional series. 
The automorphism group acts on the modules non-trivially for $R_1$ and $R_2$. 
 We collect the basic Lie algebra data of the intermediate exceptional series in Table \ref{tab:ies}. 

\begin{table}[ht]
 \begin{center}
\caption{The Lie algebra data for the intermediate exceptional series  including $IM$ and $\Dhh$. Here $\delta={h^\vee \dim(\mathfrak{g})}/{12} -|\rho|^2$ measures the discrepancy of the Freudenthal--de Vries strange formula. There are three distinguished irreducible modules $R_1$, $R_2$ and $R_3$. The automorphism group Aut acts on  modules $R_1$ and $R_2$.}\label{tab:ies}
\begin{tabular}{c|cccccccccc}
\hline
$\AAA$  & $-$ &$-$ & $-$ & $-$ & $\RR$ &$\CC$   & $\HH$  & $\SSS$ & $\OO$  \\
 \hline
  $\mathfrak{g}$ & $IM$ & $U\!A_{1+1/2} $  & $AG_{1+1/2} $  &$AD_{3+1/2}$  &  $C_{3+ {1}/{2}}$ & $A_{5+{1}/{2}}$   & $D_{6+{1}/{2}}$ & $D_{6+{1}/{2}+1/2}$ & $E_{7+{1}/{2}}$   \\  
 $a$ & $-4/3$ &  $ -1$ & $-2/3$ & $0$ &  $1$ &  $2$ & $4$ & $6$ &   $8$ \\
  $\hv $ & $2/3 $ &  $3/2  $ &  $ 7/3$  & $4 $  & $  13/2 $  &  $  9 $   & $  14 $   & $19$ &  $  24 $   \\
 $\dim(\mg)$ &  $ 1$ & $  4$ & $ 8$ & $18$  &   $  36 $  &  $  56 $   & $ 99  $   &  $144$  &    $ 190  $   \\
Aut & $0$  &  $ \ZZ_2 $  &  $ 0 $  &   $  \ZZ_3 $  &  $ 0  $   & $\ZZ_2   $   &   $0   $& $ \ZZ_2 $  & $ 0  $     \\
$\delta$ & $  - $  & $ 1/8  $  & $1/18   $  & $ 1/2 $  & $ 1/8 $  & $  0$  & $  0 $    & $-$ & $ 0 $   \\
\hline
 $\dim (R_1)$ &   $1$ & $1$ & $ 2$  &   $  2 $  &  $  6 $  &   $ 6  $  &  $  12 $   &  $12$   & $ -  $   \\ 
 $\dim (R_2)$ &    $1$ & $2$ & $5 $  &   $  6 $  &  $ 20  $  &   $ 21  $  &  $  44 $      &  $45$  & $ -  $   \\
 $\dim (R_3)$ &   $1$& $3$ &  $5 $ &   $  9 $  &  $  15 $  &   $ 21 $  &  $  33 $   &  $ {45}$    & $ 57  $   \\ \hline
\end{tabular}

 \end{center}
\end{table}

\subsection{Affine VOA of level $1$}\label{sec:level1}
For simple Lie algebra $\mg$, the affine VOA $L_{k}(\mg)$ at positive integer level $k$ is a typical example of rational $C_2$-cofinite VOAs. The 2d RCFTs of this type are called Wess--Zunimo--Witten models. It has central charge 
\begin{align}
    c=\frac{k\dim(\mg)}{k+h^\vee},
\end{align}
and conformal weights
\begin{align}
    h_\lambda=\frac{C_2(R_\lambda)}{2({h^\vee}+k)}. 
\end{align}
Assuming these formulas still hold for the intermediate cases at least for small levels, then the affine VOA of level $1$ associated with intermediate exceptional series \eqref{eq:ies} denoted as $L_1(\mg_I)$  has  central charge
\begin{align}
    c = \frac{24(2a+3)}{5(a+4)}=\frac{12(4\hv-1)}{5(\hv+6)} .
\end{align}
Except for $\mg_I=\Eh$, the $L_1(\mg_I)$ always has four or more irreducible modules. Up to degeneracy, there exist four distinct characters. We observe the four conformal weights are\footnote{The second and third conformal weights for $\Eh$ are $\frac{19}{30},\frac{5}{6}$, which are spurious.}
\begin{align}\label{eq:level1weights}
  h_i=  0,\frac{2(2a+3)}{5(a+4)},\frac{a+2}{a+4},\frac{4}{5},\qquad i=0,1,2,3.
\end{align}
The $h_{1,2,3}$ correspond to the modules $R_{1,2,3}$ of $\mg_I$ respectively and also give their quadratic Casimir invariants. It is easy to check that the above central charge $c$ and conformal weights always result in a 4th order holormophic MLDE regardless of the value of $a$. 
Precisely, the characters of $L_1(\mathfrak{g}_I)$ satisfy the MLDE
\begin{align}\label{eq:4MLDE}
    \Big[D^4+\mu_1 E_4D^2+ \mu_2 E_6D +\mu_3 E_4^2\Big]\chi=0,
\end{align}
with 
\begin{align}\nonumber
    \mu_1=   -\frac{203 a^2+724 a+1448}{900 (a+4)^2},\ \mu_2=\frac{811 a^2+3068 a+2896}{5400 (a+4)^2},\ \mu_3=-\frac{(2 a+3)^2 (6 a^2+53 a+91)}{625 (a+4)^4}.
\end{align}
The MLDE \eqref{eq:4MLDE} has some solutions satisfying the CFT-type condition: 1) unique vacuum; 2) all Fourier coefficients are non-negative integers. Such solutions were classified in e.g. \cite{Kawasetsu:2018tzs} and \cite{Kaidi:2021ent}, following the MMS classification of 2nd order holomorphic MLDE solutions \cite{Mathur:1988na}. The characters of $L_1(\mathfrak{g}_I)$ are part of the solutions satisfying the CFT-type condition. One of the advantages of intermediate Lie algebra interpretation is that the irreducible modules give suitable normalizations of the MLDE solutions just like the simple Lie algebra case. The exact formulas as modular forms for the MLDE solutions with leading Fourier coefficients normalized as $1$ have been given in \cite{Kawasetsu:2018tzs}. 

Notice the difference between conformal weights $h_2-h_1=1/5$ in \eqref{eq:level1weights}. In fact, this is necessary for the following general coset constructions for $L_1(\mathfrak{g}_I)$:
\begin{equation}\label{cosetLY}
L_1(\mathfrak{g}_I)=\frac{L_1(\mathfrak{g}_{\rm CD})}{M_{\rm eff}(5,2)},
\end{equation}
and
\begin{align}\label{cosetM}
    L_1(\mathfrak{g}_I)=L_1(\mathfrak{g}_{\rm sub})\otimes M_{\rm eff}(5,3)= \frac{L_1(\mathfrak{g}_{\rm sub})}{M(5,3)}.
\end{align}
The 
$\otimes$ in \eqref{cosetM} is not exactly tensor product, but should be understood as as an inverse action of the coset ${L_1(\mathfrak{g}_{\rm sub})}/{M(5,3)}$. We will show one example for $\Dh$ later. These coset constructions can also be regarded as the defining property of the intermediate exceptional series. 
The exact character relations of the coset \eqref{cosetLY} are\footnote{For $\mathfrak{g}_{\rm CD}=E_8$, the second relation is spurious.}
\begin{equation}\label{eq:cosetlevel1}
    \begin{aligned}
&\chi_0^{\mathfrak{g}_{\rm CD}}= \chi_0^{\mathfrak{g}_I}\phi_0 +\chi^{\mathfrak{g}_I}_{R_3}\phi_1   , \\
&\chi_{f }^{\mathfrak{g}_{\rm CD}}=  \chi^{\mathfrak{g}_I}_{R_1} \phi_1 + \chi^{\mathfrak{g}_I}_{R_2}\phi_0 .
    \end{aligned}
\end{equation}
The $\phi_0,\phi_1$ are the characters of effective Lee-Yang model ${M_{\rm eff}(5,2)}$, i.e., the renowned Roger--Ramanujan functions 
\begin{align} \label{RRf}
\phi_0=q^{-\frac{1}{60}}\prod_{n=0}^{\infty}\frac{1}{(1-q^{5n+1})(1-q^{5n+4})},\qquad
\phi_1=q^{\frac{11}{60}}\prod_{n=0}^{\infty}\frac{1}{(1-q^{5n+2})(1-q^{5n+3})}.
\end{align}
Naively, all $L_1(\mathfrak{g}_I)$ has negative fusion coefficients due to the coset with $M_{\rm eff}(5,2)$. This can be cured by making a shuffle among the vacuum and non-vacuum primaries, resembling the effective description of non-unitary minimal models. The exact character relations of coset \eqref{cosetM} are more complicated. We will give some examples in the following sections. 

As vector-valued modular forms of degree $4$, the characters of many $L_1(\mathfrak{g}_I)$ have been realized by Hecke operator in \cite{Harvey:2018rdc,Duan:2022ltz}. What we now find are the WZW model interpretation for these Hecke images. We collect some useful information of affine VOA $L_1(\mathfrak{g}_I)$ in Table \ref{tab:ieslevel1}.

\begin{table}[ht]
\def\arraystretch{1.2}
 \begin{center}
\caption{Affine VOA $L_1(\mathfrak{g})$ associated with the intermediate exceptional series including $IM$ and $\Dhh$. Here $c$ is the central charge, $\#\chi$ is the number of distinct characters, $\#\phi$ is the number of conformal primaries, $h$ are the non-vacuum conformal weights with degeneracy, $N$ is the conductor. We also add the Hecke operator interpretation when there is one. The $D_{2\rm A}$ is a rational VOA called \textit{Dihedral U2A} with central charge $6/5$ and associated to the 2A conjugacy class of the Monster group \cite{lam2000z2}.}\label{tab:ieslevel1}
\begin{tabular}{c|cccccccccccc}
\hline
$\mathfrak{g}$    & $c$ & $\#\chi$  &  $\#\phi$ &$h$  & $N$  & Hecke & Coset \\
 \hline
 $IM$   &$ 3/5 $& $ 4 $ & $ 4 $ & $\frac{1}{20},\frac{1}{4},\frac{4}{5}$ & $  40$ &   $-$ & ${L_1(A_1)}/{M_{\rm eff}(5,2)}$   \\
 $U\!A_{1+1/2} $    & $ 8/5 $ &$ 4 $ &$ 6 $   &  ${(\frac{2}{15})_2,(\frac{1}{3})_2,\frac{4}{5}} $ & $  15$ &  $-$ & ${L_1(A_2)}/{M_{\rm eff}(5,2)}$ \\
 $AG_{1+1/2} $    & $12/5$  &$ 4 $ & $ 4 $ &  $\frac{1}{5},\frac{2}{5},\frac{4}{5}$ & $  10$ &    $- $   &  ${L_1(G_2)}/{M_{\rm eff}(5,2)}$    \\
 $AD_{3+1/2}$    &     $18/5$  & $ 4 $ & $ 8 $ & $(\frac{3}{10})_3,(\frac{1}{2})_3,\frac{4}{5}  $ & $  20$ &  $\T_3D_{2\rm A}$ &  ${L_1(D_4)}/{M_{\rm eff}(5,2)}$ \\
 $C_{3+ {1}/{2}}$   &  $ 24/5  $ &$ 4 $ & $ 4 $ &${\frac25,\frac35,\frac45} $ &  $ 5$ &   $\T_2 L_1(\AG )  $  &  ${L_1(F_4)}/{M_{\rm eff}(5,2)}$ \\
 $A_{5+{1}/{2}}$     & $ 28/5  $   &$ 4 $ &$ 6 $ & $(\frac{7}{15})_2 ,(\frac23)_2,\frac45$& $  30$ & $\T_7M_{\rm sub}(6,5)$  & ${L_1(E_6)}/{M_{\rm eff}(5,2)}$ \\
 $D_{6+{1}/{2}}$   & $ 33/5  $   &  $ 4 $ & $ 4 $ & $\frac{11}{20} ,\frac34,\frac45$& $  40$ & $\T_{11} {M_{\rm eff}(5,3)}$  & ${L_1(E_7)}/{M_{\rm eff}(5,2)}$ \\
${D}_{6+{1}/{2}+{1}/{2}}$ &   $36/5$   & $ 3 $ &$ 4 $ &${\frac35,(\frac45})_2 $ & $  10$ & $\T_9 M_{\rm eff}(5,2)^2 $  &  ${L_1(E_{7+1/2})}/{M_{\rm eff}(5,2)}$  \\
 $E_{7+{1}/{2}}$   &   $ 38/5  $ & $ 2 $ & $ 2 $ &  $\frac45$ & $60$ &  $\T_{19}{M_{\rm eff}(5,2)}$ & ${L_1(E_8)}/{M_{\rm eff}(5,2)}$  \\  
   \hline
\end{tabular}

 \end{center}
\end{table}

The intermediate VOA $L_1(\mathfrak{g}_I)$ has a close relation with $W$-algebras. The affine $W$-algebra associated with a minimal nilpotent element denoted as $W_k(\mathfrak{g},f_\theta)$ has central charge \cite{Kac:2003jh}
\begin{equation}
    c_{W}=\frac{k\dim\mathfrak{g}}{k+\hv}-6k+\hv-4.
\end{equation}
For $\mathfrak{g}_{\rm CD}$ belonging to the Cvitanovi\'c--Deligne exceptional series, as $\dim(\mathfrak{g}_{\rm CD})$ and $h^\vee$ are both functions of Landsberg--Manivel's parameter $a$, it is easy to obtain the central charge for $W_{-h^\vee/6}(\mathfrak{g}_{\rm CD},f_\theta)$ as 
\begin{align}
    c=\frac{6 (7 a+8 )}{5 (a+4 )} .
\end{align}
These $W$-algebras were well studied in \cite[Theorem 8.1]{Kawasetsu:2018tzs} and proved to be simple rational and $C_2$-cofinite. It is easy to check that for the same Landsberg--Manivel's parameter $a$, the difference between the following two central charges is a constant independent of $a$:
\begin{align}
    c(L_1(\mathfrak{g}_I))-c(W_{-h^\vee/6}(\mathfrak{g}_{\rm CD},f_\theta))=\frac{6}{5}.
\end{align}
This $6/5$ in fact comes from the difference between the central charge $3/5$ of $M_{\rm eff}(5,3)$ and the central charge $-3/5$ of $M(5,3)$. The relation between $L_1(\mathfrak{g}_I)$ and $W_{-h^\vee/6}(\mathfrak{g}_{\rm CD},f_\theta)$ can be thought as they come from the same rational VOA product with $M_{\rm eff}(5,3)$ and $M(5,3)$ respectively. As $M_{\rm eff}(5,3)$ is an effective description rather than genuine VOA, this makes $L_1(\mathfrak{g}_I)$ naively have negative fusion rules. 
One advantage of the intermediate Lie algebra interpretation is that the central charge, conformal weights and module dimensions are directly linked to the characters, and it is easier to obtain the characters of $L_k(\mathfrak{g}_I)$ from MLDE or Hecke operator especially when the level $k$ is higher than $1$. On the other hand, to our knowledge, it is not known yet how to compute the characters of $W_{k}(\mathfrak{g}_{\rm CD},f_\theta)$ for $k>-h^\vee/6$.

\section{Intermediate Lie algebra $D_{6+1/2}$}

\subsection{Weyl vector and quadratic Casimir invariants}
The $D_{6+1/2}$ has rank $r=6$, dual Coxeter number $h^\vee=14$ and dimension $99$. We first define the Weyl vector for $\Dh$ as the sum of all positive roots, i.e.,
\begin{equation}\label{eq:weylvector}
\rho=\rho_{D_6}+\rho_{\bf \overline{32}}=\frac{1}{2}\Big(\Delta_+(D_6)+\frac{1}{2}\mathbf{\overline{32}}\Big).
\end{equation}
Here $\mathbf{\overline{32}}$ is the conjugate spinor module of $D_6$ with highest weight $(000001)$. 
Explicitly, we have in the $D_6$ fundamental weight basis
\begin{equation}
\rho_{D_6}=(1,1,1,1,1,1),\qquad \rho_{\bf \overline{32}} =(1,0,1,0,0,1).
\end{equation}
Thus $\rho=(2,1,2,1,1,2)$. This shows that there exist three fermionic fundamental weights $w_1,w_3$ and $w_6$. 
The comarks for  $w_1,w_3$ and $w_6$ are doubled from those of $D_6$ and we  have the comarks of $\Dh$ as
\begin{align}
  (a_0^\vee,a_1^\vee,\dots,  a_6^\vee)=(1,2,2,4,2,1,2).
\end{align}
The summation gives exactly the dual Coxeter number $\hv=14$ of $\Dh$. The dual Coxeter number can also be computed from
\begin{align}
h^\vee = h^\vee_{D_6}+   \frac{1}{2r}\sum_{w\in \mathbf{ \overline{32}}}\langle w,w\rangle_{D_6}= 10+4=14.
\end{align}
We also checked that $\Dh$ satisfies the Freudenthal--de Vries strange formula
\begin{align}
    \langle\rho,\rho\rangle_{D_6}=\frac{h^\vee \dim(\Dh)}{12}=\frac{231}{2}.
\end{align}

For irreducible integrable module $R_\lambda$ of $\Dh$ with highest weight $\lambda$, we find 
the following conjectural quadratic Casimir invariant formula
\begin{align}\label{eq:DhC2}
C_2(R_\lambda)=\langle\lambda+2\rho,\lambda\rangle_{D_6}+\langle\lambda,\lambda\rangle_{\rm odd},
\end{align}
where the bilinear form $\langle , \rangle_{\rm odd}$ is defined by $\sum_{i\le j}M_{ij}n_in_j$ with the matrix
\begin{align}\label{eq:oddDhcorrection}
M_{ij}=\begin{cases}
    1/2, & i,j\in S_{\rm odd},\, i=j, \\
    -1, &  i,j\in S_{\rm odd},\,i\neq j, \\
    0, & \mathrm{otherwise}.
\end{cases}
\end{align}
This conjectural $C_2$ formula is inspired from the one for $\Eh$ given in \cite{Lee:2023owa} and the conformal weights \eqref{eq:Dhlevel2fweights} for $L_2(\Dh)$ given by fermionic Hecke operator and fermionic MLDE. It passes many further checks from Vogel's universal Casimir invariant formulas, the indices of modules and the conformal weights for $L_k(\Dh)$ with $k\ge 3$. 
The odd correction term is slightly different from the one for $\Eh$ conjectured in \cite[Equation (4.8)]{Lee:2023owa}, such that \eqref{eq:DhC2} produces the correct conformal weights for $L_2(\Dh)$.  More explicitly, we have quadratic Casimir invariant for integrable irreducible module with highest weight $\lambda=\sum n_i w_i$ as 
\begin{align}\label{DhC2}
    C_2=&\,\frac12[n_1 (3 n_1+2 (2 n_2+n_3+2 n_4+n_5+15))+4 n_2 (n_2+2 n_3\\ \nonumber
    &+2 n_4+n_5+n_6+13)+n_3 (7 n_3+12 n_4+6 n_5+4 n_6+70)\\[+1mm] \nonumber&
    +8 n_4 (n_4+n_5+n_6+10)
    +n_5 (3 n_5+4 n_6+42)+4n_6(n_6+11 )].
\end{align}
We will use this conjectual formula to compute the conformal weights for $L_k(\Dh)$.

We draw the analogy of the Dynkin diagram for $\Dh$ in Figure \ref{fig:Dhdynkin}.  We remark that $\Dh$ is on Vogel's plane and has Vogel's parameter $(\alpha,\beta,\gamma)=(-2,6,10)$ while $D_6$ has $(\alpha,\beta,\gamma)=(-2,4,8)$.

\begin{figure}[h]
 \caption{The analogy of the Dynkin diagram for $\Dh$ and irreducible modules associated with fundamental weights. The three white circle nodes are fermionic fundamental weights.}
    \centering
   \begin{center}
\resizebox{7.7cm}{!}{\begin{tikzpicture}{xscale=1cm,yscale=1cm}
\coordinate[label=below:${\bf 12}$,label=above:$1$](B) at (-3.6,0);
\coordinate[label=below:${\bf 99}$,label=above:$2$](C) at (-1.8,0);
\coordinate[label=below:${\bf 252 }$, label=above:$3$](D) at (0,0);
\coordinate[label=below:${\bf  945 }$,label={[label distance=.6mm]45:$4$}](E) at (1.8,0);
\coordinate[label=below:${\bf  33 }$,label=above:$6$](F) at (3.6,0);
\coordinate[label=left:${\bf 44}$,label=right:$5$](H) at (1.8,1.8);
\draw (B)--(C)--(D)--(E)--(F);
\draw (E)--(H);
\draw (B) circle (.08);
\fill (C) circle (.1);
\draw (D) circle (.08);
\fill (E) circle (.1);
\draw (F) circle (.08);
\fill (H) circle (.1);
\end{tikzpicture}}
\end{center}
\label{fig:Dhdynkin}
\end{figure}

\subsection{Irreducible modules of $\Dh$}
\subsubsection{Weyl dimension formula}
For any $\Dh$ irreducible modules associated with the highest weight $\lambda=\sum_{i=2,4,5}n_iw_i$, $n_{i}\in \NN$ which we call \textit{purely bosonic weight}, we find the following direct generalization of Weyl dimension formula 
\begin{equation}\label{weyldim}
\dim(R_\lambda)=\frac{\prod_{\alpha\in\Delta_+(D_6)}\langle\lambda+\rho,\alpha\rangle\prod_{\alpha\in\frac{1}{2}{\bf \overline{32}} }\langle\lambda+\rho,\alpha\rangle}{\prod_{\alpha\in\Delta_+(D_6)}\langle\rho,\alpha\rangle\prod_{\alpha\in\frac{1}{2}{\bf \overline{32}} }\langle\rho,\alpha\rangle}.
\end{equation}
For example, for highest root $\theta=w_2$, we have
\begin{equation}
   \dim(R_{n\theta})=\frac{(2n+13)}{2^{21}
3^{10}
5^5
7^3
11^1
13^1}\prod_{j=1}^{12}(n+j)\prod_{j=3}^{10}(n+j)\prod_{j=5}^{8}(n+j) 
\end{equation}
It is easy to check this is equivalent to the $a=6$ case of \cite[Theorem 7.2]{LM}.
The dimensions of the first a few purely bosonic modules from \eqref{weyldim} are
\begin{align}\nonumber
   1, 44, 99, 891, 945, 3696, 3927, 11440, 24024, 65637, 65637, 89661, 
106964, 128700, 219912, \dots
\end{align}
Interestingly, we notice that the Weyl dimensional formula \eqref{weyldim} also works for the highest weight $\lambda$ when two \textit{distinct} fermionic fundamental weights can pair together,\footnote{We checked this is also true for $\Eh$, which was not realized in \cite{Lee:2023owa}. For example, the dimension of $\Eh$ module with highest weight $(0000111)$ in \cite[Table 2]{Lee:2023owa} was undetermined, but now we can use Weyl dimension formula to compute it as $107137485$.} that is
\begin{align}
    \lambda=(a+b)w_1+(b+c)w_3+(a+c)w_6+\sum_{i=2,4,5}n_iw_i,\qquad a,b,c,n_{i}\in \NN.
\end{align}
We checked for these highest weight, \eqref{weyldim} always gives positive integer dimensions and matches with the predictions when there is overlap with Vogel's \cite{vogel1999universal} or Langsberg--Manivel's  \cite{LMseries} decomposition formulas. For example, for $\lambda=n(w_1+w_6)$, we have 
\begin{align}
 \dim(R_\lambda)=\frac{(n + 6)^5 (n + 7)^4 (n + 8)^2 (n + 9) (n + 10) (2 n + 9) (2 n + 
   11) (2 n + 13)}{2^{22}
3^{12}
5^6
7^4
11^1
13^1
}\prod_{i=1}^5(n+i)^i   .
\end{align}
The first few dimensions are $616, 102816, 7674480, 325014228...$ In particular, $616$ matches with the dimension formula for the $C$ module for the subexcetional series given in \cite{LMseries}. 
Note that irreducible modules with this kind of highest weight are always bosonic, but reversely it is not true. For example, for bosonic module with highest weight $(200000)$, \eqref{weyldim} gives zero, but we know from  \cite{vogel1999universal} and \cite{LMseries} that the dimension should be $77$.

For fermionic modules, we have much less predictions. Similar to the $\Eh$ case, we only have Weyl type dimension formula when there is one single fermionic fundamental weight. For any highest weight $\lambda=\sum_{i=1}^6n_iw_i$ with $n_1+n_3+n_6=1$, we find the modified Weyl dimension formula is 
\begin{equation}\label{weyldimodd}
\dim(R_\lambda)=\frac{\prod_{\alpha\in\Delta_+(D_6)}\langle\lambda+\rho,\alpha\rangle\prod_{\alpha\in\frac{1}{2}{\bf \overline{32}}}\big(\langle\lambda+\rho,\alpha\rangle-\langle\lambda,\alpha\rangle_{\rm odd})}{2\prod_{\alpha\in\Delta_+(D_6)}\langle\rho,\alpha\rangle\prod_{\alpha\in\frac{1}{2}{\bf \overline{32}}}\langle\rho,\alpha\rangle}.
\end{equation}
We checked that this formula also always produces positive integers for arbitrary $n_{2,4,5}\in \NN$, and matches with the predictions when there is overlap with \cite{vogel1999universal} or  \cite{LMseries}. We collect all irreducible modules of $\Dh$ and the corresponding $D_6$ modules and relevant data in Table  \ref{tb:repsDv} for level $k\le 2$ and Table \ref{tb:repsD3v} for level $k=3$. We collect some irreducible modules for level $k\ge 4$ in Table \ref{tb:repsD4v}. The modules with $-$ mean that we cannot predict their dimensions. When a module appears in Vogel's decomposition formulas \cite{vogel1999universal} or Landsberg--Manivel's decomposition formulas \cite{LMseries}, we mark their module names respectively. We remark that Vogel's $X_3,X_3',X_3''$ dimension formulas \cite{vogel1999universal} for $\Dh$ do not give rational numbers. We cannot predict these dimensions either.

\begin{table}[ht]
\def\arraystretch{1.1}
\caption{All irreducible modules of $D_6$ and $\Dh$ with level $k\le 2$. At each level, we order the highest weights by the $C_2$ of $\Dh$.}
	\centering
	\begin{tabular}{|c|c|c|c|c|c|c|c|c|c|c|c|c|c|}
		\hline
$k$&	$\lambda$	   &  $D_6$  & $C_2$  & $I$  & $\Dh$  &  $C_2$  & $I$  & Vogel \cite{vogel1999universal} & LM \cite{LMseries} & F/B\\
		\hline
$1$  &   $100000$ & $\bf 12$ & $ 11  $ & $1$ & $\bf 12$ & $\frac{33}{2} $  & $1$  &  $ -$ &   $ -   $& F  \\ 
$1$  &     $000010$ & $\bf 32$ & $  \frac{33}{2} $ & $4$ & $\bf  44 $ & $\frac{45}{2}   $ & $  5 $ &   $ -$&   $  V  $& B\\
$1$  &  $000001$ & $\bf \overline{32}$ & $ \frac{33}{2}  $ & $ 4$ & $\bf  33 $ & $  24  $  & $ 4  $  &  $-Y_3'' $&   $  -  $ & F \\ \hline
$2$  &  $010000$ & $\bf 66 $  & $ 20  $ & $ 10 $ & $\bf  99$ & $ 28 $&  $14  $ &  $ X_1 $&   $  \mathfrak{g}  $ & B \\
$2$  &  $200000$ & $\bf  77$  & $  24 $ & $  14 $ & $\bf 77 $ & $ 36 $&  $ 14 $ & $ Y_2''  $&   $   Q $  & B\\
$2$  &  $001000$ & $\bf220  $  & $   27   $ & $ 45 $ & $\bf 252 $ & $ \frac{77}{2} $&  $  49$ &  $ -  $&   $  -  $&  F\\
$2$  &  $100010$ & $\bf \overline{352}     $  & $\frac{57}{2}  $ & $ 76  $ & $\bf 495 $ & $ 40 $&  $ 100 $ &  $  - $&   $  L  $&   F\\
$2$  &  $100001$ & $\bf  352 $  & $ \frac{57}{2}  $ & $ 76 $ & $\bf 616 $ & $ \frac{81}{2} $&  $ 126 $ & $ -  $  &   $  C  $  & B\\
$2$  &  $000100$ & $\bf 495 $  & $  32 $ & $ 120 $ & $\bf 945 $ & $ 44 $&  $ 210 $ &    $ Y_2'  $ &   $  V_2  $ & F\\ 

$2$  &  $000020$ & $\bf 462   $  & $ 36   $ & $ 126 $ & $\bf 891 $ & $ 48 $&  $216  $  & $ Y_3'  $&   $  V^{(2)}  $& B\\
$2$  &  $000011$ & $\bf  792   $  & $ 35  $ & $ 210 $ & $\bf 1188 $ & $ \frac{97}{2} $&  $ 291 $  & $ -  $&   $  -  $& F\\
$2$  &  $000002$ & $\bf \overline{462} $  & $ 36  $ & $ 126 $ & $\bf 495' $ & $ 52 $&  $ 130 $ & $ B  $ &   $ -   $& B\\
 \hline
		\end{tabular}
			\label{tb:repsDv}
		\end{table}

\begin{table}[ht]
\def\arraystretch{1.1}
\caption{All irreducible modules of $D_6$ and $\Dh$ with level $k=3$. At each level, we order the highest weights by the $C_2$ of $\Dh$.}
	\centering
	\begin{tabular}{|c|c|c|c|c|c|c|c|c|c|c|c|c|c|}
		\hline
$k$&	$\lambda$	   &  $D_6$  & $C_2$  & $I$  & $\Dh$  &  $C_2$  & $I$  & Vogel \cite{vogel1999universal} & LM \cite{LMseries} & B/F \\
		\hline
$3$  &  $110000$ & $\bf  560$  & $ 33  $ & $ 140 $ & $\bf 924 $ & $ \frac{93}{2} $&  $  217$ & $ - $ &   $  -  $  & F\\
$3$  &  $010010$ & $\bf 1728         $  & $ \frac{77}{2}  $ & $504  $ & $\bf 3696 $ & $ \frac{105}{2} $&  $ 980 $ & $ - $ &   $V\mathfrak{g}    $  & B\\
$3$  &  $010001$ & $\bf \overline{1728} $  & $   \frac{77}{2}  $ & $504  $ & $\bf 2816 $ & $ 54 $&  $ 768 $ & $-C''  $  &   $  -  $ & F\\
$3$  &  $101000$ & $\bf 2079 $  & $ 40  $ & $  630$ & $\bf 4752 $ & $ 56 $&  $ 1344 $ & $ X_2 $ &   $  \mathfrak{g}_2  $ & B \\
$3$  &  $300000$ & $\bf 352' $  & $39   $ & $104  $ & $\bf 352 $ & $ \frac{117}{2} $&  $ 104 $ & $ - $ &   $   - $  & F\\
$3$  &  $200001$ & $\bf \overline{2112} $  & $ \frac{85}{2}  $ & $ 680 $ & $-$ & $ 60 $&  $  -$ & $  -$ &   $  -  $&   F\\
$3$  &  $200010$ & $\bf 2112 $  & $  \frac{85}{2} $ & $ 680 $ & $\bf  3024$ & $ \frac{121}{2} $&  $ 924 $ & $ - $ &   $  VQ  $&   B  \\
$3$  &  $100100$ & $\bf 4928 $  & $  45 $ & $ 1680 $ & $\bf 9900 $ & $\frac{125}{2}  $&  $ 3125 $ & $ - $ &   $   - $&   F\\
$3$  &  $001010$ & $\bf \overline{4928}'         $  & $\frac{93}{2}   $ & $ 1736 $ & $\bf 8316 $ & $ 64 $&  $2688  $ & $ - $ &   $ -   $&   F\\
$3$  &  $001001$ & $\bf 4928' $  & $   \frac{93}{2}   $ & $ 1736  $ & $\bf  13200$ & $\frac{129}{2}  $&  $4300  $ & $ - $ &   $V_3    $&   B\\
$3$  &  $100011$ & $\bf 8085         $  & $  48 $ & $2940  $ & $\bf  20736$ & $66  $&  $ 6912 $ & $ C' $ &   $  VC  $&   B\\
$3$  &  $100020$ & $\bf  4752      $  & $  49   $ & $ 1764 $ & $\bf 9504 $ & $ \frac{133}{2} $&  $3192  $ & $ - $ &   $  VL  $&   F\\
$3$  &  $100002$ & $\bf \overline{4752} $  & $  49     $ & $ 1764 $ & $- $ & $\frac{137}{2}  $&  $ - $ & $ - $ &   $  -  $&   F\\
$3$  &  $000110$ & $\bf  8800        $  & $ \frac{105}{2}  $ & $ 3500 $ & $\bf  24024$ & $ \frac{141}{2} $&  $ 8554 $ & $  -$ &   $  VV_2  $&   B\\
$3$  &  $000101$ & $\bf \overline{8800} $  & $   \frac{105}{2}  $ & $ 3500 $ & $\bf  20020$ & $ 72 $&  $ 7280 $ & $ - $&   $  -  $ &   F\\
$3$  &  $000021$ & $\bf \overline{9504}         $  & $ \frac{113}{2}  $ & $ 4068 $ & $\bf  20592$ & $ 76 $&  $ 7904 $ & $ - $&   $   - $ &   F\\
$3$  &  $000030$ & $\bf  4224         $  & $\frac{117}{2}   $ & $ 1872 $ & $\bf  11440$ & $ \frac{153}{2} $&  $ 4420 $ & $ - $ &   $  V^{(3)}  $&   B \\
$3$  &  $000012$ & $\bf 9504 $  &$ \frac{113}{2}  $ & $ 4068 $ & $- $ & $\frac{157}{2}  $&  $  -$ & $ - $&   $  -  $ &   B\\
$3$  &  $000003$ & $\bf \overline{4224} $  & $   \frac{117}{2}   $ & $ 1872  $ & $-$ & $ 84$&  $ - $ & $ - $ &   $  -  $&   F\\
 \hline

		\end{tabular}
			\label{tb:repsD3v}
		\end{table}

\begin{table}[ht]
\def\arraystretch{1.1}
\caption{Some bosonic irreducible modules of $D_6$ and $\Dh$ with level $k>3$.}
	\centering
	\begin{tabular}{|c|c|c|c|c|c|c|c|c|c|c|c|c|c|}
		\hline
$k$&	$\lambda$	   &  $D_6$  & $C_2$  & $I$  & $\Dh$  &  $C_2$  & $I$  & Vogel \cite{vogel1999universal} & LM \cite{LMseries}   \\
		\hline
$4$  &  $020000$ & $\bf 1638 $  & $  44 $ & $  546$ & $\bf 3927 $ & $ 60 $&  $ 1190 $ &   $  Y_2$  &   $\mathfrak{g}^{(2)}    $     \\

$ 4$  &  $210000$ & $\bf 2860 $  & $ 48  $ & $ 1040 $ & $\bf 5049 $ & $ 68 $&  $ 1734 $ & $B'  $ &   $  \mathfrak{g}Q  $   \\

 $ 4$  &  $110001$ & $\bf 13728 $  & $ \frac{105}{2}  $ & $ 5460 $ & $\bf 40392 $ & $ \frac{145}{2} $&  $ 14790 $ & $ - $ &   $  \mathfrak{g}C  $  \\
$ 4$  &  $010100$ & $\bf  21021$  & $  56 $ & $ 8918 $ & $\bf 65637 $ & $ 76 $&  $ 25194 $ & $ B'' $ &   $  \mathfrak{g}V_{2}  $   \\

$4 $  &  $010020$ & $\bf  21450$  & $ 60  $ & $ 9750 $ & $\bf 65637' $ & $ 80 $&  $ 26520 $ & $ - $ &   $  \mathfrak{g}V^{(2)}  $  \\

$ 4$  &  $002000$ & $\bf  14014$  & $ 60  $ & $ 6370 $ & $\bf  -$ & $ 84 $&  $ - $ & $  X_3$ &   $  -  $  \\
$ 4$  &  $200100$ & $\bf 27456 $  & $ 60  $ & $ 12480 $ & $\bf  -$ & $ 84 $&  $ - $ & $ X_3' $ &   $ V_2Q   $   \\

$ 4$  &  $200002$ & $\bf \overline{27027} $  & $  64  $ & $ 13104  $ & $\bf  102816$ & $  88$&  $ 45696  $ & $  - $ &   $  -   $  \\

$ 4$  &  $100101$ & $\bf 82368 $  & $ \frac{133}{2}   $ & $ 41496  $ & $\bf  323136$ & $\frac{181}{2}  $&  $  147696 $ & $-   $ &   $     $  \\

 $4 $  &  $001011$ & $\bf  90090$  & $ 68  $ & $ 46410 $ & $\bf 353430 $ & $ 92 $&  $ 164220 $ & $  -$ &   $  VV_3  $  \\
 $4 $  &  $100021$ & $\bf 91520 $  & $ \frac{141}{2}  $ & $ 48880  $ & $\bf 340340 $ & $\frac{189}{2}  $&  $ 162435  $ & $ - $ &   $  V^{(2)}C  $   \\
$ 4$  &  $000200$ & $\bf 55055 $  & $ 72  $ & $ 30030 $ & $\bf 219912 $ & $ 96 $&  $ 106624 $ & $ - $ &   $  V_2^{(2)}  $   \\
$4 $  &  $000120$ & $\bf 84942 $  & $ 76  $ & $  48906$ & $\bf  329868$ & $ 100 $&  $166600  $ & $  -$ &   $  V_2V^{(2)}  $  \\
$4 $  &  $000040$ & $\bf 28314 $  & $ 84  $ & $ 18018 $ & $\bf  106964$ & $ 108 $&  $ 58344 $ & $ - $ &   $    V^{(4)}$   \\

\hline
$5 $  &  $020010$ & $\bf 36960 $  & $  \frac{129}{2} $ & $ 18060  $ & $\bf 128700 $ & $ \frac{173}{2} $&  $ 56225 $ & $ - $ &   $  V\mathfrak{g}^{(2)}  $  \\
$ 5$  &  $111000$ & $\bf  57344$  & $ 66  $ & $ 28672 $ & $\bf 219648 $ & $ 90 $&  $ 99840 $ & $ C $ &   $ \mathfrak{gg}_2   $  \\

$5 $  &  $010030$ & $\bf 174240 $  & $ \frac{169}{2}  $ & $ 111540 $ & $\bf 756756 $ & $ \frac{221}{2} $&  $ 422331  $ & $ - $ &   $  \mathfrak{g}V^{(3)}  $   \\
\hline
$6 $  &  $030000$ & $\bf 23100 $  & $ 72  $ & $ 12600 $ & $\bf 89661 $ & $ 96 $&  $ 43472 $ & $ Y_3 $ &   $ \mathfrak{g}^{(3)}   $   \\
 
 \hline

		\end{tabular}
			\label{tb:repsD4v}
		\end{table}

\subsubsection{Decomposition $D_6\subset D_{6+1/2} \subset E_7 $}
First consider $D_6\subset D_{6+1/2}$. It is easy to find from the coset construction \eqref{cosetM} that
\begin{align}
   \bf 99&=\bf66+\overline{32}+1,\\
   \bf 44&=\bf 32 +12\\
   \bf 33&=\bf \overline{32}+1. 
\end{align}
Here $\bf 32$ is the spinor module. To obtain more module decompositions, we apply the method used in \cite[Section 5.3]{Lee:2023owa} for $\Eh$. Simply speaking, we require the multiplicities in the module decomposition in $D_6\subset D_{6+1/2} \subset E_7 $ as small as possible. In fact, almost all of the multiplicities should be $0$ or $1$. Together with the condition that both sides of the decomposition should have the same dimensions and the same indices, this puts very strong constraints on the form of the module decomposition. It turns out for most modules with low levels, we can uniquely determine the  decompositions in $D_6\subset D_{6+1/2} \subset E_7 $. For more level $2$ modules, we find the decompositions
\begin{align}
\bf 252 & =\bf 220+ 32 ,\\
    \bf 495& =\bf \overline{352}+77+66, \\
   \bf  616 &=\bf 352 +220+32+12, \\
   \bf 945 &=\bf 495+\overline{352}+66+\overline{32}  ,\\
     \bf 891& =\bf 462+\overline{352}+77, \\
   \bf 1188 &=\bf 792 + 352+32+12,\\
   \bf 495'& =\bf \overline{462}+\overline{32}+1, 
\end{align}
For level $3$ modules, we find the decompositions
\begin{align}
    \bf 924& =\bf 560 + 352 + 12, \\
    \bf 3696& =\bf 1728 + 792 + 560 + 352 + 220 + 32 + 12, \\
    \bf 2816& ={\bf \overline{1728} + \overline{462} + 495 + 66} + 2\cdot\mathbf{\overline{32}  + 1}, \\
    \bf 4752& =\bf 2079 + \overline{1728} + 495 + \overline{352} + 66 + \overline{32}, \\
    \bf 3024& =\bf 2112 + 352’ + 560,\\
    \bf 9900& =\bf 4928 + 2112 + 1728 + 560 + 352 + 220,\\
    \bf 9504& =\bf 4752 + 2112 + 1728 + 560 + 352',\\
    \bf 24024& =\bf 8800 + 4752 + 4928 + 2112 + 1728 + 792 + 560 + 352,\\
    \bf 11440& =\bf 4224+4752+2112 + 352'.
\end{align}
For level $4$ modules, we find the decomposition
\begin{align}
    \bf 3927 & =\bf 1638 +\overline{ 1728} + \overline{462} + 66 + \overline{32} + 1.
\end{align}
For level $6$ modules, we find the decomposition
\begin{align}
    \bf 89661 & =\bf 23100+ \overline{36960}+\overline{21450}+\overline{4224}+\overline{1728}+1638+\overline{462}+66+\overline{32}+1.
\end{align}

Now consider $D_{6+1/2} \subset E_7$. We determine  the following module decompositions
\begin{align}
\bf 56&=\bf 44 +12,\\
   \bf 133&=\bf 99+33+1,\\
    \bf 912& =\bf 616+252+44 ,\\
     \bf 1463 & =\bf 891+495 +77,\\
      \bf 1539& =\bf 945 + 495 + 99,\\
       \bf 6480& =\bf 3696 + 924 + 1188 + 616 + 44 + 12,\\
       \bf 7371 & =\bf 3927  + 2816 + 495' + 99 + 33 + 1,\\
       \bf 8645& =\bf 4752 + 2816 + 945 + 99 + 33,\\
       \bf 24320& =\bf 11440 + 9504 + 3024 + 352,\\
       \bf 51072& =\bf 24024 + 9504 + 9900 + 3024 + 3696 + 924  .
\end{align}
It is easy to check that in all above decompositions, both sides have the same dimensions and the same indices.

\subsubsection{Tensor product decomposition}
Following Vogel's universal decomposition formulas \cite{vogel1999universal}, for double tensor products, we have for $\mg=\bf 99$,
\begin{align}
    \mathrm{Sym}^2\mathfrak{g}&= \mathbf{3927}  + \mathbf{945}+ \mathbf{77}+\mathbf{1},\\
    \mathrm{Alt}^2\mathfrak{g} &=  \mathbf{ 4752}  + \mathbf{99} .  
\end{align}
For triple tensor products, we have 
\begin{align}
    (3)\mathfrak{g} &= 2\times\mathbf{99+4752+495'+5049+65637+89661+891-33},\\
    (21)\mathfrak{g} &= 2\times\mathbf{99}+2\times\mathbf{4752+3927+945+77+495'+5049+65637}\\ \nonumber
    &\phantom{=}\,+\mathbf{219648+20736-2816} ,\\
    (111)\mathfrak{g} &= \bf  1+ 4752+3927+945+77+ \mathfrak{g}_3   .
\end{align}
The $(111)\mathfrak{g} $ case need special care. The $ \mathfrak{g}_3$ is a reducible module with dimension $147147$, which is consistent with the dimension formula in \cite{LMseries}. We believe for $\Dh$, $ \mathfrak{g}_3$ should decompose to three irreducible modules. 

For the distinguishing module $V=\mathbf{44}$, by \cite{LMseries} we have
\begin{align}
(2)\mathbf{44}&=\mathbf{ 99+891},\\
    (11)\mathbf{44}&=\mathbf{945+1},\\
    (3)\mathbf{44}&=\mathbf{44+11440+3696 },\\
    (21)\mathbf{44}&=\mathbf{44+616+3696+24024},\\
    (111)\mathbf{44}&=\mathbf{13200+44},\\
    (4)\mathbf{44}&=\mathbf{1 + 945 + 891 + 106964 + 65637 + 3927  },\\
    (31)\mathbf{44}&=\mathbf{ 945 + 2 891 + 2 99 + 20736 + 65637 + 65637 + 4752 + 329868 },\\
    (22)\mathbf{44}&=\mathbf{1 + 2 945 + 20736 + 65637 + 77 + 3927 + 219912  },\\
    (211)\mathbf{44}&=\mathbf{ 945 + 891 + 99 + 20736 + 65637 + 4752 + 495 +353430},\\
    (1111)\mathbf{44}&=V_4+\mathbf{ 945+1}.
\end{align}
Here the symmetric tensor products of $V$ is inspired by the following universal formula for $C_3,A_5,D_6,E_7$ given in \cite{LMseries}
\begin{align}
    \sum_{k=0}^\infty t^k\mathrm{Sym}^kV=\frac{1}{(1-tV)(1-t^2\mathfrak{g})(1-t^3V)(1-t^4)(1-t^4V_2)}.
\end{align}
The $V_4$ is reducible and $\dim V_4=134805$. At least, $V_4$ should contain $R_{(010002)}$ and $R_{(002000)}$ and possibly more. 
As comparison, for $D_6$, $V_4 =\mathbf{14014 + 21450}.$

Following \cite{LMseries}, we further have the tensor product decompositions
\begin{align}
    \mathbf{99}\times \mathbf{44} & =\mathbf{44+616+3696},\\
    \mathbf{ 3927}\times \mathbf{44} & =\mathbf{128700 +3696+ 40392 },\\
    \mathbf{945 }\times \mathbf{44} & =\mathbf{  616 + 3696 + 13200 + 24024 + 44 },\\
    \mathbf{891 }\times \mathbf{44} & =\mathbf{ 3696 + 24024 + 11440 + 44   },\\
    \mathbf{ 616}\times \mathbf{44} & =\mathbf{ 945 + 99 + 20736 + 4752 + 495 + 77 	  },\\
    \mathbf{ 3696}\times \mathbf{44} & =\mathbf{  945 + 891 + 99 + 20736 + 65637 + 65637 + 4752 + 3927  },\\
    \mathbf{ 13200}\times \mathbf{44} & =\mathbf{ 945 + 20736 + 65637 + 4752 + 495 + 353430 }+ V_4,\\
    \mathbf{ 24024}\times \mathbf{44} & =\mathbf{ 945 + 891 + 20736 + 65637 + 65637' + 353430 + 329868 + 219912 },\\
    \mathbf{ 11440}\times \mathbf{44} & =\mathbf{ 891 + 65637' + 329868 + 106964  }.
\end{align}
and\footnote{There is a missing $Q$ for $(2)V_2$ in \cite[Section 6]{LMseries}.} 
\begin{align}
    (2) \mathbf{891} & =\mathbf{1 + 945 + 65637 + 106964 + 3927 + 219912 },\\
   (11) \mathbf{891} & =\mathbf{891+99+65637+329868  },\\
   (2) \mathbf{945} & =\mathbf{ 219912 + 134805 + 65637' + 3927 + 2 945 + 1 + 20736+77 },\\
   (11) \mathbf{945} & =\mathbf{891 + 99 + 20736 + 65637 + 4752 + 495 + 353430 },
   \end{align}
and 
\begin{align}
    \mathbf{ 945}\times \mathbf{99} & =\mathbf{  945 + 99 + 891 + 20736 + 65637 + 4752 + 495 },\\
    \mathbf{ 891}\times \mathbf{99} & =\mathbf{ 945 + 891 + 20736 + 65637'  },\\
    \mathbf{11440 }\times \mathbf{99} & =\mathbf{  11440 + 756756 + 24024+340340 },
    \end{align}
and 
\begin{align}
    \mathbf{ 945}\times \mathbf{891} & =\mathbf{ 945 + 891 + 99 + 20736 + 65637 + 65637 + 4752 + 353430 + 329868   }.
\end{align}

\subsection{Affine VOA $L_k(\Dh)$}
Consider the affine VOA $L_k(\Dh)$ with positive integer level $k$. We observe that they exhibit similar properties as $L_k(\Eh)$, that is, there exists an upper bound $k_{\rm max}$ such that the affine VOA is rational. To be precise, we show that $L_k(\Dh)$ is rational only for $k=1,2,3$. For $L_1(\Dh)$ and $L_2(\Dh)$, we compute their characters and find some coset constructions.

\subsubsection{Affine VOA $D_{6+1/2}$ at level $1$}
Based on the general discussion in Section \ref{sec:level1}, the affine VOA $L_1(D_{6+1/2})$ has central charge and conformal weights
\begin{align}\label{eq:Dhlevel1}
    c=\frac{33}{5},\qquad h_{\lambda}=0, \frac{11}{20} ,\frac34,\frac45.
\end{align}
The characters of this affine VOA as a vector-valued modular form of degree 4 have appeared in the holomorphic modular bootstrap in e.g. \cite[Table 6]{Kaidi:2021ent}. Such a vector-valued modular form was also realized as a $\T_{11}$ Hecke operation on the $M_{\rm eff}(5,3)$ \cite[Section 5.3]{Duan:2022ltz}. Therefore, on the character level we have
\begin{align}
    L_1(D_{6+1/2})=\T_{11}  M_{\rm eff}(5,3).
\end{align}
By Hecke operator, we compute the characters as
\begin{equation}\label{eq:chiDhlevel1}
    \begin{aligned}
\chi_0=\,&q^{-\frac{11}{40}} (1+99 q+1122 q^2+7425 q^3+37191 q^4+ \dots),\\
\chi_{\frac{11}{20}}=\,&q^{\frac{11}{40}} (12+264 q+2024 q^2+11484 q^3+51348 q^4+ \dots),\\
\chi_{\frac{3}{4}}=\,&q^{\frac{19}{40}} (44+660 q+4764 q^2+25388 q^3+110220 q^4+ \dots),\\
\chi_{\frac{4}{5}}=\,&q^{\frac{21}{40}} (33+451 q+3234 q^2+16929 q^3+73062 q^4+ \dots).
    \end{aligned}
\end{equation}
The conductor $N=40$. The $S$-matrix is
\begin{align}\label{Dh1Smatrix}
    S=\left(
\begin{array}{cccc}
\alpha_+  & \alpha_- & \alpha_+ & \alpha_- \\
 \alpha_- & \alpha_+ & -\alpha_- & -\alpha_+ \\
 \alpha_+ & -\alpha_- & -\alpha_+ & \alpha_- \\
 \alpha_- & -\alpha_+ & \alpha_- & -\alpha_+ \\
\end{array}
\right),\qquad \alpha_\pm=\sqrt{\frac{2}{5} \left(\frac{5}{8}\pm \frac{\sqrt{5}}{8}\right)}.
\end{align}
By Verlinde formula, this $S$-matrix would lead to negative fusion coefficients. 

The flavored characters of $L_1(D_{6+1/2})$ have the following Fourier expansion
\begin{equation}
\begin{aligned}
    \chi_0=\,&q^{-\frac{11}{40}} (1+\mathbf{99}q+(\mathbf{945}+\mathbf{77}+\mathbf{99}+1) q^2+(\mathbf{4752 + 495'}\\  &\phantom{--}+\mathbf{ 891 + 945 + 77 }+ 1 + 3\cdot \mathbf{99 - 33}) q^3+ \dots),\\
\end{aligned}
\end{equation}
and
 \begin{align}\label{eq:chiDhlevel1flavored}
\chi_{\frac{11}{20}}=\,&q^{\frac{11}{40}} (\mathbf{12}+(\mathbf{252}+\mathbf{12}) q+(\mathbf{1188} + \mathbf{924} + 2\cdot \mathbf{252} + 2\cdot \mathbf{12} - \mathbf{616}) q^2+ \dots),\\
\chi_{\frac{3}{4}}=\,&q^{\frac{19}{40}} (\mathbf{44}+(\mathbf{616}+\mathbf{44}) q+(\mathbf{3696} + 2\cdot\mathbf{ 616} + 2\cdot\mathbf{ 44} -\mathbf{252}) q^2+\dots),\\
\chi_{\frac{4}{5}}=\,&q^{\frac{21}{40}} (\mathbf{33}+(\mathbf{495 + 33 - 77}) q+(\mathbf{2816} + 2\cdot\mathbf{ 495} + 3\cdot\mathbf{ 33} -\mathbf{ 495'} - \mathbf{99 }- \mathbf{77}) q^2+\dots).
    \end{align}
Here all the Fourier coefficients are the linear combinations of $\Dh$ irreducible modules. 
Note $\mathbf{12}$ and $\mathbf{33}$ are fermionic. We observe that in the Fourier coefficients of $\chi_0$ and $\chi_{\frac34}$, all bosonic modules have positive signs and all fermionic modules have negative signs. On the other hand, in the Fourier coefficients of $\chi_{\frac{11}{20}}$ and $\chi_{\frac45}$, it is exactly the opposite. Similar phenomenon was observed for the flavored characters of $L_1(\Eh)$ in \cite{Lee:2023owa}.

For $L_1(\Dh)$, we have the following coset construction
\begin{equation}\label{coset1}
L_1(\Dh)=\frac{L_1(D_6)}{M(5,3)}=L_1(D_6)\otimes M_{\rm eff}(5,3).    
\end{equation}
$M_{\rm eff}(5,3)$ has central charge $\frac35$, and conformal weights $0,\frac{1}{20},\frac14,\frac45$. Use $M$ short for $M_{\rm eff}(5,3)$, we find the precise character relations are 
\begin{align}
    \chi^{\Dh }_0&= \chi^{D_6}_0\chi^{M}_0+ \chi^{D_6}_{c}\chi^{M}_{\frac14} ,\\
    \chi^{\Dh }_{\frac{11}{20}}&= \chi^{D_6}_{v}\chi^{M}_{\frac{1}{20}}+ \chi^{D_6}_{s}\chi^{M}_{\frac45} ,\\
    \chi^{\Dh }_{\frac{3}{4}}&= \chi^{D_6}_{v}\chi^{M}_{\frac{1}{4}}+ \chi^{D_6}_{s}\chi^{M}_{0},  \\
    \chi^{\Dh }_{\frac{4}{5}}&=  \chi^{D_6}_{0}\chi^{M}_{\frac{4}{5}}+ \chi^{D_6}_{c}\chi^{M}_{\frac{1}{20}}  .
\end{align}
Here $v,s,c$ denote the vector, spinor, conjugate spinor $D_6$ modules $\bf 12, 32, \overline{32}$ respectively with corresponding conformal weights $0,\frac12,\frac34,\frac34$. When the level is $1$, we often use $\mathfrak{g}$ in the upper script short for $L_1(\mg)$.  
One can  check that the flavored characters of $L_1(\Dh)$ are consistent with the above coset relations by decomposing all $\Dh$ modules to $D_6$ modules. 

We also have the coset construction
\begin{equation}\label{coset2}
L_1(\Dh)=\frac{L_1(E_7)}{M_{\rm eff}(5,2)}.    
\end{equation}
The precise character relations with Roger-Ramanujan functions \eqref{RRf} are
\begin{align}
    \chi_0^{E_7 }&= \chi^{\Dh}_0\phi_0+ \chi^{\Dh}_{\frac45}\phi_1 ,\\
    \chi_{\frac34}^{E_7 }&= \chi^{\Dh}_{\frac34}\phi_0+ \chi^{\Dh}_{\frac{11}{20}}\phi_1   .
\end{align}
The $L_1(\Dh)$ is vertex subalgebra of lattice VOA $L_1(E_7)$. 
By \cite[Theorem 2.2]{Kawasetsu}, we have the following expression of the $L_1(\Dh)$ characters 
\begin{align}\label{eq:nahmDh}
\chi_0=\,&q^{-\frac{11}{40}}\cdot \sum_{\mathbf{k}\in Q_{E_7},k_1\ge 0} \frac{q^{ {(\mathbf{k}, \mathbf{k})}/{2}}}{(q)_{k_1}(q)^6_\infty},
\end{align}
\begin{align}
\chi_{\frac{11}{20}}=\,&q^{-\frac{19}{40}}\cdot \sum_{\mathbf{k}\in (P\backslash Q)_{E_7},k_1\ge 1} \frac{q^{{(\mathbf{k}, \mathbf{k})}/{2}}}{(q)_{k_1-1}(q)^6_\infty},
\end{align}
\begin{align}
\chi_{\frac34}=\,& q^{-\frac{11}{40}}\cdot \sum_{\mathbf{k}\in (P\backslash Q)_{E_7},k_1\ge 0} \frac{q^{ {(\mathbf{k}, \mathbf{k})}/{2}}}{(q)_{k_1}(q)^6_\infty},
\end{align}
\begin{align}
    \chi_{\frac{4}{5}}=\,&q^{\frac{21}{40}} \cdot \sum_{\mathbf{k}\in Q_{E_7},k_1\ge 1} \frac{q^{{(\mathbf{k}, \mathbf{k})}/{2}}}{(q)_{k_1-1}(q)^6_\infty}.
\end{align}
Finally we remark that the characters of $L_1(D_{6+1/2})$ coincide with the characters of $W_{-3}(E_7,f_\theta)$ studied in \cite{Kawasetsu:2018irs}.

\subsubsection{Affine VOA $D_{6+1/2}$ at level $2$}\label{sec:Dhlevel2}
The affine VOA $L_2(\Dh)$ has central charge $c=99/8$. 
The conjectural formula \eqref{DhC2} for the quadratic Casimir invariants gives the conformal weights from small to big as
\begin{align}\label{eq:Dhlevel2}
    \left\{0,\frac{33}{64},\frac{45}{64},\frac{3}{4},\frac{7}{8},\frac{9}{8},\frac{77}{64},\frac{5}{4},\frac{81}{64},\frac{11}{8},\frac{3}{2},\frac{97}{64},\frac{13}{8}\right\}.
\end{align}
The denominator $N=64$. It is worthwhile to remark that the rational affine VOA $L_2(D_6)$ has 
central charge $11$ and $13$ primaries with conformal weights from small to big as
\begin{align}
 0,\frac{11}{24}, \left(\frac{11}{16}\right)_2, \frac56,1,\frac98,\left(\frac{19}{16}\right)_2,\frac43,\frac{35}{24},\left(\frac32\right)_2 .
\end{align}
This WZW model is fermionizable. In particular, it can be realized as a fermionic Hecke image $\TF_{11}SM(6,4)$ \cite{Lee:2022yic}. 

We find that $L_2(\Dh)$ is also fermionizable, which has four irreducible Neveu-Schwarz modules. It can be described as the fermionic Hecke image $\TF_{11}$ of effective supersymmetric minimal model $SM_{\rm eff}(16,2)$. 
It can also be obtained as a solution of 4th order fermionic MLDE with index $l=0$, which is the generalization of MLDE on $SL(2,\ZZ)$ to the congruence subgroup of level $2$. Indeed, a $c=\frac{99}{8}$ solution for such MLDE was found in \cite{Duan:2022kxr} which has Neveu-Schwarz and Ramond conformal weights as
\begin{align}\label{eq:Dhlevel2fweights}
    h_{\rm NS}=0,\frac{3}{4},\frac{7}{8},\frac98,\qquad h_{
    \rm R}=\frac{33}{64} ,\frac{
45}{
64} ,\frac{
77}{
64 },{\frac{
81}{
64}}.  
\end{align}
From both fermionic MLDE and fermionic Hecke operator, the conformal weight $h_{(100001)}=81/64$ does not need odd correction, i.e. the quadratic Casimir invariant of $R_{(100001)}$ should be $\frac{81}{2}$ which equals to just $\langle \lambda+2\rho,\lambda \rangle$. This is one main reason that the odd correction function \eqref{eq:oddDhcorrection} is different from that of $\Eh$.

From the   fermionic Hecke operator \cite{Lee:2022yic}, we can obtain the following fermionic characters. The characters of the four irreducible Neveu-Schwarz modules are
\begin{align}\label{D6hlevel2NSchi}
    \chi_0^{\rm NS}&=\chi_0+\chi_{\frac32}=q^{-\frac{33}{64}}( 1+99 q+891 q^{3/2}+5049 q^2+22572 q^{5/2}+
   \dots ),\\
   \chi_{\frac34}^{\rm NS}&=\chi_{\frac34}+\chi_{\frac54}=q^{\frac{15}{64}}(33+495 \sqrt{q}+3267 q+16280 q^{3/2}+67518 q^2+\dots),\\
    \chi_{\frac78}^{\rm NS}&=\chi_{\frac78}+\chi_{\frac{11}{8}}=q^{\frac{23}{64}}(99+945 \sqrt{q}+5874 q+27918 q^{3/2}+111078 q^2+ \dots),\\ \label{D6hlevel2NSchi4}
     \chi_{\frac98}^{\rm NS}&=\chi_{\frac98}+\chi_{\frac{13}{8}}=q^{\frac{39}{64}}(77+495 \sqrt{q}+2574 q+11253 q^{3/2}+42075 q^2+ \dots).
\end{align}
These are called NS characters in the fermionic CFT literature. They form a vector-valued modular form of $\Gamma_{\theta}$ that is a congruence subgroup of level $2$ generated by $T^2$ and $S$. The exact $S$-matrix of \eqref{D6hlevel2NSchi}--\eqref{D6hlevel2NSchi4} is
\begin{align}
    S= \frac{1}{\sqrt{2}}\left(
\begin{array}{cccc}
 \sin \left(\frac{3 \pi }{16}\right) & \cos \left(\frac{3 \pi }{16}\right) & \cos \left(\frac{\pi }{16}\right) & \sin \left(\frac{\pi }{16}\right) \\
 \cos \left(\frac{3 \pi }{16}\right) & -\sin \left(\frac{3 \pi }{16}\right) & \sin \left(\frac{\pi }{16}\right) & -\cos \left(\frac{\pi }{16}\right) \\
 \cos \left(\frac{\pi }{16}\right) & \sin \left(\frac{\pi }{16}\right) & -\cos \left(\frac{3 \pi }{16}\right) & \sin \left(\frac{3 \pi }{16}\right) \\
 \sin \left(\frac{\pi }{16}\right) & -\cos \left(\frac{\pi }{16}\right) & \sin \left(\frac{3 \pi }{16}\right) & \cos \left(\frac{3 \pi }{16}\right) \\
\end{array}
\right)  .
\end{align}
The fusion rules of the NS sector can be computed by Verlinde formula as
\begin{align}
    \psi_{\frac34}\times \psi_{\frac34}& =\psi_{0} -2\psi_{\frac34}+2\psi_{\frac78}+2\psi_{\frac98},\quad \psi_{\frac34}\times \psi_{\frac78}= 2\psi_{\frac34}-\psi_{\frac98},\\
    \psi_{\frac34}\times \psi_{\frac98} &=\psi_{0}+ 2\psi_{\frac34}-\psi_{\frac78}-2\psi_{\frac98},\quad 
    \psi_{\frac78}\times \psi_{\frac78}=\psi_{0} +\psi_{\frac78}+\psi_{\frac98},\\
    \psi_{\frac78}\times \psi_{\frac98} &=-\psi_{\frac34}+\psi_{\frac78}+\psi_{\frac98},\quad 
    \psi_{\frac98}\times \psi_{\frac98}=\psi_{0} -2\psi_{\frac34}+\psi_{\frac78}+\psi_{\frac98}.
\end{align}
The negative signs can be cured if one chooses the NS primary with weight $9/8$ as the effective vacuum. 
The $\widetilde{\rm NS}$ characters in the fermionic CFT literature are given by simply reversing the sign of the half-integer coefficients in \eqref{D6hlevel2NSchi}--\eqref{D6hlevel2NSchi4}. On the other hand, 
the characters of the four irreducible Ramond twisted modules are
\begin{align}\label{D6hlevel2Rchi}
    \chi_{\frac{33}{64}}^{\rm R}&=\chi_{\frac{33}{64}} +\chi_{\frac{97}{64}}=  12+2376 q+60192 q^2+786368 q^3+7135920 q^4+ 
   \dots ,\\
   \chi_{\frac{45}{64}}^{\rm R}&=\chi_{\frac{45}{64}} =q^{\frac{ 3}{16}}(  44+4356 q+93456 q^2+1123848 q^3+9681804 q^4+
   \dots ),\\
   \chi_{\frac{77}{64}}^{\rm R}&=\chi_{\frac{77}{64}} =q^{\frac{ 11}{16}}(  252+9240 q+141372 q^2+1412532 q^3+10770188 q^4+ 
   \dots ),\\ \label{D6hlevel2Rchi4}
   \chi_{\frac{81}{64}}^{\rm R}&=\chi_{\frac{81}{64}} =q^{\frac{ 3}{4}}(  616+20592 q+304776 q^2+2983728 q^3+22438152 q^4+ 
   \dots ).
\end{align}
These are called R characters in the fermionic CFT literature.\footnote{The last three R characters sometimes would be multiplied by a factor $\sqrt{2}$ in physics literature.} The $\widetilde{\rm NS}$ characters and R characters transform to each other under the $S$-action of $SL(2,\ZZ)$. There is one remaining 
$\widetilde{\rm R}$ character 
\begin{align}\label{D6hlevel2Rtchi}
  \chi_{\frac{97}{64}}-\chi_{\frac{33}{64}}=12.   
\end{align}
It is $SL(2,\ZZ)$ invariant as required by the supersymmetry. This shows that after fermionization $L_2(\Dh)$ resembles many renowned $N=1$ SVOAs such as $L_6(A_1)$, $L_3(A_3)$, $L_2(A_5)$, see e.g. \cite{johnson2020supersymmetry}. 
The bosonic characters of all $13$ irreducible modules of $L_2(\Dh)$ can be easily solved from the above relations between fermionic characters and affine characters. They satisfy a 13th order MLDE with index $36$.

Coset construction is very useful in the study of rational VOAs. We find two coset constructions for the fermionization of $L_2(\Dh)$. Following the notation of \cite{Lee:2022yic}, the first one is
\begin{align}\label{cosetA}
\frac{ \cF(E_7)_2}{\cF(\Dh)_2} =F^{-1}\otimes SM_{\rm sub}(80,2).  
\end{align}
Here the supersymmetric minimal model $SM_{\rm eff}(80,2)$ has effective central charge $c=\frac{57}{40}$ and $20$ effective Neveu-Schwarz conformal weights from small to big as 
\begin{align}
    \left\{0,\frac{1}{40},\frac{3}{40},\frac{3}{20},\frac{1}{4},\frac{3}{8},\frac{21}{40},\frac{7}{10},\frac{9}{10},\frac{9}{8},\frac{11}{8},\frac{33}{20},\frac{39}{20},\frac{91}{40},\frac{21}{8},3,\frac{17}{5},\frac{153}{40},\frac{171}{40},\frac{19}{4}\right\}.
\end{align}
We construct a non-diagonal $\Gamma_\theta$ modular invariant of $SM_{\rm eff}(80,2)$ by introducing the following eight extended Neveu-Schwarz primaries with characters
\begin{align}
    \chi^{SM}_0&=\chi^{SM_{\rm eff}(80,2)}_0-\chi^{SM_{\rm eff}(80,2)}_3,\\
    \chi^{SM}_{\frac{1}{40}}&=\chi^{SM_{\rm eff}(80,2)}_{\frac{1}{40}}+\chi^{SM_{\rm eff}(80,2)}_{\frac{21}{40}},\\
    \chi^{SM}_{\frac{3}{20}}&=\chi^{SM_{\rm eff}(80,2)}_{\frac{3}{20}}+\chi^{SM_{\rm eff}(80,2)}_{\frac{33}{20}},\\
    \chi^{SM}_{\frac{1}{4}}&=\chi^{SM_{\rm eff}(80,2)}_{\frac{1}{4}}+\chi^{SM_{\rm eff}(80,2)}_{\frac{19}{4}},\\
    \chi^{SM}_{\frac{3}{8}}&=\chi^{SM_{\rm eff}(80,2)}_{\frac{3}{8}}-\chi^{SM_{\rm eff}(80,2)}_{\frac{11}{8}},\\
    \chi^{SM}_{\frac{9}{10}}&=\chi^{SM_{\rm eff}(80,2)}_{\frac{9}{10}}+\chi^{SM_{\rm eff}(80,2)}_{\frac{17}{5}},\\
    \chi^{SM}_{\frac{9}{8}}&=\chi^{SM_{\rm eff}(80,2)}_{\frac{9}{8}}+\chi^{SM_{\rm eff}(80,2)}_{\frac{21}{8}},\\
    \chi^{SM}_{\frac{91}{40}}&=\chi^{SM_{\rm eff}(80,2)}_{\frac{91}{40}}-\chi^{SM_{\rm eff}(80,2)}_{\frac{171}{40}}.
\end{align}
Note that even though there are negative signs in the above combinations, all Fourier coefficients of the $\sqrt{q}$ series are still non-negative integers. As the $S$-matrix of $SM_{\rm eff}(80,2)$ is known, it is then straightforward to verify that the above eight characters have a closed transformation under the $S$ action of $SL(2,\ZZ)$. One can further verify that the following combination, i.e. the summation of the norm square of all extended NS characters
\begin{align}\label{SM802sub}
    Z=\big|\chi^{SM}_0\big|^2+\big|\chi^{SM}_{\frac{1}{40}} \big|^2+\big|  \chi^{SM}_{\frac{3}{20}}\big|^2+\big|\chi^{SM}_{\frac{1}{4}} \big|^2+\big|\chi^{SM}_{\frac{3}{8}} \big|^2+\big| \chi^{SM}_{\frac{9}{10}}\big|^2+\big|  \chi^{SM}_{\frac{9}{8}}\big|^2+\big|\chi^{SM}_{\frac{91}{40}}\big |^2
\end{align}
is invariant under $T^2$ and $S$ actions of $SL(2,\ZZ)$. This can be viewed as a D-type non-diagonal $\Gamma_\theta$ modular invariants of  $SM_{\rm eff}(80,2)$ from simple current extension.
Then we find and check the precise character relations of the coset construction \eqref{cosetA} as
\begin{align}
 \chi_0^F   \chi^{ \cF(E_7)_2}_0&=\chi^{ \cF(\Dh)_2}_0\chi^{SM}_0+ \chi^{ \cF(\Dh)_2}_{\frac{3}{4}}\chi^{SM}_{\frac{1}{4}}+ \chi^{ \cF(\Dh)_2}_{\frac{7}{8}}\chi^{SM}_{\frac{9}{8}}+ \chi^{ \cF(\Dh)_2}_{\frac{9}{8}}\chi^{SM}_{\frac{3}{8}}   ,\\
  \chi_0^F  \chi^{ \cF(E_7)_2}_{\frac{9}{10}}&=  \chi^{ \cF(\Dh)_2}_{0}\chi^{SM}_{\frac{9}{10}}+ \chi^{ \cF(\Dh)_2}_{\frac{3}{4}}\chi^{SM}_{\frac{3}{20}}+ \chi^{ \cF(\Dh)_2}_{\frac{7}{8}}\chi^{SM}_{\frac{1}{40}}+ \chi^{ \cF(\Dh)_2}_{\frac{9}{8}}\chi^{SM}_{\frac{91}{40}}.
\end{align}

We find one more coset construction for the fermionization of $L_2(\Dh)$ as
\begin{align}\label{cosetB}
\frac{\cF(\Dh)_1^{\otimes 2}}{\cF(\Dh)_2} = SM_{\rm sub}(16,10).    
\end{align}
This resembles the Maverick cosets found in the 90s, but a supersymmetric version. It is worthy to mention that we also find this type of coset construction for $\Eh$ in \cite{Lee:2023owa}. To elaborate this coset, we first show that $(\Dh)_1^{\otimes 2}$ allows fermionization to a theory with three NS characters, i.e., a simple current extension to a $N=1$ SVOA. We find the three NS characters are related to the $(\Dh)_1^2$ characters by the following relations
\begin{align}\label{FDh121}
    \chi_{\mathrm{NS},0}^{\cF(\Dh)_1^{\otimes2}} &=\chi_{0}^{\Dh} \chi_{0}^{\Dh}+\chi_{\frac34}^{\Dh}\chi_{\frac34}^{\Dh} ,\\ \label{FDh122}
    \chi_{\mathrm{NS},\frac{4}{5}}^{\cF(\Dh)_1^{\otimes2}} &=\chi_{0}^{\Dh} \chi_{\frac45}^{\Dh}+\chi_{\frac{11}{20}}^{\Dh}\chi_{\frac34}^{\Dh} ,\\ \label{FDh123}
     \chi_{\mathrm{NS},\frac{11}{10}}^{\cF(\Dh)_1^{\otimes 2}} &=\chi_{\frac{11}{20}}^{\Dh} \chi_{\frac{11}{20}}^{\Dh}+\chi_{\frac45}^{\Dh}\chi_{\frac45}^{\Dh} .
\end{align}
This in fact solves a puzzle on a putative $c=\frac{66}{5}$ solution appearing in the modular bootstrap of 3rd order fermionic MLDE in \cite{Bae:2021mej}. We regard this solution exactly as the Neveu-Schwarz characters of $\cF(\Dh)_1^2$. We checked \eqref{FDh121}-\eqref{FDh123} agree with the $q$-series of the NS-sector solution with central charge $c=\frac{66}{5}$ given in \cite[Table 8]{Bae:2021mej}. Besides, it is easy to see that the second character \eqref{FDh122} has degeneracy two. We further write down the Ramond (R) characters as
\begin{align}
\label{FDh121R}
     \chi_{\mathrm{R},\frac{11}{20}}^{\cF(\Dh)_1^{\otimes 2}} &=\chi_{0}^{\Dh}\chi_{\frac{11}{20}}^{\Dh}+\chi_{\frac34}^{\Dh}\chi_{\frac45}^{\Dh},\\
     \label{FDh122R}
    \chi_{\mathrm{R},\frac34}^{\cF(\Dh)_1^{\otimes2}} &=\chi_{0}^{\Dh} \chi_{\frac34}^{\Dh} ,\\ \label{FDh123R}
    \chi_{\mathrm{R},\frac{27}{20}}^{\cF(\Dh)_1^{\otimes2}} &=\chi_{\frac{11}{20}}^{\Dh} \chi_{\frac45}^{\Dh}  .
\end{align}
We checked \eqref{FDh121R}-\eqref{FDh123R} agree with the $q$-series of the  R-sector solution with $c=\frac{66}{5}$ given in \cite[Table 11]{Bae:2021mej} up to constant factors. 
There exist one more $\widetilde{\rm R}$ character 
\begin{align}\label{FDh12Rtilde}
     \chi_{\widetilde{\mathrm{R}},\frac{11}{20}}^{\cF(\Dh)_1^{\otimes 2}} =\chi_{0}^{\Dh}\chi_{\frac{11}{20}}^{\Dh}-\chi_{\frac34}^{\Dh}\chi_{\frac{4}{5}}^{\Dh} =12.
\end{align}
It is easy to prove the $\widetilde{\rm R}$ character is indeed constant from the $S$-matrix of $L_1(\Dh)$ given in \eqref{Dh1Smatrix}. This is also exactly required by supersymmetry.

We then construct a $\Gamma_\theta$ modular invariant of supersymmetric minimal model $SM(16,10)$. The $SM(16,10)$ has in total 34 NS characters. We choose 20 of them with NS conformal weights  
\begin{align}
   \left\{-\frac{1}{40},0,\frac{1}{20},\frac{1}{8},\frac{9}{40},\frac{7}{20},\frac{27}{40},\frac{7}{8},\frac{11}{10},\frac{27}{20},\frac{77}{40},\frac{9}{4},\frac{119}{40},\frac{19}{5},\frac{17}{4},\frac{209}{40},\frac{55}{8},\frac{81}{10},\frac{81}{8},14\right\}
\end{align}
to define the following extended NS primaries with characters
\begin{align}   \chi^{SM}_0&=\chi^{SM(16,10)}_{0}+\chi^{SM(16,10)}_{14},\\
    \chi^{SM}_{\frac{1}{8}}&=\chi^{SM(16,10)}_{\frac{1}{8}}+\chi^{SM(16,10)}_{\frac{81}{8}},\\
    \chi^{SM}_{\frac{7}{8}}&=\chi^{SM(16,10)}_{\frac{7}{8}}+\chi^{SM(16,10)}_{\frac{55}{8}},\\
    \chi^{SM}_{\frac{9}{4}}&=\chi^{SM(16,10)}_{\frac{9}{4}}+\chi^{SM(16,10)}_{\frac{17}{4}},\\
    \chi^{SM}_{\frac{11}{10}}&=\chi^{SM(16,10)}_{\frac{11}{10}}+\chi^{SM(16,10)}_{\frac{81}{10}},\\
    \chi^{SM}_{\frac{9}{40}}&=\chi^{SM(16,10)}_{\frac{9}{40}}+\chi^{SM(16,10)}_{\frac{209}{40}},\\
    \chi^{SM}_{-\frac{1}{40}}&=\chi^{SM(16,10)}_{-\frac{1}{40}}+\chi^{SM(16,10)}_{\frac{119}{40}},\\
    \chi^{SM}_{\frac{7}{20}}&=\chi^{SM(16,10)}_{\frac{7}{20}}+\chi^{SM(16,10)}_{\frac{27}{20}}.
\end{align}
We also denote $\chi^{SM}_{h}=\chi^{SM(16,10)}_{h}$ for $h=\frac{1}{20},\frac{27}{40},\frac{77}{40},\frac{19}{5}$. Since the $S$-matrix for all $34$ NS characters of $SM(16,10)$ is known, 
we can check that the following combination
\begin{align}\label{SM1610sub}
Z=&\,\big|\chi^{SM}_0\big|^2+\big|\chi^{SM}_{\frac{1}{8}} \big|^2+\big|  \chi^{SM}_{\frac{7}{8}}\big|^2+\big|\chi^{SM}_{\frac{9}{4}} \big|^2+\big|\chi^{SM}_{\frac{11}{10}} \big|^2+\big| \chi^{SM}_{\frac{9}{40}}\big|^2+\big|  \chi^{SM}_{-\frac{1}{40}}\big|^2+\big|\chi^{SM}_{\frac{7}{20}}\big |^2\\ \nonumber
&\,+2 \big|\chi^{SM}_{\frac{1}{20}} \big|^2+2 \big|\chi^{SM}_{\frac{27}{40}} \big|^2+2 \big|\chi^{SM}_{\frac{77}{40}} \big|^2+2 \big|\chi^{SM}_{\frac{9}{5}} \big|^2
\end{align}
is invariant under $T^2$ and $S$ actions of $SL(2,\ZZ)$. This gives the sub-theory we denote as $SM_{\rm sub}(16,10)$ in the coset \eqref{cosetB}. This can also be viewed as a D-type non-diagonal $\Gamma_\theta$ modular invariants of  $SM(16,10)$ from simple current extension.  Then we find and check the following exact character relations for the coset \eqref{cosetB}:
\begin{align}\label{cosetB1}
 \chi_{\mathrm{NS},0}^{\cF(\Dh)_1^{\otimes 2}} &=\chi_{\mathrm{NS},0}^{\cF(\Dh)_2}\chi^{SM}_{0}-\chi_{\mathrm{NS},\frac{3}{4}}^{\cF(\Dh)_2}\chi^{SM}_{\frac{ 9}{4 }}+\chi_{\mathrm{NS},\frac{7}{8}}^{\cF(\Dh)_2}\chi^{SM}_{\frac{1}{8}} +\chi_{\mathrm{NS},\frac{9}{8}}^{\cF(\Dh)_2}\chi^{SM}_{\frac{7}{8}} ,\\ \label{cosetB2}
    \chi_{\mathrm{NS},\frac{4}{5}}^{\cF(\Dh)_1^{\otimes 2}} &=-\chi_{\mathrm{NS},0}^{\cF(\Dh)_2}\chi^{SM}_{\frac{9}{5}}+\chi_{\mathrm{NS},\frac{3}{4}}^{\cF(\Dh)_2}\chi^{SM}_{\frac{1}{20}}-\chi_{\mathrm{NS},\frac{7}{8}}^{\cF(\Dh)_2}\chi^{SM}_{\frac{77}{40}} -\chi_{\mathrm{NS},\frac{9}{8}}^{\cF(\Dh)_2}\chi^{SM}_{\frac{27}{40}} ,\\ \label{cosetB3}
     \chi_{\mathrm{NS},\frac{11}{10}}^{\cF(\Dh)_1^{\otimes 2}} &=\chi_{\mathrm{NS},0}^{\cF(\Dh)_2}\chi^{SM}_{\frac{ 11}{10 }}-\chi_{\mathrm{NS},\frac{3}{4}}^{\cF(\Dh)_2}\chi^{SM}_{\frac{ 7}{20 }}+\chi_{\mathrm{NS},\frac{7}{8}}^{\cF(\Dh)_2}\chi^{SM}_{\frac{9}{40}} +\chi_{\mathrm{NS},\frac{9}{8}}^{\cF(\Dh)_2}\chi^{SM}_{-\frac{1}{40}}  .
\end{align}
Here \eqref{cosetB2} has degeneracy two on both sides which is consistent.

We expect that $L_2(D_{6+1/2})$ is closely related to the $W$-algebra $W_{-2}(E_7,f_\theta)$. The latter has central charge $c_W=75/8$. The characters and coset constructions we find for $L_2(D_{6+1/2})$ should be useful to study the properties of $W_{-2}(E_7,f_\theta)$. For example, the coset of $W$-algebras
\begin{align}
   \frac{W_{-3}(E_7,f_\theta)\otimes W_{-3}(E_7,f_\theta)}{ W_{-2}(E_7,f_\theta)} 
\end{align}
has central charge $c=57/40$. We conjecture that it is  isomorphic to the effective $N=1$ supersymmetric minimal model $SM_{\rm eff}(16,10)$.

\subsubsection{Affine VOA $D_{6+1/2}$ at level $3$} 
The conjectural rational VOA $L_3(\Dh)$ has central charge $c=\frac{297}{17}$. The 32 conformal weights computed from the conjectural $C_2$ formula \eqref{DhC2} are
\begin{equation}\label{eq:Dhlevel3}
\begin{aligned}
h_\lambda=\    &  0,\frac{33}{68},\frac{45}{68},\frac{12}{17},\frac{14}{17},\frac{18}{17},\frac{77}{68},\frac{20}{17},\frac{81}{68},\frac{22}{17},\frac{93}{68},\frac{24}{17},\frac{97}{68},\frac{26}{17},\frac{105}{68},\frac{27}{17},\frac{28}{17},\\
   &\frac{117}{68},\frac{30}{17},\frac{121}{68},\frac{125}{68},\frac{32}{17},\frac{129}{68},\frac{33}{17},\frac{133}{68},\frac{137}{68},\frac{141}{68},\frac{36}{17},\frac{38}{17},\frac{9}{4},\frac{157}{68},\frac{42}{17} .
\end{aligned}
\end{equation}
The conductor $N=136 $. Clearly $L_3(\Dh)$ has no super structure and no non-diagonal modular invariants. 

\subsubsection{Higher levels}\label{sec:Dhhighlevel}
To discuss the rationality of $L_k(\Dh)$ and also $L_k(\Ah)$, we need to exploit the MLDE indices. Here we briefly review the technique. 
A degree $d$ MLDE of $SL(2,\ZZ)$ has the form
\begin{align}
    \Big[D^d+\sum_{i=0}^{d-1}\phi_i(\tau)D^{i}\Big]\chi(\tau)=0.
\end{align}
Here $D$ is the Serre derivative and $\phi_i(\tau)$ is a meromorphic modular form of weight  $2d-2i$ on ${SL}(2,\ZZ)$. To measure how meromorphic the MLDE is, there exists a non-negative integer $l$ called the index. In general, 
the MLDE has $d$ number of independent solutions as series $q^{\alpha_i}\sum_{j=0}^\infty a_{ij}q^j$. By a Wronskian analysis \cite{Mathur:1988na}, the index $l$ is related to the degree $d$ and the exponents $\alpha_i$ by 
\begin{align}\label{eq:ldef}
\frac{l}{6}=\frac{d(d-1)}{12}-\sum_{i=1}^{d}\alpha_i.
\end{align}
All characters of a rational VOA should form a vector-valued modular form of weight $0$. Each character has the well-known exponent $\alpha_i=-\frac{c}{24}+h_i$. 
If all $d$ characters of a rational VOA satisfy a MLDE of degree $d$, then the MLDE index $l$ computed from \eqref{eq:ldef} must be a non-negative integer. This puts some constraints on the number of characters $d$, central charge $c$ and conformal weights $h_i$ together. If a VOA has $d$, $c$ and $h_i$ resulting in a non-integer index $l$ by \eqref{eq:ldef}, then we deduce that the VOA cannot be rational. By this method, we found in \cite{Lee:2023owa} that $L_k(\Eh)$ cannot be rational for $k>5$.

If the irreducible modules of $\Dh$ are labeled by the $D_6$ Dynkin diagram, the number $r(k)$ of irreducible modules at level $k$ is generated by 
\begin{align}
 \sum_{k=0}^\infty r(k)x^k=   \frac{1}{(1-x)^4(1-x^2)^3}=  1+4 x+13 x^2+32 x^3+71 x^4+140 x^5+259 x^6+  \dots
\end{align}
Assuming the our conjectural $C_2$ formula \eqref{DhC2} for $\Dh$ is correct, it is easy to compute all conformal weights and the MLDE index $l(k)$ for $L_k(\Dh)$ with arbitrary positive integer level $k$. We collect the indices of the MLDE for the characters of $L_k(D_6)$ and $L_k(\Dh)$ in Table \ref{tb:lkD6}. We can see that for $k\ge 4$, the index $l(k)$ of $L_k(\Dh)$ ceases to be integer. This contradicts with the fact that all characters of a RCFT should form a vector-valued modular form of weight $0$ and satisfy a MLDE with non-negative integer index. Therefore, $L_k(\Dh)$ cannot be rational for $k\ge 4$. This is parallel to the main statement for $\Eh$ in \cite{Lee:2023owa} that $L_k(\Eh)$ cannot be rational for $k\ge 6$.

\begin{table}[h]
\def\arraystretch{1.1}
  \caption{The number of characters $r(k)$ and MLDE indices $l(k)$ for $L_k(D_6)$ and $L_k(\Dh)$.}\label{tb:lkD6}
  \centering
  \begin{tabular}{c|c|c|c|c|c|c|c}\hline
level &  $k$  & $1$ & $2$ & $3$ & $4$& $5$& $6$  \\ \hline
& $r(k)$  &  $3  $ & $10$ & $22 $& $  49$& $ 91$ & $ 168$ \\  
$D_6$ & $l(k)$ &  $   0$ &  $  15 $ &  $ 135  $ &  $  870 $ &  $3390   $&  $   12420$  \\ \hline
& $r(k)$  &  $ 4 $& $13$ & $32$ & $71$ & $140$ & $259$ \\ 
$D_{6+1/2}$ & $l(k)$ &  $  0 $ &  $ 36  $&  $ 340  $ &  $ \frac{12101}{6}  $&  $\frac{162856}{19}   $ &  $ \frac{123291}{4}  $    \\ \hline
     \end{tabular}
\end{table}

\subsection{$D_{6+1/2}$ as gauge algebra}\label{sec:Dhgauge}
It was suggested in \cite{Lee:2023owa} that the intermediate Lie algebra $\Eh$ may serve as a gauge algebra for supersymmetric gauge theories in dimension four, five and six with eight supercharges, which might be geometrically engineered from certain exotic local Calabi-Yau threefolds. Similarly we expect $\Dh$ can also serve as a gauge algebra at least in dimension four and five. Assuming the celebrated 3d monopole formula of Benvenuti--Hanany--Mekareeya \cite{Benvenuti:2010pq} still holds, the {5d one $\Dh$ instanton Nekrasov partition function} should have the following expression 
\begin{equation}\label{Dh1inst}
\begin{split}
  Z_1^{\rm Nek}&=v^{h^\vee-1}\sum_{n=0}^\infty v^{2n}\dim(n\theta)\\ 
  &= v^{13}(1+99v^2+ 3927v^4+ 89661v^6+ 1387386v^8+ 15991118v^{10}+\dots)\\
&=  \frac{1}{(v-v^{-1})^{13}}\Big(v^{\pm13}+73 v^{\pm11}+1678 v^{\pm9}+17134 v^{\pm7} \\
& \quad +90025 v^{\pm5}+262977 v^{\pm3}+445302 v^{\pm1} \Big).
\end{split}
\end{equation}
This is also the Hilbert series for the putative one $\Dh$ instanton moduli space. The rational function \eqref{Dh1inst} obtained from the infinite summation is palindromic, i.e., symmetric for $v\leftrightarrow v^{-1}$. This shows that it satisfies the criterion of Stanley \cite{stanley1978hilbert} for the Cohen–Macaulay ring of the moduli space to be Gorenstein, as in the $\Eh$ case \cite{Lee:2023owa}. This is an important necessary condition for a valid one-instanton Nekrasov partition function. It would be very interesting if one can construct the Nekrasov partition function with all gauge fugacities.

Moreover, the conjectural one $\Eh$ instanton moduli space is expected to be the associated variety of the affine VOA $V_{-5}(\Eh)$ \cite{Lee:2023owa}, following the results on $V_{-{h^\vee}/{6}-1}({\mathfrak{g}})$ for the Cvitanovi\'c--Deligne  exceptional series in \cite{AM,AK}. It would be interesting to study what VOA associated with $\Dh$ at negative levels could be related to instanton moduli space and 4d $N=2$ SCFTs.

\section{Intermediate Lie algebra $A_{5+1/2}$}
\subsection{Weyl vector and quadratic Casimir invariants}
The intermediate Lie algebra $\Ah$ has rank $r=5$, dual Coxeter number $h^\vee=9$ and dimension $56$. We define the Weyl vector for $\Ah$ as the sum of all positive roots, i.e.,
\begin{equation}\label{eq:weylvectorA}
\rho=\rho_{A_5}+\rho_{\bf 20}=\frac{1}{2}\Big(\Delta_+(A_5)+\frac{1}{2}\mathbf{20}\Big).
\end{equation}
Here $\bf 20$ is the $A_5$ module with highest weight $(00100)$. Explicitly, we have in the fundamental weight basis
\begin{equation}
\rho_{A_5}=(1,1,1,1,1),\qquad \rho_{\bf 20} =(1,0,1,0,1).
\end{equation}
Thus $\rho=(2,1,2,1,2)$. This shows that there exist three fermionic fundamental weights $w_1,w_3$ and $w_5$ for $\Ah$. The comarks of $\Ah$ then become 
\begin{align}
    (a_0^\vee,a_1^\vee,\dots,a_5^\vee)=(1,2,1,2,1,2).
\end{align}
The summation gives exactly the dual Coxeter number $h^\vee =9$. The dual Coxeter number can also be computed from
\begin{align}
h^\vee = h^\vee_{A_5}+   \frac{1}{2r}\sum_{w\in \mathbf{ 20}}\langle w,w\rangle_{A_5}= 6+3= 9.
\end{align}
We also checked that $\Ah$ satisfies the Freudenthal--de Vries strange formula
\begin{align}
    \langle\rho,\rho\rangle_{A_5}=\frac{h^\vee \dim(\Ah)}{12}=42.
\end{align}

Inspired by the results for $\Dh$ and $\Eh$, we conjecture the following quadratic Casimir invariant $C_2$ formula for the irreducible module of $\Ah$:
\begin{align}\label{eq:AhC2formal}
C_2(R_\lambda)=\langle\lambda+2\rho,\lambda\rangle_{A_5}+\langle\lambda,\lambda\rangle_{\rm odd},
\end{align}
where we use the same odd correction \eqref{eq:oddDhcorrection} as in $\Dh$. One main reason is that the adjoint module $R_{(10001)}$ of $\Ah$ has quadratic Casimir invariant $C_2=18$ which receives zero odd correction. For the highest weight $\lambda=\sum_{i=1}^5n_iw_i$, the main part 
\begin{align}
    \langle\lambda+2\rho,\lambda\rangle_{A_5}=\,&\frac{1}{6}(5 n_{1}^2+2 n_{1} (4 n_{2}+3 n_{3}+2 n_{4}+n_{5}+24)+8 n_{2}^2+4 n_{2} (3 n_{3}+2 n_{4}+n_{5}+18)\\ \nonumber
    &+9 n_{3}^2+12 n_{3} n_{4}+6 n_{3} n_{5}+84 n_{3}+8 n_{4}^2+8 n_{4} n_{5}+72 n_{4}+5 n_{5}^2+48 n_{5}).
\end{align}
We notice that \eqref{eq:AhC2formal} works perfectly well when one of three fermionic  fundamental weights does not contribute, i.e. one of $n_1,n_3,n_5$ vanishes. However, when all $n_{1,3,5}$ are nonzero, one needs to choose arbitrary two, say $n_1,n_3$ and reduce both by $\min(n_1,n_3)$. This makes the odd correction for $(10101)$ is $+1/2$ instead of $-3/2$. The reason for this reduction is unclear to us. Nevertheless for level $k\le 3$, this procedure only affects the module with the highest weight $(10101)$. 
This conjectural $C_2$ formula passes many checks from the indices of irreducible modules and the conformal weights of affine VOAs. We draw the analogy of the Dynkin diagram of $\Ah$ in Figure \ref{fig:Ahdynkin}. The irreducible modules associated with each fundamental weight will be discussed later. 

At last, we remark that $\Ah$ is not on Vogel's plane. The naive anticipation of Vogel's parameters $(\alpha,\beta,\gamma)=(-2,4,7)$ actually gives the simple Lie algebra $B_5$ with dimension $55$, instead of $\Ah$. In comparison, $A_5$ has Vogel's parameters $(\alpha,\beta,\gamma)=(-2,2,6)$.

\begin{figure}[h]
 \caption{The analogy of Dynkin diagram for $\Ah$ and irreducible modules associated with fundamental weights. The three circled nodes are fermionic fundamental weights.}
    \centering
   \begin{center}
\resizebox{7.5cm}{!}{\begin{tikzpicture}{xscale=1cm,yscale=1cm}
\coordinate[label=below:${\bf 6}$,label=above:$1$](B) at (-3.6,0);
\coordinate[label=below:${\bf 21}$,label=above:$2$](C) at (-1.8,0);
\coordinate[label=below:${\bf  21'}$, label=above:$3$](D) at (0,0);
\coordinate[label=below:${\bf  \overline{21}}$,label=above:$4$](E) at (1.8,0);
\coordinate[label=below:${\bf  \overline{6}}$,label=above:$5$](F) at (3.6,0);
\draw (B)--(C)--(D)--(E)--(F);
\draw (B) circle (.08);
\fill (C) circle (.1);
\draw (D) circle (.08);
\fill (E) circle (.1);
\draw (F) circle (.08);
\end{tikzpicture}}
\end{center}
   
    \label{fig:Ahdynkin}
\end{figure}

\subsection{Irreducible modules of $\Ah$}
\subsubsection{Weyl dimension formula}
The Weyl dimension formula firstly works for purely bosonic modules. For $\Ah$ irreducible modules with highest weight $\lambda=\sum_{i=2,4}n_iw_i$, $n_i\in \NN$,    we have  
\begin{equation}\label{weyldimAh}
\dim(R_\lambda)=\frac{\prod_{\alpha\in\Delta_+(A_5)}\langle\lambda+\rho,\alpha\rangle\prod_{\alpha\in\frac{1}{2}{\bf 20} }\langle\lambda+\rho,\alpha\rangle}{\prod_{\alpha\in\Delta_+(A_5)}\langle\rho,\alpha\rangle\prod_{\alpha\in\frac{1}{2}{\bf 20} }\langle\rho,\alpha\rangle}.
\end{equation}
For $0\le n_{2,4}\le 4$, the above formula gives the following module dimensions
\begin{align}\nonumber
    \left(
\begin{array}{ccccc}
 1 & 21 & 210 & 1386 & 6930   \\
 21 & 384 & 3465 & 21120 & 99099   \\
 210 & 3465 & 28875 & 165165 & 735735   \\
 1386 & 21120 & 165165 & 896896 & 3825822   \\
 6930 & 99099 & 735735 & 3825822 & 15731352   \\
\end{array}
\right).
\end{align}
These are exactly consistent with the dimension formulas of distinguished modules for Severi series in \cite{LM02}. Surprisingly, we find that the Weyl dimensional formula \eqref{weyldimAh} also work for highest weight $\lambda$ with two \textit{distinct} fermionic fundamental weights, such as 
$\lambda=m\theta+ n_2w_2+n_4w_4$, $m,n_i\in \NN$ where highest root $\theta=w_1+w_5$. 
For example, for $\lambda=m\theta$, from \eqref{weyldimAh} we find 
\begin{align}\label{dimAhntheta}
 \dim(R_{m\theta})= \frac{1}{2^{	10}
3^5
5^3
7
} (m+1) (m+2)^2 (m+3)^3 (m+4)^3 (m+5)^3 (m+6)^2 (m+7) .
\end{align}
It is easy to show that this always produces integers for $m\in \NN$.  
The first few $\dim(R_{m\theta})$ are 
$$
56, 1176, 14112, 116424, 731808, 3737448, 16195608, 61408347,
208416208...,
$$
which correctly reproduce the dimension $56$ of $\Ah$. We also checked that for arbitrary highest weight $\lambda=m_1\theta+m_2(w_1+w_3)+m_3(w_3+w_5)+ n_2w_2+n_4w_4$, $m_i,n_i\in \NN$, \eqref{weyldimAh} always produces positive integers. The validity of this generalization may be owing to the observation that two distinct fermionic fundamental weights  combined together behave like a bosonic fundamental weight.

For any highest weight $\lambda=\sum_{i=1}^5n_iw_i$ with $n_1+n_3+n_5=1$, we find a modified Weyl dimension formula as 
\begin{equation}\label{weyldimoddAh}
\dim(R_\lambda)=\frac{\prod_{\alpha\in\Delta_+(A_5)}\langle\lambda+\rho,\alpha\rangle\prod_{\alpha\in\frac{1}{2}{\bf 20}}\big(\langle\lambda+\rho,\alpha\rangle-\langle\lambda,\alpha\rangle_{\rm odd})}{2\prod_{\alpha\in\Delta_+(A_5)}\langle\rho,\alpha\rangle\prod_{\alpha\in\frac{1}{2}{\bf 20}}\langle\rho,\alpha\rangle}.
\end{equation}
We find that the odd correction here should still be \eqref{eq:oddDhcorrection} as in $\Dh$. 
We checked that this modified Weyl dimension formula always gives positive integers. For example, for highest weight $\lambda=w_3+ n_2w_2+n_4w_4$, \eqref{weyldimoddAh} gives the dimension
\begin{align}
    \frac{1}{2^{12}
3^6
5^3
7
}\prod_{i=1}^6(n_2+i)(n_4+i)\cdot \prod_{i=5}^7(n_2+n_4+i)\cdot \prod_{i=7}^9(n_2+n_4+i)
\end{align}
We remark that there exist five irreducible modules of $\Ah$ with dimension $21$, which are $\bf 21$, $\bf \overline{21}$, $\bf 21'$, $\bf 21''$ and $\bf \overline{21}''$. The $\bf  {21}''$ is the module denoted by $J$ in \cite{LMseries}.\footnote{There is a typo in the dimension formula of $J$ in \cite{LMseries}. The correct one should be $$\dim J=\frac{3m(m+1)(8-m)(3m+2)}{2(m+4)(m+6)}.$$} We collect all irreducible modules of $\Ah$ and the corresponding $A_5$ modules and relevant data in Table  \ref{tb:repsA} for level $k\le 2$ and Table \ref{tb:repsA3} for level $k=3$. For higher levels, we only have very limited predictions. We collect some irreducible modules for level $4$ in Table \ref{tb:repsA4}. The modules with $-$ mean that we cannot predict their dimensions.

\begin{table}[ht]
\def\arraystretch{1.1}
\caption{All irreducible modules of $A_5$ and $\Ah$ with level $k\le 2$. At each level, we order the highest weight $\lambda$ by the $C_2$ of $A_5$.}
	\centering
	\begin{tabular}{|c|c|c|c|c|c|c|c|c|c|c|c|c|c|}
		\hline
$k$&	$\lambda$	   &  $A_5$  & $C_2$  & $I$  & $\Ah$  &  $C_2$  & $I$   & $\rm Aut$ & LM \cite{LMseries} & F/B\\
		\hline
$1$  &   $10000$ & $\bf 6$ & $ \frac{35}{6}   $ & $\frac12$ & $\bf 6$ & $ \frac{28}{3} $  & $\frac12$  & $\mathbb{Z}_2$ & $-  $ & F \\ 
$1$  &     $01000$ & $\bf 15$ & $  \frac{28}{3} $ & $  2$ & $\bf  21$ & $ \frac{40}{3}   $ & $  \frac52 $ & $\mathbb{Z}_2$ & $V$ & B \\
$1$  &  $00100$ & $\bf 20$ & $ \frac{21}{2}  $ & $3 $ & $\bf   21'$ & $ 16   $  & $  3  $  &  $0$ & $-  $ & F \\ \hline
$2$  &  $10001$ & $\bf  35 $  & $ 12  $ & $  6 $ & $\bf  56$ & $ 18 $&  $9  $ & $0$ & $\mathfrak{g}$& B \\
$2$  &  $20000$ & $\bf  21 $  & $\frac{40}{3}  $ & $4  $ & $\bf 21'' $ & $\frac{64}{3}  $&  $ 4 $ & $\mathbb{Z}_2$ & $J$& B\\

$2$  &  $01001$ & $\bf 84 $  & $ \frac{95}{6}  $ & $ 19 $ & $\bf  120$ & $ \frac{70}{3} $&  $ 25 $ & $\mathbb{Z}_2$ &$-  $ & F\\
$2$  &  $11000$ & $\bf 70 $  & $ \frac{33}{2}  $ & $ \frac{33}{2} $ & $\bf 105 $ & $24  $&  $ \frac{45}{2} $ & $\mathbb{Z}_2$ &    $-  $  & F\\
$2$  &  $00101$ & $\bf 105 $  & $ \frac{52}{3}  $ & $ 26 $ & $\bf  210$ & $\frac{76}{3}  $&  $ \frac{95}{2} $ & $\mathbb{Z}_2$ & $V_2^*$ & B\\
$2$  &  $01010$ & $\bf  189$  & $  20 $ & $ 54 $ & $\bf 384 $ & $ 28 $&  $ 96 $ & $0$ & $VV^*$ & B\\
$2$  &  $00020$ & $\bf 105' $  & $ \frac{64}{3}  $ & $ 32 $ & $\bf 210' $ & $ \frac{88}{3} $&  $ 55 $ & $\mathbb{Z}_2$ & $-  $ & B\\
$2$  &  $00110$ & $\bf 210 $  & $ \frac{131}{6}  $ & $\frac{131}{2}  $ & $\bf 336 $ & $ \frac{94}{3} $&  $  94$ & $\mathbb{Z}_2$ &  $-  $  & F\\
$2$  &  $00200$ & $\bf  175$  & $  24 $ & $ 60 $ & $\bf 196 $ & $36  $&  $63  $ & $0$ & $-  $ & B \\  \hline

		\end{tabular}
			\label{tb:repsA}
		\end{table}

\begin{table}[ht]
\def\arraystretch{1.1}
\caption{All irreducible modules of $A_5$ and $\Ah$ with level $k=3$. At each level, we order the highest weights by the $C_2$ of $A_5$.}
	\centering
	\begin{tabular}{|c|c|c|c|c|c|c|c|c|c|c|c|c|c|}
		\hline
$k$&	$\lambda$	   &  $A_5$  & $C_2$  & $I$  & $\Ah$  &  $C_2$  & $I$   & $\rm Aut$ & LM & F/B \\
		\hline
   $  3 $  &  $20001$ & $\bf  120 $  & $ \frac{119}{6}  $ & $ 34 $ & $\bf 231 $ & $ \frac{88}{3} $&  $ \frac{121}{2} $ & $  \ZZ_2  $ & $-$ & F \\ 
 $  3 $  &  $30000$ & $\bf  56 $  & $ \frac{45}{2}  $ & $ 18 $ & $\bf  56'$ & $ 36 $&  $ 18 $ & $  \ZZ_2 $ & $-$ & F\\ 
 
  $  3 $  &  $11001$ & $\bf 384  $  & $ \frac{70}{3}  $ & $ 128 $ & $\bf 924 $ & $ \frac{100}{3} $&  $ 275 $ & $   \ZZ_2$ & $\mathfrak{g}V$ & B \\ 
  
 $  3 $  &  $20010$ & $\bf  280 $  & $  24 $ & $ 96 $ & $\bf  350$ & $ 36 $&  $ \frac{225}{2} $ & $  \ZZ_2 $& $J V^*$ &  B \\

$  3 $  &  $10101$ & $\bf  540 $  & $\frac{49}{2}   $ & $ 189 $ & $\bf 784 $ & $ 36  $&  $ 252 $ & $  0 $ & $-$ & F \\ 
 
 $   3$  &  $00012$ & $\bf  210' $  & $ \frac{76}{3}  $ & $ 76 $ & $\bf 330 $ & $ \frac{112}{3} $&  $ 110 $ & $  \ZZ_2 $ &  $J^*V^*$ & B \\

 $   3$  &  $00102$ & $\bf 336  $  & $\frac{155}{6}   $ & $ 124 $ & $\bf 825 $ & $ \frac{112}{3} $&  $ 275 $ & $ \ZZ_2  $ & $-$ & F \\

$   3$  &  $11010$ & $\bf  840 $  & $ \frac{167}{6}  $ & $ 334 $ & $\bf 1848 $ & $ \frac{118}{3} $&  $ 649 $ & $ \ZZ_2  $ & $-$ & F \\

 $ 3  $  &  $10020$ & $\bf  560 $  & $ \frac{57}{2}  $ & $ 228 $ & $\bf 1155 $ & $ 40 $&  $ \frac{825}{2} $ & $  \ZZ_2 $ & $-$ & F  \\  

 $   3$  &  $10110$ & $\bf 1050  $  & $ \frac{88}{3}  $ & $ 440 $ & $\bf  3234 $ & $ \frac{124}{3} $&  $ \frac{2387}{2} $ & $  \ZZ_2 $ & $V_2V^*$ & B \\

 $   3$  &  $00021$ & $\bf 420  $  & $ \frac{179}{6}  $ & $ 179 $ & $\bf 924' $ & $ \frac{124}{3} $&  $ 341 $ & $ \ZZ_2  $ & $-$ & F \\

 $   3$  &  $11100$ & $\bf  896 $  & $ 30  $ & $ 384 $ & $\bf 2640 $ & $ 42 $&  $990  $ & $ \ZZ_2  $ & $V_2V$ & B \\

  $   3$  &  $10200$ & $\bf  840' $  & $ \frac{191}{6}  $ & $382  $ & $\bf - $ & $\frac{136}{3}  $&  $ - $ & $  \ZZ_2 $ & $-$ & F \\

 $   3$  &  $02010$ & $\bf  1176 $  & $ \frac{100}{3}  $ & $560  $ & $\bf  3465$ & $ \frac{136}{3} $&  $ \frac{2805}{2} $ & $\ZZ_2$ &  $ V^{(2)}V^*  $ & B \\

$  3 $  &  $01110$ & $\bf  1960 $  & $ \frac{69}{2}  $ & $ 966 $ & $\bf 4851 $ & $48  $&  $ 2079 $ & $  0 $ & $-$  & F \\

 $   3$  &  $02100$ & $\bf 1176'  $  & $  \frac{215}{6} $ & $602  $ & $\bf 2772 $ & $\frac{148}{3}  $&  $ 1221 $ & $ \ZZ_2  $ & $-$ & F\\


 $ 3  $  &  $03000$ & $\bf  490 $  & $  36 $ & $ 252 $ & $\bf  1386$ & $ 48 $&  $ 594 $ & $  \ZZ_2 $ & $V^{(3)}$ & B \\   
 
$  3 $  &  $01200$ & $\bf  1470 $  & $  \frac{112}{3} $ & $ 784 $ & $\bf - $ & $\frac{160}{3}  $&  $-  $ & $ \ZZ_2  $ & $-$ & B \\  

  $   3$  &  $00300$ & $\bf  980 $  & $ \frac{81}{2}  $ & $ 567 $ & $\bf  -$ & $ 60 $&  $ - $ & $  0 $ & $-$ & F \\

\hline

		\end{tabular}
			\label{tb:repsA3}
		\end{table}

\begin{table}[ht]
\def\arraystretch{1.1}
\caption{Some irreducible modules of $A_5$ and $\Ah$ with level $k=4$. }
	\centering
	\begin{tabular}{|c|c|c|c|c|c|c|c|c|c|c|c|c|c|}
		\hline
$k$&	$\lambda$	   &  $A_5$  & $C_2$  & $I$  & $\Ah$  &  $C_2$  & $I$   & $\rm Aut$ & LM \cite{LMseries} & F/B  \\
		\hline

$  4 $  &  $20002$ & $\bf  405 $  & $  28 $ & $ 162 $ & $\bf  1176$ & $  40$&  $ 420 $ & $ 0  $ & $\mathfrak{g}^{(2)}$ & B \\

  $ 4  $  &  $00004$ & $\bf 126  $  & $  \frac{100}{3} $ & $ 60 $ & $\bf  126$ & $ \frac{160}{3} $&  $ 60 $ & $ \ZZ_2  $ & $J^{*(2)}$ & B  \\

$  4 $  &  $10030$ & $\bf 2520''  $  & $ \frac{263}{6}  $ & $ 1578 $ & $\bf  7392$ & $\frac{178}{3}  $&  $ 3916 $ & $ \ZZ_2  $ & $-$ & F \\ 

$  4 $  &  $04000$ & $\bf  1764' $  & $ \frac{160}{3}  $ & $ 1344 $ & $\bf 6930 $ & $ \frac{208}{3} $&  $ 4290 $ & $ \ZZ_2  $ & $V^{(4)}$ & B \\

\hline

		\end{tabular}
			\label{tb:repsA4}
		\end{table}

\subsubsection{Decomposition $A_5\subset A_{5+1/2} \subset E_6$}
First consider $A_5\subset A_{5+1/2}$. It is easy to find the following decompositions from the coset construction \eqref{cosetM} 
\begin{align}
\bf 56&=  \bf 35+20+1  ,\\
   \bf 21&=\bf 15+\overline{6},\\
    \bf 21'&=\bf 20+1.
\end{align}
Similar for the complex conjugated modules. Following the method described for $\Dh$, we also find the following unique decompositions for the level $2$ modules
\begin{align}
    \bf   120 &=\bf 84+\overline{21}+\overline{15},\\
     \bf 105 &=\bf 70+35,\\   
    \bf   210 &=\bf 105+\overline{84}+15+\overline{6} ,\\    
    \bf  384 &= \bf189+70+\overline{70}+35+20\\
    \bf   210' &=\bf  105'+\overline{84}+21 ,\\  
    \bf 336&= \bf 210+\overline{105}+\overline{15}+6,\\
    \bf 196&=\bf 175+20+1.
\end{align}
For level $3$, we find the module decompositions
\begin{align}
\bf 231 &=\bf 120+\overline{105}+6,\\
\bf 924 &= \bf  384 + \overline{210} + \overline{120} + 105 + \overline{84} + 15 + \overline{6},\\ 
\bf 350 &=\bf 280+70, \\
 \bf   784 & =\bf 540 + 189 + 35 + 20,\\
 \bf 330&= \bf 210'+\overline{120},\\
 \bf 825 &= \bf 336+\overline{384}+84+\overline{21},\\
 \bf 1848&= \bf  840+  \overline{384} + \overline{210}' + 120 + \overline{105} + \overline{105}' + 84,\\
 \bf 1155 &= \bf 560+280+189+70+56,\\
 \bf 3234 &= {\bf  1050 + \overline{840} + 384 + \overline{336} + \overline{210}} + 2\cdot \bf \overline{84} + 105 + 105' + 21 + 15,\\
 \bf 924'&= \bf 420 +\overline{384} + 120,\\
 \bf 2640 &= \bf 896 + 560 + 540 + 280 + 189 + 70 + 70 + 35,\\
 \bf 3465&= \bf 1176+\overline{840}+\overline{420}+384+\overline{210}+210'+\overline{120}+105,\\
 \bf 1386 & =\bf 490+\overline{560}+\overline{280}+\overline{56}'.
\end{align}
For level $4$, we find the module decompositions
\begin{align}
    \bf 1176 &=\bf 405+540 + 175 + 35 + 20 + 1,\\
  \bf  6930 &=\bf 1764' + \overline{2520}'' + 1800 + \overline{720} + 126,\\
  \bf 7392 &= \bf 2520'' + \overline{1800} + \overline{1176} + 840 + 720 + \overline{210}' + \overline{126}.
\end{align}

For $ A_{5+1/2}\subset E_6$, we find the module decompositions
\begin{align}
\bf 78 &=\bf 56+ 21'+1 ,\\
   \bf 27&=\bf \overline{21}+6.
\end{align}
Similar for the complex conjugation. Besides, we find
\begin{align}
{\bf 351} &=\bf \overline{210} + 120+\overline{21},\\
   {\bf 351'}&=\bf \overline{210}'+ 120+\overline{21}''\\
   \bf 650 &=\bf  384 +105+\overline{105}+56. 
\end{align}
Here $E_6$ modules $\bf 351$ and $\bf 351'$ have highest weights $(000100)$ and $(000020)$ respectively. We further find the module decompositions
\begin{align}
    {\bf 2430} &={\bf 1176+784}+2\cdot \bf 196+56+21'+1,\\
    \bf 2925 &= 2\cdot \bf  784 + 350+\overline{350} + 384 + 196 + 56 +21',\\
    \bf 3003 & = {\bf \overline{1386} + 1155 + 350}+2\cdot\mathbf{56'},\\
    \bf 1728& = \bf \overline{924} + 336 + \overline{210} + 231 + \overline{21} + 6 ,\\
    \bf 5824& = \bf 2640 + 1155 + 784 + 350+\overline{350} + 384 + 105 + 56,\\
    \bf 7371& =\bf \overline{3234} + 1848 + \overline{924}+ 825 + \overline{210} + \overline{210}'  +120 ,\\
    \bf 7722& =\bf \overline{3465}+ 1848 + 924' + \overline{924}  + \overline{330} + 231  .
\end{align}
One can check that in all above decompositions, both sides have the same dimensions and the same indices.

\subsubsection{Tensor product decomposition}
For the tensor product of $\Ah$, we find the following novel decompositions
\begin{align}
    (2)\bf 56 &= \bf  1176+ 384+ 56+1-21',\\
    (11)\bf 56 &= \bf 350+\overline{350}+    56+ 784.
\end{align}
It is clear that Vogel's universal decomposition formulas fail for $\Ah$. 
As predicted for Severi series by \cite{LMseries}, we have
\begin{align}
    \mathbf{56}\times \mathbf{21}&=\mathbf{ 21+210+924 +21'' },\\
     \mathbf{21'}\times \mathbf{\overline{21}}&=\mathbf{ 336+120+6 -21'' }   .
\end{align}
We also find the tensor product decomposition for the fermionic module $\bf 21'$ as
 \begin{align}
    (2)\bf 21' &={\bf 196}+\bf  56-21'  ,\\
    (11)\bf 21' &=\bf 384+56+1-105-\overline{105}-21' .
\end{align}

\subsection{Affine VOA $L_k(A_{5+1/2})$}
Consider the affine VOA $L_k(\Ah)$ with positive integer level $k$. We would like to show that $L_k(\Ah)$ is rational only for $k=1,2$. We can determine the conformal weights for $L_1(\Ah)$ and $L_2(\Ah)$. But for characters, we are only able to compute for $L_1(\Ah)$. 

\subsubsection{Affine VOA $A_{5+1/2}$ at level $1$}
The affine VOA $L_1(\Ah)$ has central charge $c=\frac{28}{5}$ and conformal weights with degeneracy
\begin{align}\label{eq:Ahlevel1}
    h_{\lambda}=0, \left(\frac{7}{15}\right)_2 ,\left(\frac23\right)_2,\frac45.
\end{align}
This VOA inherits a $\ZZ_3$ automorphism from $\Ah$. It acts on all six conformal primaries by the following two orbits
$   \big(\phi_0,\phi_{\frac23},\phi_{\frac23}\big),\big(\phi_{\frac{7}{15}},\phi_{\frac{7}{15}},\phi_{\frac{4}{5}}\big).$ 
The characters of this VOA as a vector-valued modular form of degree $4$ have appeared in the holomorphic modular bootstrap in e.g. \cite[Table 6]{Kaidi:2021ent}. Such a vector-valued modular form was also realized as a $\T_{7}$ Hecke operation on the $M_{\rm sub}(6,5)$ \cite[Section 5.5]{Duan:2022ltz}. The Hecke relation also suggests that the conformal primaries with weights $\frac{7}{15}$ and $\frac23$ have degeneracy $2$. 
By Hecke operator, we compute the characters of $L_1(\Ah)$ as
    \begin{align}
\chi_0=\,&q^{-\frac{7}{30}} (1+56 q+476 q^2+2632 q^3+11270 q^4 + \dots),\\
\chi_{\frac{7}{15}}=\,&q^{\frac{7}{30}} (6+105 q+672 q^2+3297 q^3+12978 q^4+ \dots),\\
\chi_{\frac{2}{3}}=\,&q^{\frac{13}{30}} ( 21+252 q+1533 q^2+7098 q^3+27210 q^4+ \dots),\\
\chi_{\frac{4}{5}}=\,&q^{\frac{17}{30}} ( 21'+196 q+1196 q^2+5264 q^3+19957 q^4+  \dots).
    \end{align}
The exact formulas as modular forms for these characters can be found in \cite[Appendix B, solution (i)]{Kawasetsu:2018tzs}  up to some constant factors. 
We can also determine 
the flavored characters of $L_1(A_{5+1/2})$ such as 
\begin{align}\label{eq:chiAhlevel1flavor}
\chi_0=\,&q^{-\frac{7}{30}} (1+\mathbf{56} q+(\mathbf{384}+2\cdot \mathbf{56}+1-\mathbf{21'}) q^2+  \dots).
    \end{align}
The $L_1(A_{5+1/2})$ is a vertex subalgebra of lattice VOA $L_1(E_6)$.  By \cite[Theorem 2.2]{Kawasetsu}, the characters of $L_1(A_{5+1/2})$ have similar expressions like \eqref{eq:nahmDh} for $\Dh$. Besides, the characters of $L_1(A_{5+1/2})$ coincide with the characters of $W_{-2}(E_6,f_\theta)$ studied in \cite{Kawasetsu:2018irs}.

\subsubsection{Affine VOA $A_{5+1/2}$ at level $2$}
We conjecture that the affine VOA $L_2(A_{5+1/2} )$ is rational and  has central charge $ c=\frac{112}{11}$ and conformal weights with degeneracy
\begin{align}\label{eq:Ahlevel2}
h_{\lambda}=  0,\left(\frac{14}{33}\right)_2,\left(\frac{20}{33}\right)_2,\frac{8}{11},\frac{9}{11},\left(\frac{32}{33}\right)_2,\left(\frac{35}{33}\right)_2,\left(\frac{12}{11}\right)_2,\left(\frac{38}{33}\right)_2,\frac{14}{11},\left(\frac{4}{3}\right)_2,\left(\frac{47}{33}\right)_2 ,\frac{18}{11} .
\end{align}
The conductor $N=33$. These conformal weights are computed from the conjectural quadratic Casimir invariant formula \eqref{eq:AhC2formal}. 
The $L_2(A_{5+1/2} )$ has a $\mathbb{Z}_3$ automorphism, that is a $\mathbb{Z}_3$ parafermionic structure, resembling the affine VOA $L_2(E_6)$. The $\mathbb{Z}_3$ automorphism acts on all 21 conformal primaries by the following seven orbits
\begin{align}\label{Z3orbit1}
\big(\phi_0,\phi_{\frac43},\phi_{\frac43}\big),\big(\phi_{\frac{20}{33}},&\,\phi_{\frac{20}{33}},\phi_{\frac{14}{11}}\big),\big(\phi_{\frac{8}{11}},\phi_{\frac{35}{33}},\phi_{\frac{35}{33}}\big),\big(\phi_{\frac{9}{11}},\phi_{\frac{38}{33}},\phi_{\frac{38}{33}}\big),\big(\phi_{\frac{32}{33}},\phi_{\frac{32}{33}},\phi_{\frac{18}{11}}\big),\\ \label{Z3orbit2}
&\big(\phi_{\frac{14}{33}},\phi_{\frac{14}{33}},\phi_{\frac{12}{11}}\big),\big(\phi_{\frac{12}{11}},\phi_{\frac{47}{33}},\phi_{\frac{47}{33}}\big).
\end{align}
In each $\ZZ_3$ orbit, the differences of the conformal weights are always multiples of $1/3$. The two orbits in \eqref{Z3orbit2} are connected as they split the degeneracy two of $\phi_{{12}/{11}}$. Inspired by the fact that $L_2(A_{5} )$ can realized by a fermionic Hecke operation $\TF_7$ on a unitary $N=1$ supersymmetric minimal model $SM(8,6)$ \cite{Lee:2022yic}, we conjecture that the characters of $L_2(A_{5+1/2} )$ can realized by a Hecke operation $\T_7$ on certain theory with central charge $16/11$. This may need a $\ZZ_3$ parafermionic generalization of the Hecke operator. We also expect the coset constructions
 \begin{align}
     \frac{L_2(E_6)}{ L_2(\Ah)}\quad \textrm{and}\quad \frac{L_1(\Ah)\otimes L_1(\Ah)}{ L_2(\Ah)}
 \end{align}
exist with central charge $74/77$ and $56/55$ respectively. They should be related to some minimal models with $\mathbb{Z}_3$ symmetry. Unfortunately, as we do not have the characters of $L_2(\Ah)$, we could not determine the precise cosets.

\subsubsection{Higher levels}
Assuming the $\Ah$ irreducible modules are one-to-one corresponding to the $A_5$ irreducible modules, then
 the number $r(k)$ of distinct affine characters (i.e. modulo degeneracy) at each level $k$ is generated by 
\begin{align}
 \sum_{k=0}^\infty r(k)x^k=   \frac{1+x^2}{(1-x)^6 (1+x)^2}=  1+4 x+13 x^2+32 x^3+70 x^4+   \dots
\end{align}
Further assuming our conjectural $C_2$ formula \eqref{eq:AhC2formal} is correct, then we can easily compute the indices $l(k)$ of the MLDEs for the characters of $L_k(\Ah)$, which are in Table \ref{tb:lkA5}. Remarkably, the indices $l(k)$ for $L_k(A_5)$ and $L_k(\Ah)$ become fractional for level $k>2$. Note that the index $l(k)$ as a non-negative integer is only a necessary condition for the rationality. Therefore we conjecture that $L_k(\Ah)$ cannot be rational for $k\ge 3$.

\begin{table}[h]
\def\arraystretch{1.1}
  \caption{The number $r(k)$ of distinct characters and the MLDE indices $l(k)$ for $L_k(A_5)$ and $L_k(\Ah)$.}\label{tb:lkA5}
  \centering
  \begin{tabular}{c|c|c|c|c|c}\hline
level &  $k$  & $1$ & $2$ & $3$ & $4$    \\ \hline
& $r(k)$  &  $4  $ & $ 13  $ & $ 32$& $  70$  \\  
$A_5$ & $l(k)$ &  $   0$ &  $ 36  $ &  $ 340   $ &  $  1955 $     \\   
$A_{5+1/2}$ & $l(k)$ &  $  0 $ &  $  36  $&  $ \frac{679}{2}  $ &  $   \frac{25385}{13} $ \\ \hline
     \end{tabular}
\end{table}

\subsection{$A_{5+1/2}$ as gauge algebra}\label{sec:Ahgauge}
Similar with the discussion on $\Dh$ in Section \ref{sec:Dhgauge}, we would like to argue that $A_{5+1/2}$ may also serve as a gauge algebra. Again assuming the 3d monopole formula \cite{Benvenuti:2010pq} still holds, the {5d one $\Ah$ instanton Nekrasov partition function} should have the following expression 
\begin{align}\label{Ah1inst}
  Z_1^{\rm Nek}&=v^{h^\vee-1}\sum_{n=0}^\infty v^{2n}\dim(n\theta)\\ \nonumber
  &= v^{8}(1+56 v^2+1176 v^4+14112 v^6+116424 v^8+731808 v^{10}+\dots)\\ \nonumber
&= \frac{1}{(v-v^{-1})^{8}}\big(v^{\pm8}+40 v^{\pm6}+400 v^{\pm4}+1456   v^{\pm2}+2212 \big)  .
\end{align}
Here we used the $\dim(n\theta)$ formula \eqref{dimAhntheta} for $\Ah$. The rational function \eqref{Ah1inst} is again palindromic, i.e., symmetric for $v\leftrightarrow v^{-1}$. This satisfies the necessary condition for the Cohen–Macaulay ring of the one $\Ah$ instanton moduli space to be Gorenstein, as in the $\Dh$ case in Section \ref{sec:Dhgauge}.

\section{The other intermediate Lie algebras}

\subsection{$C_{3+1/2}$}
\begin{figure}[h]
 \caption{The analogy of the Dynkin diagram for $\Ch$ and irreducible representations associated with fundamental weights.}
    \centering
   \begin{center}
\resizebox{4.1cm}{!}{\begin{tikzpicture}{xscale=1cm,yscale=1cm}
\coordinate[label=below:${\bf 6}$,label=above:$1$](B) at (-3.6,0);
\coordinate[label=below:${\bf 20}$,label=above:$2$](C) at (-1.8,0);
\coordinate[label=below:${\bf  15}$, label=above:$3$](D) at (0,0);
\draw (B)--(C);
\draw[<-] (C)--(D);
\draw (B) circle (.08);
\draw (C) circle (.08);
\draw (D) circle (.08);
\end{tikzpicture}}
\end{center}
   
    \label{fig:Chdynkin}
\end{figure}
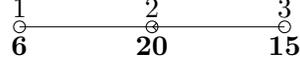

This intermediate Lie algebra has rank $r=3$, dual Coxeter number  $h^\vee={13}/{2}$ and dimension $36$. We define the Weyl vector for $\Ch$ as the sum of all positive roots, i.e.,
\begin{equation}\label{eq:weylvectorC}
\rho=\rho_{C_3}+\rho_{\bf 14'}=\frac{1}{2}\Big(\Delta_+(C_3)+\frac{1}{2}\mathbf{14'}\Big).
\end{equation}
Here $\mathbf{14'}$ is the $C_3$ module with highest weight $(001)$. Explicitly, we have in the fundamental weight basis
\begin{equation}
\rho_{C_3}=(1,1,1),\qquad \rho_{\bf 20} =(2,0,\frac12),\qquad \rho=(3,1,\frac32).
\end{equation}
The norm $|\rho|^2= {155}/{8}$ does not satisfy the Freudenthal--de Vries strange formula. The dual Coxeter should increase from the $4$ of $C_3$ by $
\frac{1}{2r}\sum_{w\in \mathbf{ 14'}}|w|^2
={5}/{2}.
$
Thus the dual Coxeter number of $\Ch$ is $4+5/2=13/2$ as required. The dual Coxeter number can also understood from the comarks which scales as the $\rho$ component as 
\begin{align}
  (a_0^\vee,a_1^\vee,a_2^\vee,a_3^\vee)= (1,3,1,\frac32)  .
\end{align}
The summation again gives the dual Coxeter number $13/2$.   We draw the analogy of the Dynkin diagram for $\Ch$ in Figure \ref{fig:Chdynkin}. 
All three fundamental weights should be regarded as fermionic. We do not find any Weyl-type dimension formula or $\dim(k\theta)$ formula for $\Ch$. $\Ch$ is not on Vogel's plane either. Nevertheless, $\Ch$ fits in the Severi-section series such that some modules follow from \cite{LM}. We collect some irreducible modules of $\Ch$ and the corresponding $C_3$ modules in Table \ref{tb:repsC}.

\begin{table}[ht]
\def\arraystretch{1.1}
\caption{All irreducible modules of $C_3$ and $\Ch$ with level $k=1$ and some irreducible modules with $k\le 2$. At each level, we order the highest weights by the $C_2$ of $\Ch$.}
	\centering
	\begin{tabular}{|c|c|c|c|c|c|c|c|c|c|c|c|c|c|}
		\hline
$k$&	$\lambda$	   &  $C_3$  & $C_2$  & $I$  & $\Ch$  &  $C_2$  & $I$  &LM \cite{LMseries} & F/B \\
		\hline
$1$  &   $100$ & $\bf 6$ & $ 7/2  $ & $1/2$ & $\bf 6 $ & $ 6 $  & $1/2$  & $-$ & F \\ 
$1$  &     $010$ & $\bf 14$ & $  6 $ & $ 2 $ & $\bf 20 $ & $  9  $ & $  5/2 $ & $V$ & F \\
$1$  &  $001$ & $\bf 14'$ & $ 15/2  $ & $ 5/2$ & $\bf  15 $ & $  12  $  & $  5/2  $ &  $-$  & F \\ \hline
$2$  &  $200$ & $\bf  21 $  & $  8 $ & $4  $ & $\bf 36 $ & $ 13 $&  $ 13/2 $ & $\mathfrak{g}$ & B\\
$2$  &  $110$ & $\bf  64 $  & $ \frac{21}{2}  $ & $ 16  $ & $\bf 99 $ & $16  $&  $ 22 $ & $-$ & B\\
$2$  &  $101$ & $\bf  70 $  & $ 12  $ & $  20 $ & $\bf 154 $ & $ 18 $&  $ \frac{77}{2} $ & $V_2$ & B \\
$2$  &  $020$ & $\bf  90 $  & $  14 $ & $  30 $ & $\bf  189$ & $ 20 $&  $ \frac{105}{2} $ & $V^{(2)}$ & B \\ \hline
		\end{tabular}
			\label{tb:repsC}
		\end{table}

Consider the module decomposition under $C_3\subset C_{3+1/2} \subset F_4$. For $C_3\subset C_{3+1/2}$, we find the following unique decompositions
\begin{align}
\bf 20&=\bf 14+ 6,\\
   \bf 15&=\bf 14'+1,\\
   \bf 36&=\bf 21+  14'+1,\\
   \bf 99& =\bf 64+21+14,\\
   \bf 154& =\bf 70+64+14+6 \\
   \bf 189 &= \bf 90+64+21+14' . 
\end{align}
For $C_{3+1/2} \subset F_4$, we find the following unique decompositions
\begin{align}
\bf 52&=\bf 36+15+1,\\
   \bf 26&=\bf 20+6,\\
   \bf 273  &=\bf 154 + 99+20,\\
   \bf 324 &=\bf 189+99+36. 
\end{align}

Now consider the tensor product decomposition of irreducible modules of $\Ch$. As predicted for Severi-section series by \cite{LMseries}, 
\begin{align}
    (2)\bf 20 &= \bf  189+20+1  , \\
    (11)\bf 20 &= \bf 154 + 36. 
\end{align}
and
\begin{align}
    \mathbf{6}\times \mathbf{ 20} &=  \bf 99 +15+6.
\end{align}
We further obtain the tensor product of fermionic modules
\begin{align}
    (2)\bf 6 &= \bf 36 -15    , \\
    (11)\bf 6 &= \bf 20+1-6 . 
\end{align}

The affine VOA $L_1(\Ch)$ has central charge and conformal weights
\begin{align}
    c=\frac{24}{5},\quad\textrm{and}\quad  h_\lambda=0,\frac25,\frac35,\frac45.
\end{align}
The characters of $L_1(\Ch)$  can be expressed in terms of Roger-Ramanujan functions \eqref{RRf} and have the following Fourier expansion
\begin{equation}\label{eq:chiChlevel1}
    \begin{aligned}
\chi_0&= \phi_0^{12}+24\phi_0^7\phi_1^5-6\phi_0^2\phi_1^{10}=q^{-\frac{1}{5}} ( 1+36 q+240 q^2+1144 q^3+  \dots),\\
\chi_{\frac{2}{5}}&=6\phi_0^{10}\phi_1^2+24\phi_0^5\phi_2^7-\phi_1^{12}=q^{\frac{1}{5}} (6+84 q+461 q^2+1980 q^3+ \dots),\\
\chi_{\frac{3}{5}}&=20\phi_0^9\phi_1^3+ 15\phi_0^4\phi_1^8=q^{\frac{2}{5}} ( 20+195 q+1020 q^2+4170 q^3+ \dots),\\
\chi_{\frac{4}{5}}&= 15\phi_0^8\phi_1^4-20\phi_0^3\phi_1^9=q^{\frac{3}{5}} ( 15+100 q+540 q^2+2040 q^3+  \dots).
    \end{aligned}
\end{equation}
This vector-valued modular form of degree 4 has appeared in the study on holomorphic MLDE in e.g. \cite[Table 6]{Kaidi:2021ent}. Such a vector-valued modular form are also realized as a generalized $\T_{2}$ Hecke image of $L_1(\AG)$ denoted as $M_{12/5}$ in \cite[Section 5.2]{Duan:2022ltz}.

\subsection{$AD_{3+1/2}$}
This intermediate Lie algebra has rank $r=3$, dual Coxeter number $h^\vee=4$ and dimension $18$. The dual Coxeter number should increase from the $2$ of $A_1^3$ by
$
\frac{1}{2r}\sum_{w\in \mathbf{ (2,2,2)}}|w|^2= 2.
$
Thus $h^\vee$ of $\AD$ is $2+2=4$ as expected. For the Weyl vector of $\AD$, we can choose the half weights of $\bf (2,2,2)$ as the first $A_1$ only has positive weight. Then we have
\begin{align}
    \rho= (1,1,1)+(2,0,0)=(3,1,1).
\end{align}
This gives $|\rho|^2=11/2$, which does not satisfy the Freudenthal--de Vries strange formula. It is also easy to check that $AD_{3+1/2}$ is not on Vogel's plane. Besides, all three fundamental weights should be regarded as fermionic. We do not find any Weyl-type dimension formula or $\dim(n\theta)$ formula in this case. We have very limited information on the modules and collect some small irreducible modules of $\AD$ in Table \ref{tb:repsAD}.

 \begin{table}[ht]
\def\arraystretch{1.1}
\caption{All irreducible modules of $A_1^3$ and $\AD$ with level $k=1$ and the adjoint module.}
	\centering
	\begin{tabular}{|c|c|c|c|c|c|c|c|c|c|c|c|c|}
		\hline
$k$&	$\lambda$	   &  $A_1^3$  & $C_2$  & $6I$  & $\AD$  &  $C_2$  & $6I$   & $\rm Aut$   \\
		\hline
$1$  &   $100$ & ${\bf (2,1,1)}$ & $  \frac32   $ & $  1 $ & ${\bf 2}_a$ & $ 3  $  & $  1 $  & $\mathbb{Z}_3$ \\ 
$1$  &     $110$ & ${\bf (2,2,1)}  $ & $  3  $ & $  4 $ & ${\bf  6}_{ab}$ & $   5  $ & $  5   $ & $\mathbb{Z}_3$  \\
$1$  &  $111$ & ${\bf (2,2,2)}$ & $  \frac92 $ & $ 12 $ & $\bf   9 $ & $  8   $  & $  12  $  &  $0$  \\ \hline
$2$  &  $200+020+002$ & ${\bf (3,1,1)+(1,3,1)+(1,1,3)} $  & $ 4  $ & $ 12  $ & $\bf 18 $ & $ 8 $&  $ 24 $ & $0 $   \\  
\hline

		\end{tabular}
			\label{tb:repsAD}
		\end{table}

Consider the module decomposition under $A_1^3\subset \AD \subset D_4$. All three Lie algebras have $\ZZ_3$ automorphism. 
For $A_1^3\subset \AD$, we find 
\begin{align}
    {\bf 6}_{ab}& =\bf (2,2,1)+(1,1,2)  , \\
    \bf 9& =\bf (2,2,2)+(1,1,1)  , \\
    \bf 18& =\bf (3,1,1)+(1,3,1)+(1,1,3)+(2,2,2)+(1,1,1)   .
\end{align}
For $ \AD\subset D_4$, we find 
\begin{align}
    \bf 8_v& ={\bf 6}_{bc}+{\bf 2}_a  , \\
    \bf 28& =\bf 18+9+1     .
\end{align}
Similar for $\bf 8_s$ and $\bf 8_c$ by triality.  It is easy to check the indices match in the above module decompositions.

Consider the affine VOA $L_1(\AD)$. The central charge is $c=18/5$ and the conformal weights with degeneracy are
$0,(\frac{3}{10})_3,(\frac{1}{2})_3,\frac{4}{5}  .$
The characters are
\begin{equation}\label{eq:chiAChlevel1}
    \begin{aligned}
\chi_0=\,&q^{-\frac{3}{20}} ( 1+18 q+81 q^2+306 q^3+909 q^4+ \dots),\\
\chi_{\frac{3}{10}}=\,&q^{\frac{3}{20}} (2+18 q+78 q^2+262 q^3+774 q^4+ \dots),\\
\chi_{\frac{1}{2}}=\,&q^{\frac{7}{20}} (6+40 q+162 q^2+534 q^3+1532 q^4+  \dots),\\
\chi_{\frac{4}{5}}=\,&q^{\frac{13}{20}} ( 9+34 q+153 q^2+450 q^3+1284 q^4  +\dots).
    \end{aligned}
\end{equation}
They are related to the characters of the familiar $D_{2\rm A}$ VOA by a $\T_3$ Hecke operator \cite[Section 5.3]{Duan:2022ltz}. The exact formulas as modular forms for these characters can be found in \cite[Appendix B, solution (g)]{Kawasetsu:2018tzs} up to some constant factors. Finally, we conjecture that all $L_k(\AD)$ for $k>1$ are not rational.

\subsection{$AG_{1+1/2}$}
This intermediate Lie algebra has rank $1$, dimension $8$ and dual Coxeter number  $7/3$.  
Recall in $A_1\times A_1 \subset G_2 $, the adjoint and fundamental of $G_2$ decompose as
\begin{align}
\bf  14 &=\bf  (3,1)+(1,3)+(4,2),\\
\bf  7 &=\bf   (3,1)+(2,2).
\end{align}
The intermediate Lie algebra $AG_{1+1/2}$ is flitrated by the second $A_1$. Thus for $A_1 \subset \AG $, we have the module decompositions
\begin{align}
    {\bf 5}& =\bf 4+1   , \\
    \bf 5' & =\bf 3+2  , \\
     \bf 8& =\bf  3+4 +1   .
\end{align}
The last decomposition indicates that the dual Coxeter increases by
$
\frac{1}{2}\sum_{w\in \mathbf{4}}|w|^2= 5/3.
$
As the first $A_1$ has the bilinear form scaled by a third, which has dual Coxeter number $2/3$, the intermediate Lie algebra $\AG$ has dual Coxeter number $2/3+5/3=7/3$, satisfying the linear relation with Landsberg--Manivel's parameter $a$. For $ \AG\subset G_2$, we find the module decompositions
\begin{align}
    \bf 7& =\bf  5' +2  , \\
    \bf 14& =\bf  8+5+1        .
\end{align}
We do not have a Weyl-type dimension formula for $\AG$.

The characters of affine VOA $L_1(\AG)$ can be computed from MLDE. They can be expressed in terms of Roger-Ramanujan functions \eqref{RRf} and have the following Fourier expansion
\begin{equation}\label{eq:chiAGhlevel1}
    \begin{aligned}
\chi_0&=\phi_0^6+2\phi_0\phi_1^5=q^{-\frac{1}{10}} ( 1+8 q+23 q^2+68 q^3+154 q^4 +\dots),\\
\chi_{\frac{1}{5}}&=2\phi_0^5\phi_1-\phi_1^6=q^{\frac{1}{10}} (2+9 q+32 q^2+76 q^3+186 q^4+ \dots),\\
\chi_{\frac{2}{5}}&=5\phi_0^4\phi_1^2=q^{\frac{3}{10}} ( 5+20 q+60 q^2+150 q^3+350 q^4+ \dots),\\
\chi_{\frac{4}{5}}&=5\phi_0^2\phi_1^4=q^{\frac{7}{10}} ( 5+10 q+35 q^2+80 q^3+185 q^4 +\dots).
    \end{aligned}
\end{equation}
The four characters correspond to the modules $\bf 1,2,5',5$ of $\AG$ respectively. 
We do not find a proper construction for $L_k(\AG)$ beyond level $1$.

\subsection{$U\!A_{1+1/2}$}
This intermediate Lie algebra has dimension $4$ and dual Coxeter number $3/2 $. Recall under $A_2\to  A_1\times U_1$, $A_2$ module decompose as $\mathbf{8}=\mathbf{1}_0+ \mathbf{2}_{3\oplus(-3)} + \mathbf{3}_0$ and $\mathbf{3}=\mathbf{1}_{-2}+ \mathbf{2}_{1}$, where the subscripts are the $U_1$ charges. Thus, the adjoint of $\UA$ comes from the $U_1$ charges $ \mathbf{4}  = 2(0)\oplus  3\oplus (-3)$. The dual Coxeter number of $\UA$ is  just half of $A_2$ as required. For $U_1\subset\UA $, we have module decompositions
\begin{align}
    \mathbf{3} & = 0\oplus  3\oplus (-3),\\
    \mathbf{2} & = (-2)\oplus  1,\\
    \mathbf{1}' &= 1.
\end{align}
Note the $\bf 1'$ of $\UA$ is different from the trivial module $\bf 1$ of $\UA$. The former has $U_1$ charge $1$ while the latter has $U_1$ charge $0$. For $\UA\subset A_2$, we have  
\begin{align}
    \mathbf{8} & =  \bf 4+3+1,\\
     \mathbf{3} & =  \bf 2+1' .
\end{align}
Similar for $\bf\bar{3}$ of $A_2$. The characters of affine VOA $L_1(\UA)$ have the following Fourier expansion
\begin{equation}\label{eq:chiUAhlevel1}
    \begin{aligned}
\chi_0=\,&q^{-\frac{1}{15}} ( 1+4 q+8 q^2+20 q^3+37 q^4+ \dots),\\
\chi_{\frac{2}{15}}=\,&q^{\frac{1}{15}} (1+2 q+7 q^2+12 q^3+26 q^4+  \dots),\\
\chi_{\frac{1}{3}}=\,&q^{\frac{4}{15}} (2+5 q+12 q^2+23 q^3+46 q^4+     \dots),\\
\chi_{\frac{4}{5}}=\,&q^{\frac{11}{15}} ( 3+4 q+10 q^2+20 q^3+38 q^4+  \dots).
    \end{aligned}
\end{equation}
The four characters correspond to the modules $\bf 1,1',2,3$ of $\UA$ respectively. The exact formulas as modular forms for these characters can be found in \cite[Appendix B, solution (e)]{Kawasetsu:2018tzs}  up to some constant factors. The $L_1(\UA)$ can also be understood from the coset
\begin{align}
    L_1(\UA)=\frac{L_1(U_1)}{M(5,3)}.
\end{align}
This example is special so we provide more details. The $L_1(U_1)$ has central $1$ and corresponds to a free boson CFT on a circle with radius $R=\sqrt{6}$. It is denoted as $U(1)_3$ in \cite[Section 5.6]{Duan:2022ltz} and has conformal weights with degeneracy $  0,(
\frac{1}{12})_2,(\frac
1
3
)_2,
\frac34$. Recall $M(5,3)$ has conformal weights $-\frac{1}{20},0,\frac15,\frac34$. We find the precise character relation for the above coset as
\begin{align}
    \chi_0^{U_1}&=\chi_0^{\UA}\chi^{M(5,3)}_0 -\chi_{\frac{4}{5}}^{\UA}\chi^{M(5,3)}_{\frac15},\\
    \chi_{\frac{1}{12}}^{U_1}&=\chi_{\frac{2}{15}}^{\UA}\chi^{M(5,3)}_{-\frac{1}{20}} -\chi_{\frac{1}{3}}^{\UA}\chi^{M(5,3)}_{\frac34},\\
    \chi_{\frac{1}{3}}^{U_1}&=\chi_{\frac{1}{3}}^{\UA}\chi^{M(5,3)}_{0} -\chi_{\frac{2}{15}}^{\UA}\chi^{M(5,3)}_{\frac15},\\
    \chi_{\frac{3}{4}}^{U_1}&=\chi_{\frac{4}{5}}^{\UA}\chi^{M(5,3)}_{-\frac{1}{20}} -\chi_{0}^{\UA}\chi^{M(5,3)}_{\frac{3}{4}}.
\end{align}

Finally, we do not find a proper construction for $L_k(\UA)$ beyond level $1$.

\subsection{$IM$}\label{sec:IM}
The formal algebra $IM$ is intermediate between $0\subset IM\subset A_1$. It has a formal dual Coxeter number $2/3$ and dimension $1$. It is better understood from its level-one affine VOA $M_{\rm eff}(5,3)$ which has a well-known coset construction
\begin{align}
    M_{\rm eff}(5,3)=\frac{L_1(A_1)}{M_{\rm eff}(5,2)},
\end{align}
as required by the relation \eqref{cosetLY} between intermediate  and Cvitanovi\'c--Deligne exceptional series. We define the affine VOA associated to $IM$ at level $k$ as effective minimal model $ M_{\rm eff}(3k+2,3).$ 
Recall effective minimal model $M(p,q)$ has effective central charge $1-6/(pq)$. Thus the central charge as a formal WZW model  $k/(k+\frac23)$ is consistent with the effective central charge of minimal model $ M_{\rm eff}(3k+2,3).$ 

The level $2$ case, i.e., $M_{\rm eff}(8,3)$ with effective central charge $3/4$ is a renowned minimal model with simple current extension to a $N=1$ SVOA. To be precise, it can be fermionized to the $N=1$ supersymmetric minimal model $SM_{\rm eff}(8,2)$, which is the simplest non-unitary supersymmetric minimal model and often called the supersymmetric Lee-Yang model ($SLY$). It has two Neveu-Schwarz primaries with NS weight $0,1/4$, and the NS characters are 
\begin{align}
    \chi_{\mathrm{NS},0}^{SLY}&=q^{-\frac{1}{32}} \prod _{n=0}^{\infty} \frac{1}{\big(1-q^{\frac{1}{2} (8 n+1)}\big) \big(1-q^{\frac{1}{2} (8 n+4)}\big)\big(1-q^{\frac{1}{2} (8 n+7)}\big)},\\
 \chi_{\mathrm{NS},\frac{1}{4}}^{SLY}&=q^{\frac{7}{32}} \prod _{n=0}^{\infty} \frac{1}{\big(1-q^{\frac{1}{2} (8 n+3)}\big) \big(1-q^{\frac{1}{2} (8 n+4)}\big) \big(1-q^{\frac{1}{2} (8 n+5)}\big)}.
\end{align}

We notice that $L_2(IM)$ admits similar coset constructions as $L_2(\Dh)$ resembling the coset \eqref{cosetA}. For example consider the coset $L_2(A_1)/L_2(IM)$. The $L_2(A_1)$ has central charge $3/2$ and a well-known fermionization to a tensor product of three free fermions, with the following character relations
\begin{align}
    \chi_{\mathrm{NS},0}^{F^3}=\big(\chi_{\mathrm{NS},0}^{F}\big)^3=\chi^{L_2(A_1)}_0+\chi^{L_2(A_1)}_{\frac{1}{2}} ,\quad \textrm{where}\quad \chi_{\mathrm{NS},0}^{F}=\sqrt{\theta_3/\eta}.
\end{align}
Then using the notattion of \cite{Lee:2022yic}, we find the following super coset
\begin{align}
   \frac{\cF(A_1)_2}{\cF(IM)_2}= \frac{F^3}{SM_{\rm eff}(8,2)}=SM_{\rm eff}(8,2).
\end{align}
We checked the precise NS character relation for this coset is
\begin{align}
\chi_{\mathrm{NS},0}^{F^3}=\big(\chi_{\mathrm{NS},0}^{SLY}\big)^2+\big(\chi_{\mathrm{NS},\frac{1}{4}}^{SLY}\big)^2. 
\end{align}

It is also interesting to consider $L_1(IM)^{\otimes 2}/L_2(IM)$ resembling the coset \eqref{cosetB}. Here we just point out that $L_1(IM)^{\otimes 2}$, i.e., $M_{\rm eff}(5,3)^{\otimes 2}$ allows fermionization to a subtheory of supersymmetric minimal model $SM_{\rm eff}(20,2)$. This subtheory, which is  a D-type non-diagonal $\Gamma_\theta$ modular invariant of $SM_{\rm eff}(20,2)$ was constructed in \cite[Section 5.5]{Lee:2022yic} and denoted as $SM_{\rm sub}(20,2)$. It has Neveu-Schwarz weights with degeneracy $0,\frac{1}{10},(\frac{3}{10})_2$. The precise relation between the NS characters of the subtheory and those of $SM_{\rm eff}(20,2)$ can be found in \cite[Equation (5.11)]{Lee:2022yic}. The fermionization of $M_{\rm eff}(5,3)^{\otimes 2}$ to $SM_{\rm sub}(20,2)$ is established by the following character relations
\begin{align}
    \chi_{\mathrm{NS},0}^{SM_{\rm sub}(20,2)} & = (\chi_0^{IM})^2+(\chi_{\frac14}^{IM})^2 ,\\
     \chi_{\mathrm{NS},\frac{1}{10}}^{SM_{\rm sub}(20,2)} & = (\chi_{\frac{1}{20}}^{IM})^2+(\chi_{\frac45}^{IM})^2 ,\\
      \chi_{\mathrm{NS},\frac{3}{10}}^{SM_{\rm sub}(20,2)} & = \chi_0^{IM}\chi_{\frac45}^{IM}+ \chi_{\frac{1}{20}}^{IM}\chi_{\frac14}^{IM}.
\end{align}
Here $\chi_{h}^{IM}$ is the character of $M_{\rm eff}(5,3)$ with $h=0,\frac{1}{20},\frac{1}{4},\frac45$. Similar relations can be established for the Ramond characters of ${SM_{\rm sub}(20,2)}$.

\section{Theta blocks of critical weight}\label{sec:theta-blocks}
As natural extensions of finite-dimensional semi-simple Lie algebras, affine Kac--Moody algebras have elegant properties. For example, they generate rational affine vertex operator algebras whose irreducible modules are the integrable highest weight representations indexed by dominant integral weights. The characters of these representations define holomorphic functions of several variables, which have nice modularity and can be calculated by the Kac--Weyl character formulas \cite{KP84}. The common denominator of such character formulas is given by the Macdonald--Weyl denominator identity \cite{Mac72}. The infinite sum side of the denominator identity concerns the affine Weyl group, while the infinite product side involves all positive roots. We know from \cite{GSZ19} that these infinite products define holomorphic Jacobi forms of singular weight. Unlike finite-dimensional semi-simple Lie algebras, the vertex operator algebras generated by intermediate Lie algebras seem to be irrational in high level. Therefore, the characters of the integrable highest weight representations of the affine intermediate Lie algebras (if they exist) are not always modular forms. However, the intermediate Lie algebras still yield exceptional Jacobi forms following the philosophy of denominator identities. Let’s explain it in detail below.

We first review Jacobi forms having infinite product expansions. Let $h$ be a positive rational number and $L$ be an integral positive-definite lattice with bilinear form $\latt{-,-}$. Let $D$ be a positive integer and $\alpha_1$, $\alpha_2$, ..., $\alpha_r$ be vectors of $L\otimes\QQ$ which satisfy 
\begin{equation}\label{eq:dual-Coxeter}
\sum_{j=1}^r \latt{\alpha_j, \mathfrak{z}}^2 = h \latt{\mathfrak{z}, \mathfrak{z}}
\end{equation}
for any $\mathfrak{z}\in L\otimes\CC$. Following Gritsenko--Skoruppa--Zagier \cite{GSZ19} we introduce the function
\begin{equation}\label{eq:theta-block-alpha}
\vartheta_{\alpha}(\tau,\mathfrak{z})=\eta(\tau)^D\prod_{j=1}^r\frac{\vartheta(\tau, \latt{\alpha_j,\mathfrak{z}})}{\eta(\tau)^3},
\end{equation}
where $\eta$ and $\vartheta$ are respectively the Dedekind eta function and the odd Jacobi theta function:
\begin{align*}
\eta(\tau)&=q^{\frac{1}{24}}\prod_{n=1}^\infty(1-q^n), \quad \tau\in \HH, \; q=e^{2\pi i\tau},\\
\vartheta(\tau,z)&=-q^{\frac{1}{8}}\zeta^{-\frac{1}{2}}\prod_{n=1}^\infty(1-q^{n-1}\zeta)(1-q^n\zeta^{-1})(1-q^n), \quad z\in \CC, \; \zeta= e^{2\pi iz}.
\end{align*}
Let $K$ be the integral lattice
$$
\{ x\in L\otimes \QQ : \latt{x,\alpha_j} \in \ZZ \quad \text{for $1\leq j \leq r$} \}
$$
equipped with the bilinear form
$$
(x,x):=h\latt{x,x}=\latt{\alpha_1,x}^2+\latt{\alpha_2,x}^2+\cdots +\latt{\alpha_r,x}^2. 
$$
The function $\vartheta_{\alpha}$ satisfies the functional equations
\begin{align*}
\vartheta_{\alpha} \left( \frac{a\tau +b}{c\tau + d},\frac{\mathfrak{z}}{c\tau + d} 
\right)& = \upsilon_\eta(A)^D (c\tau + d)^{D/2-r} 
\exp{\left(i \pi \frac{c(\mathfrak{z},\mathfrak{z})}{c 
\tau + d}\right)} \vartheta_{\alpha} ( \tau, \mathfrak{z} ),\\
\vartheta_{\alpha} (\tau, \mathfrak{z}+ x \tau + y)&= 
(-1)^{(x,x)+(y,y)}\exp{\Big(-i \pi \big( (x,x)\tau +2(x,\mathfrak{z})\big)\Big)} 
\vartheta_{\alpha} (\tau, \mathfrak{z} ),
\end{align*}
for $A=\begin{psmallmatrix}
a & b \\
c & d
\end{psmallmatrix} \in \mathrm{SL}_2(\ZZ)$ and $x,y\in K$, 
and it has the Fourier expansion
\begin{equation*}
\vartheta_{\alpha}( \tau, \mathfrak{z} )=\sum_{ \substack{n\in \frac{D}{24} + \NN \\ \ell \in K^\bullet } }f(n,\ell)q^n\zeta^\ell, \quad \zeta^\ell=e^{2\pi i (\ell,
\mathfrak{z})},
\end{equation*}
where $v_\eta$ is the multiplier system of $\eta$ and $K^\bullet$ is the shadow of $K$ defined as
$$
K^\bullet = \{ x \in K\otimes\QQ : (x,y) - (y,y)/2 \in \ZZ \quad \text{for all $y\in K$} \}.
$$
In other words, $\vartheta_{\alpha}$ is a weak Jacobi form of weight $D/2-r$ and lattice index $K$ \cite{EZ85, GSZ19}. This type of Jacobi form is called a \textit{theta block}. 

It is a non-trivial question in which condition $\vartheta_{\alpha}$ is a \textit{holomorphic} Jacobi form, that is, it is holomorphic at infinity. This is equivalent to ask when 
\begin{equation}\label{eq:boundary}
2n\geq (\ell, \ell)    
\end{equation} 
holds for any $n\in \frac{D}{24} + \NN$ and any $\ell\in K^\bullet$ satisfying $f(n,\ell)\neq 0$. By the theta decomposition, if $\vartheta_{\alpha}$ is a holomorphic Jacobi form then 
$D-2r$ is at least the rank of $K$ (note $\rank(K)=\rank(L)$). The possible smallest weight $\frac{1}{2}\rank(K)$ is called the \textit{singular} weight. The second smallest weight $\frac{1}{2}\rank(K)+\frac{1}{2}$ is called the \textit{critical} weight. 

Let $\mathfrak{g}$ be a finite-dimensional simple Lie algebra. We choose $\alpha$ as the positive roots of $\mathfrak{g}$ and fix $D$ as the dimension of $\mathfrak{g}$. Then these $\alpha$ satisfy an identity of type \eqref{eq:dual-Coxeter} with $h$ equal to the dual Coxeter number of $\mathfrak{g}$, and $\rank(K)$ equals the rank of $\mathfrak{g}$. By \cite{GSZ19} (or \cite[Section 2.2]{SWW23}), the associated theta block is a holomorphic Jacobi form of singular weight, that is, it satisfies the boundary condition \eqref{eq:boundary} and its weight is $\rank(\mathfrak{g})/2$. Recently, it was proved in \cite{Wan23} that every holomorphic theta block of singular weight can be constructed in this way. Therefore, there is a one-to-one correspondence between finite-dimensional semi-simple Lie algebras and holomorphic theta blocks of singular weight. 

Let $\mathfrak{g}_I$ be an intermediate Lie algebra between $\mathfrak{g}_1$ and $\mathfrak{g}_2$, i.e. $\mathfrak{g}_1 \subset \mathfrak{g}_I \subset \mathfrak{g}_2$. In this case, we take $\alpha$ as the positive roots of $\mathfrak{g}_I$, that is, the positive roots and the weights of $\mathfrak{g}_1$, which correspond to the decomposition of the adjoint representation of $\mathfrak{g}_I$ into the irreducible representations of $\mathfrak{g}_1$ (see \eqref{eq:positive-roots}). We normalize the bilinear form of $\mathfrak{g}_1$ such that it has the dual Coxeter number given by the universal formula for the subexceptional series \eqref{u1DCsub}. Then \eqref{eq:dual-Coxeter} holds for these $\alpha$ and $h$ equal to the dual Coxeter number of $\mathfrak{g}_I$ (i.e. half the sum of the dual Coxeter numbers of $\mathfrak{g}_1$ and $\mathfrak{g}_2$; see Table \ref{tab:ies}). Again, we fix $D$ as the dimension of $\mathfrak{g}_I$. Recall that
$$
\dim(\mathfrak{g}_I) = \#\{ \text{positive roots of $\mathfrak{g}_I$} \} \times 2 + \rank(\mathfrak{g}_1) + 1. 
$$
Thus the associated theta block $\vartheta_\alpha$ is a weak Jacobi form of critical weight, that is, the weight is $\big(\rank(\mathfrak{g}_1)+1\big)/2$. We further propose the following conjecture.
\begin{conj}
The theta blocks of type \eqref{eq:theta-block-alpha} associated with the intermediate Lie algebras in \eqref{eq:ies} are holomorphic Jacobi forms of critical weight and lattice index.
\end{conj}

By the second functional equation, the Fourier coefficients of $\vartheta_\alpha$ satisfy that 
$$
f(n,\ell) = (-1)^{(x,x)} f\big(n+(\ell,x)+(x,x)/2,\ell+x\big)
$$
for all $x\in K$. It follows that up to sign the Fourier coefficients $f(n,\ell)$ depend only on the class of $\ell$ modulo $K$ and on the hyperbolic norm $2n-(\ell,\ell)$. Therefore, in order to prove that $\vartheta_\alpha$ is holomorphic at infinity, it suffices to verify the boundary condition \eqref{eq:boundary} holds for any $n< \delta_K$ and any $\ell\in K^\bullet$ satisfying $f(n,\ell)\neq 0$, where 
\begin{equation}\label{eq:delta}
\delta_K=\max\big\{ \min\{ (v,v)/2: v\in x+K \} \; : \; x\in K^\bullet  \big\}.    
\end{equation}
We can verify this for $\Ah$, $\Ch$, $\AD$, $\AG$ and $\UA$ by calculating the constant $\delta_K$ and the necessary Fourier coefficients. For $\Eh$ and $\Dh$, we have not verified due to limited computational power. We discuss them case by case in the following.
\begin{enumerate}
\item The theta block associated with $\UA$ is just 
\begin{align}
    \eta(\tau) \vartheta(\tau, z).
\end{align}
It is a Jacobi cusp form of weight $1$ and rank-one lattice index $\ZZ$, because it is the tensor product of a Jacobi form of singular weight and index $\ZZ$ and a Jacobi cusp form of weight $1/2$ and index $0$.  

\item The theta block associated with $\AG$ is clearly
\begin{align}
  \frac{\vartheta(\tau,2z)\vartheta(\tau,z)\vartheta(\tau,3z)}{\eta(\tau)}.
\end{align}
This is a pullback of the singular-weight theta block associated with $A_2$:
\begin{align}
\frac{\vartheta(\tau,z_1)\vartheta(\tau,z_2)\vartheta(\tau,z_1+z_2)}{\eta(\tau)}.
\end{align}
It follows that the theta block associated with $\AG$ is a Jacobi cusp form of weight $1$ and rank-one lattice index $\ZZ(14)$. 

\item The theta block associated with $\AD$ has the form
\begin{equation}
\eta^{-3}\vartheta(2z_1)\vartheta(2z_2)\vartheta(2z_3)\vartheta(z_1+z_2+z_3)\vartheta(z_1+z_2-z_3)\vartheta(z_1-z_2+z_3)\vartheta(z_1-z_2-z_3),   
\end{equation}
where $(z_1,z_2,z_3)\in \CC^3$. This theta block is a holomorphic Jacobi form of weight $2$ and rank-three lattice index $\ZZ^3(8)$, and it turns out to be reducible. In fact, it can be viewed as the product of the singular-weight theta blocks associated with $A_3$ and $A_1$. This is not a cusp form. 

\item The theta block associated with $\Ch$ has the form
\begin{equation}
\begin{split}
&\eta^{-12}\vartheta(2z_1)\vartheta(2z_2)\vartheta(2z_3)\vartheta(z_1)\vartheta(z_2)\vartheta(z_3)\vartheta(z_1+z_2)\vartheta(z_1+z_3)\vartheta(z_2+z_3)\vartheta(z_1-z_2)\\
&\cdot\vartheta(z_1-z_3)\vartheta(z_2-z_3)\vartheta(z_1+z_2+z_3)\vartheta(z_1+z_2-z_3)\vartheta(z_1-z_2+z_3)\vartheta(z_1-z_2-z_3),     
\end{split}  
\end{equation}
where $(z_1,z_2,z_3)\in \CC^3$. This is a Jacobi form of weight $2$ and rank-three lattice index $\ZZ^3(13)$. The corresponding constant $\delta_K$ is $39/8$. To verify that the theta block is holomorphic at infinity, we only need to check the boundary condition for $n<39/8$. From the leading Fourier expansion below 
\begin{equation}
\begin{split}
&q^{\frac{3}{2}}\Big( Z_{|w|\le \frac{155}{52}}+Z_{  |w|\le\frac{259}{52}  }q+Z_{ |w|\le \frac{363}{52}  }q^2+Z_{  |w|\le \frac{467}{52}  }q^3\\
&\, +Z_{  |w|\le  \frac{571}{52} }q^4+Z_{  |w|\le  \frac{675}{52}  }q^5+Z_{  |w|\le  \frac{779}{52} }q^6+\mathcal{O}(q^7)\Big)   , 
\end{split}    
\end{equation}
we conclude that the theta block associated with $\Ch$ is a holomorphic Jacobi form. Moreover, it is a Jacobi cusp form. In the Fourier expansion above, $Z_{|w|\le t}$ stand for non-zero Fourier coefficients $f(n,\ell)\zeta^\ell$ with $(\ell,\ell)\leq t$.

\item In the case of $\Ah$, we have the embedding $A_5(9)<K$, which yields that $\delta_K\leq 27/4$.  By direct computation, the associated theta block has the leading Fourier expansion
\begin{equation}
\begin{split}
&q^{\frac{7}{3}}\Big( Z_{\frac{70}{36}\le|w|\le \frac{168}{36}}+Z_{\frac{112}{36}\le |w|\le \frac{240}{36}}q+Z_{\frac{70}{36}\le |w|\le \frac{310}{36}}q^2+Z_{\frac{184}{36}\le |w|\le \frac{384}{36}}q^3\\
&\,+Z_{\frac{118}{36}\le |w|\le \frac{456}{36}}q^4 + Z_{\frac{136}{36}\le |w|\le \frac{528}{36} }q^5+Z_{\frac{150}{36}\le |w|\le \frac{600}{36}}q^6+\mathcal{O}(q^7)\Big).    
\end{split}    
\end{equation}
We then confirm that the theta block associated with $\Ah$ is a holomorphic Jacobi form of weight $3$ and rank-five lattice index $A_5(9)$. Remark that this theta block is of type $\vartheta^{25}/\eta^{19}$ and it is not a cusp form. 

\item The constant $\delta_K$ associated with $\Dh$ is $21/2$. The theta block associated with $\Dh$ is of weight $7/2$ and of type $\vartheta^{46}/\eta^{39}$. To verify that it is holomorphic at infinity, we have to calculate its Fourier expansion up to $q^{21/2}$.  

\item The constant $\delta_K$ associated with $\Eh$ is $18$. The theta block associated with $\Eh$ is of weight $4$ and of type $\vartheta^{91}/\eta^{83}$. To verify that it is holomorphic at infinity, we have to calculate its Fourier expansion up to $q^{18}$.  
\end{enumerate}

It is well-known that holomorphic Jacobi forms of singular weight and lattice index $K$ are $\CC$-linear combinations of Jacobi theta functions associated with $K$. According to \cite[Theorem 2.8]{BS23}, holomoprhic Jacobi forms of critical weight and index $K$ can be constructed as the pullback of singular-weight holomorphic Jacobi forms indexed by lattices of type $K\oplus \ZZ(N)$. Obviously, both the pullback of singular-weight holomorphic theta blocks and the product of $\eta$ and singular-weight holomorphic theta blocks will induce holomorphic theta blocks of critical weight, such as the theta blocks (1)-(3) above. The holomorphic theta blocks of critical weight which cannot be constructed by either of the two ways are called \textit{non-trivial}. Note that the theta blocks (4)-(7) above are non-trivial. 

Following the same philosophy, we study the denominator of some other (formal) intermediate Lie algebras by adding roots. We check that the theta block associated with $D_{3+1/2}$ is not holomorphic at infinity. Note that $A_{3+1/2}$ is just $B_3$ if we regard the decomposition $\bf 21=15+6$, its theta block is the same as the one of $B_3$, thus holomorphic. The $D_{4+1/2}$ case is just $B_4$ up to an $\mathbb{Z}_3$ automorphism if we regard $\bf 36=28+8$, thus its theta block is holormophic. The $D_{5+1/2}$ case naively has $\bf 61 = 45 +\overline{16} $, we find that its theta block is not holomorphic, even the leading order has norms much bigger than the holomorphic bound. This shows the speciality of $\Dh$. 

By the discussion above, non-trivial holomorphic theta blocks of critical weight are very exceptional, and may be closely related to intermediate Lie algebras. It is an interesting question to extend the previous correspondence between finite-dimensional semi-simple Lie algebras and holomorphic theta blocks of singular weight in this context. It is also interesting to check if the theta blocks associated with $\Dhh$ and $X_1$ in Appendix \ref{app:4} are holomorphic Jacobi forms. It is also a challenging question to determine the infinite sum expansion of these theta blocks in the context of denominators of affine intermediate Lie algebras.

\section{Appendix I}\label{app:I}

\subsection{An exotic case: $\Dhh$} \label{app:4}
We denote the exotic algebra in the extended Freudenthal--Tits magic square $\mathfrak{m}(\mathbb{S},\mathbb{S})$ as $\Dhh$. This algebra has Landsberg--Manivel's parameter $a=6$, dual Coxeter number $h^\vee=19$ and dimension $144$, and does not lie on the Vogel's plane.  
It fills a hole between $\Dh$ and $\Eh$ in the intermediate exceptional series. $\Dhh$ has two conjugated modules $\bf 45$ and $\bf \overline{45}$. It also has a module $\bf 12$ which directly comes from the module $\bf 12$ of $\Dh$ and the vector module $\bf 12$ of $D_6$. Let us first consider $  D_{6+1/2} \subset \Dhh$. It is easy to find the module decompositions
\begin{align}
   \bf 144&=\bf99+44+1,\\
   \bf 45&=\bf 33 +12\\
   \bf \overline{45}&=\bf 44+1. 
\end{align}
The first decomposition also gives a way to compute the dual Coxeter number. Recall in $D_6\subset \Dh$, we have module decomposition $\bf 44=\bf 32+12$. Then
\begin{align}
    h^\vee=h^\vee_{\Dh}+\frac{1}{2r} \sum_{w\in \mathbf{  {32}}}\langle w,w\rangle_{D_6}+\frac{1}{2r}\sum_{w\in \mathbf{  {12}}}\langle w,w\rangle_{D_6} =14+4+1=19.
\end{align}
Here the rank $r$ of $\Dhh$ is still $6$. For $\Dhh\subset \Eh$, we have the module decompositions
\begin{align}
   \bf 190&=\bf144+\overline{45}+1,\\
   \bf 57&=\bf 45 +12 . 
\end{align}
We do not find more irreducible modules for $\Dhh$. The previous analogy of the Weyl dimension formula does not work in this case.

Consider the affine VOA associated with $D_{6+1/2+1/2}$ at level $1$. The central charge $c={36}/{5}$. The characters of the irreducible modules are
\begin{equation}\label{eq:chiDhhlevel1}
    \begin{aligned}
\chi_0=\,&q^{-\frac{3}{10}} (  1+144 q+1926 q^2+14160 q^3+77499 q^4 + \dots) ,\\
\chi_{{3}/{5}}=\,&q^{\frac{3}{10}} (12+297 q+2520 q^2+15478 q^3+74340 q^4+ \dots),\\
\chi_{{4}/{5}}=\,&q^{\frac{1}{2}} ( 45+760 q+6030 q^2+34740 q^3+161670 q^4+  \dots).
    \end{aligned}
\end{equation}
This solution of holomorphic MLDE of degree $3$ was called $\bf III_4$ in \cite{Das:2022uoe} and conjectured to be the characters of an intermediate VOA. The module with conformal weight $4/5$ has degeneracy two. This suggests that the Dynkin diagram of $\Dhh$ has a $\ZZ_2$ automorphism, which is also one reason we use $+1/2+1/2$ in the subscript. The above characters of $L_1(\Dhh)$ can be written explicitly  in terms of Roger-Ramanujan functions \eqref{RRf}  as
\begin{equation}\label{eq:chiDhhlevel1exact}
    \begin{aligned}
\chi_0=\,& \phi_0^{18}+126\phi_0^{13}\phi_1^5+117\phi_0^8\phi_1^{10}-12\phi_0^3\phi_1^{15} ,\\
\chi_{{3}/{5}}=\,& 12\phi_0^{15}\phi_1^3+117\phi_0^{10}\phi_1^8-126 \phi_0^5\phi_1^{13}+  \phi_1^{18},\\
\chi_{{4}/{5}}=\,&45\phi_0^{14}\phi_1^4+130\phi_0^{9}\phi_1^9-45\phi_0^4\phi_1^{14}  .
    \end{aligned}
\end{equation}
One could get the $L_1(\Dhh)$ as a coset of $L_1(E_{7+1/2})$ by $M(5,2)_{\rm eff}$ and the product of $L_1(D_{6+1/2})$ with $M(5,3)_{\rm eff}$,
\begin{equation}
   L_1(D_{6+1/2+1/2})= \frac{L_1(E_{7+1/2})}{M_{\rm eff}(5,2)}=L_1(D_{6+1/2})\otimes M_{\rm eff}(5,3). 
\end{equation}  
One might suspect that $L_k(D_{6+1/2+1/2})$ is rational if and only if $k=1,2,3,4$, similar with the $\Dh$ and $\Eh$ cases. Unfortunately, we could not find a proper construction for any $L_k(D_{6+1/2+1/2})$ for $k\ge 2$.

\subsection{An exotic case: $X_1$ in Vogel's projective plane}\label{app:5}
The $X_1$ algebra in Vogel's plane has dimension $156$ and dual Coxeter number $20$. The Vogel's triple $(\alpha,\beta,\gamma)=(-2,8,14)$. This exotic algebra has been investigated in e.g. \cite{Mkrtchyan:2012es}. This algebra is not related to the magic square or the magic triangle. Nevertheless, we observe that it is related to a mysterious solution of third order holomorphic MLDEs.

This following known vector-valued modular form of degree three has all positive Fourier coefficients:
\begin{align}\label{eq:chiX1}
    \chi_0 & =q^{-13/42}(1+156 q+2236 q^2+17056 q^3+96577 q^4+\dots ),\\
    \chi_{4/7} &= q^{11/42}( 13+364 q+3302 q^2+21112 q^3+105144 q^4+\dots  ),\\
    \chi_{6/7} &= q^{23/42}(  78+1288 q+10465 q^2+61308 q^3+290992 q^4+\dots ). 
\end{align}
This solution of holomorphic MLDE was called $\bf III_5$ in \cite{Das:2022uoe} and conjectured to be the characters of an intermediate VOA. We make the observation that this vector-valued modular form can be exactly regarded as the characters of all three irreducible modules of affine VOA $L_1(X_1)$. It allows a coset construction
\begin{align}
    L_1(X_1)=\frac{L_1(E_8)}{M_{\rm eff}(7,2)}.
\end{align}
It also has a Hecke operator interpretation \cite{Duan:2022ltz}
\begin{align}
    L_1(X_1)=\T_{13}{M_{\rm eff}(7,2)}.
\end{align}
The Hecke relation allows us to write the characters \eqref{eq:chiX1} of $L_1(X_1)$ as degree $13$ homogeneous polynomials of the $M_{\rm eff}(7,2)$ characters.

It is easy to see $X_1\subset E_8$ from the coset construction and the $E_8$ module decomposes as
\begin{align}
    \bf 248= 156+78+13+1.
\end{align}
One might suspect 
$D_7\subset X_1$ because of the module decompositions
\begin{align}
  \bf 156 & = \bf 91+\overline{64}  +1,\\
  \bf 78 &=\bf 64+ 14 .
\end{align}
However, $X_1$ module $\bf 13$  is not decomposable into the direct sum of $D_7$ modules. We expect that $X_1$ is intermediate between $D_6$ and $E_8$, which is rather different from the intermediate exceptional series we study in this paper.

\bigskip

\noindent
\textbf{Acknowledgements}
KS would like to thank T. Creutzig, C. Dong, Z. Duan, A. Hanany, V. Kac, D. Ridout, W. Yan and J. Yang for discussions and especially T. Arakawa and K. Kawasetsu for the comment on the relation with $W$-algebras.  KS is supported in part by the Olle Engkvists Stiftelse Grant No.2180108. KS also thanks MPIM Bonn and Isaac Newton Institute for Mathematical Sciences in Cambridge (during the program \textit{Black holes: bridges between number theory and holographic quantum information}, by EPSRC grant no EP/R014604/1) for support and hospitality where work on this paper was undertaken. KL is supported in part by KIAS Grants PG006904  and the National Research Foundation of Korea (NRF) Grant funded by the Korea government (MSIT)
(No.2017R1D1A1B06034369). KL also thanks KITP for the program \textit{What is String Theory? Weaving Perspectives Together}. This research was supported in part by grant NSF PHY-2309135 to the Kavli Institute for Theoretical Physics (KITP).

\bibliographystyle{plainnat}
\bibliofont

\end{document}